\documentclass[12pt]{article}
\usepackage{amssymb,epsfig}
\usepackage{amsmath}

\renewcommand{\thefootnote}{\fnsymbol{footnote}}
\newcommand{\beq}{\begin{equation}}
\newcommand{\eeq}{\end{equation}}
\newcommand{\ov}{\overline}
\newcommand{\pa}{\partial}
\newcommand{\al}{\alpha}

\begin{document}

\title{
\begin{flushright}
\ \\*[-80pt] 
\begin{minipage}{0.2\linewidth}
\normalsize
\end{minipage}
\end{flushright}
{\huge \bf 
Supersymmetry breaking, \\ R-symmetry, \\
and conformal dynamics  
\vspace*{7\baselineskip}
\\*[20pt]}}

\author{
{\LARGE
Yuji~Omura} 
\\*[20pt]}

\date{
}
\begin{titlepage}
\maketitle
\thispagestyle{empty}
\end{titlepage}

\newpage

\begin{center}
\centerline{\small \bf Abstract}
\begin{minipage}{0.9\linewidth}
\medskip 
\medskip 
\small
We study $N=1$ global and local supersymmetric theories with a 
continuous global $U(1)_R$ symmetry. We discuss conditions for supersymmetry (SUSY) breaking and vacuum structures of R-symmetric SUSY models.
Especially we find the conditions for R-symmetry breaking and runway vacua in global supersymmetric theories. 
We introduce explicit R-symmetry breaking 
terms into such models in global and local supersymmetric theories.
Such explicit R-symmetry breaking terms can lead 
to a SUSY preserving minimum.
We classify explicit R-symmetry breaking terms 
by the structure of newly appeared SUSY stationary 
points as a consequence of the R-breaking effect, which could make 
the SUSY breaking vacuum metastable. 
Based on the generic argument, we propose the scenario that conformal dynamics causes approximate R-symmetry and metastable SUSY breaking vacua.
At a high energy scale, the superpotential in our model is not R-symmetric, 
and has a supersymmetric minimum.
However, conformal dynamics suppresses several operators 
along the renormalization group (RG) flow toward the infrared fixed point.
Then we can find an approximately R-symmetric superpotential, 
which has a metastable SUSY breaking vacuum, 
and the supersymmetric vacuum moves far away from 
the metastable supersymmetry breaking vacuum.
Furthermore, we find that conformal dynamics also leads approximate R-symmetry in softly SUSY breaking theories, even in more complicated models such as the duality cascade. We investigate the RG flow of 
SUSY breaking terms as well as supersymmetric couplings in the duality cascade of softly broken supersymmetric theories.
It is found that the magnitudes of SUSY breaking terms are 
suppressed in most regimes of the RG flow through 
the duality cascade and approximate R-symmetry is realized at a low energy scale. We also show the possibility that cascading would be terminated 
by the gauge symmetry breaking, which is induced by 
the so-called B-term. Finally, we find some models to arrive at standard-model-like models and to cause gauge symmetry breaking corresponding to electro-weak symmetry breaking.
\end{minipage}
\end{center}
\newpage

\tableofcontents
\clearpage

\renewcommand{\thefootnote}{\arabic{footnote}}
\setcounter{footnote}{0}

\section{Introduction}
Supersymmetric extension of the standard model is a promising 
candidate for the physics around TeV scale. Supersymmetry (SUSY) can 
stabilize the huge hierarchy between the weak scale and the 
Planck scale, and supersymmetric models with R-parity have the 
lightest superparticle which is a good candidate for the dark matter. 
In addition, the minimal SUSY standard model realizes 
the unification of three gauge couplings at a scale 
$M_{GUT} \sim 2 \times 10^{16}$ GeV. That may suggest some 
underlying unified structure in the nature.

In our real world, the SUSY must be broken with 
certain amount of the gaugino and scalar masses. The dynamical 
SUSY breaking has a strong predictability of the structure 
of such SUSY particle masses. It was shown by Nelson and 
Seiberg (NS)~\cite{Nelson:1993nf} that a global $U(1)_R$ symmetry is 
necessary for a spontaneous F-term SUSY breaking at 
the ground state of generic models with a global SUSY. 
The spontaneous $U(1)_R$ symmetry breaking predicts an appearance of massless Goldstone mode, R-axion.
\footnote{
See for recent works on R-symmetry breaking, e.g. 
Refs.~\cite{Intriligator:2007py,Shih:2007av,Ferretti:2007ec,Essig:2007xk,
Cho:2007yn,Abel:2007jx,Ray:2007wq,Sun:2008va,Amariti:2008uz,Komargodski:2009jf,Giveon:2008ne,Marques:2008va} 
and references therein.}

Recently, it has been argued by Intriligator, Seiberg and Shih 
(ISS)~\cite{Intriligator:2007py} that the SUSY breaking 
vacuum we are living can be metastable for avoiding the 
light R-axion and also obtaining gaugino masses, 
and that such situation can be realized by a tiny size of 
explicit $U(1)_R$ breaking effects, whose representative 
magnitude is denoted by $\epsilon$. In the limit $\epsilon \rightarrow 0$, there would be no SUSY vacuum.
However, explicit R-symmetry breaking terms with a tiny, but finite size of $\epsilon$ can lead to 
a SUSY minimum. Such newly appeared 
SUSY minimum could be far away from the
SUSY breaking minimum, which is found in 
the R-symmetric model without explicit R-symmetry 
breaking terms.
Furthermore, such R-symmetry breaking terms 
would not have significant effects on the 
original SUSY breaking minimum, because 
R-symmetry breaking terms are tiny.
The distance between the original SUSY breaking 
minimum and the newly appeared SUSY preserving 
minima may be estimated as ${\cal O}(1/\epsilon)$ 
in the field space.
Thus, if R-symmetry breaking terms, i.e., 
the size of $\epsilon$, are sufficiently small, 
the original SUSY breaking minimum would be a 
long-lived metastable vacuum.

On the other hand, an introduction of gravity into 
SUSY theories requires that the SUSY 
must be a local symmetry, i.e., supergravity. In supergravity, 
the structure of the scalar potential receives a gravitational 
correction, and also the background geometry of our spacetime 
is determined by the equation of motion depending upon the 
vacuum energy. 
In the above global SUSY model with metastable SUSY breaking 
vacuum, some fields have large vacuum values at the 
SUSY preserving vacuum.
In such a case, supergravity effects might be sizable.
Another important motivation to consider supergravity is
to realize the almost vanishing vacuum energy.
The global SUSY model always has positive vacuum energy 
at the SUSY breaking minimum.
Supergravity effects could realize almost vanishing 
vacuum energy.

F-flat conditions have supergravity corrections. 
Thus, the supergravity model with global $U(1)_R$ symmetry 
would have aspects different from the global SUSY model.
Furthermore, adding R-symmetry breaking terms 
would have different effects between global and 
local SUSY theories.
In section \ref{sec:Generic arguments about supersymmetry breaking}, we study in detail generic aspects of 
global and local SUSY theories with R-symmetry 
and generic behaviors caused by adding explicit 
R-symmetry breaking terms.
We reconsider the above argument for the 
dynamical SUSY breaking and its metastability by comparing global and local SUSY theories, based on \cite{Abe:2007ax}.

The important keypoint is to realize 
the almost vanishing vacuum energy.
That is impossible in the SUSY breaking vacuum of 
global SUSY models, and that is a challenging 
issue in supergravity models.
The vacuum energy may be tuned to vanish, e.g., by 
the constant superpotential term, which 
is a sizable R-symmetry breaking term.
That would affect all of vacuum structure such as 
metastability of SUSY breaking vacua and 
presence of SUSY preserving vacua.
Here we study this vacuum structure by 
using several concrete models, where we start R-symmetric 
models and add certain classes of R-symmetry breaking 
terms such that the vanishing vacuum energy is 
realized.

In section~\ref{sec:global-SUSY}, we study a generic aspect of R-symmetric models within the framework of global SUSY, such as spontaneous SUSY breaking, R-symmetry breaking, flat directions and runaway directions. In section~\ref{sec:rsugra explicit R-breaking in global SUSY},
we consider the generalized O'Raifeartaigh (OR) 
model~\cite{O'Raifeartaigh:1975pr} following \cite{Intriligator:2007py}.
We introduce 
explicit R-breaking terms into the model and analyze in detail 
the newly appeared SUSY vacua as a consequence of the R-symmetry 
breaking effects. We also examine the stability of the original 
SUSY breaking vacuum under such R-breaking terms. 

In section~\ref{sec:rsugra}, we consider supergravity models with 
R-symmetry. We extend the argument by NS to the local SUSY theories 
and study the supergravity OR model. 
In this section, we also show a special SUSY 
stationary point, which does not obey the NS condition, and the 
associated SUSY breaking vacuum in a certain class of R-symmetric 
supergravity models. We introduce explicit R-breaking terms into 
the supergravity OR model in section~\ref{sec:r-breaking-sugra} 
and classify them.

In section~\ref{sec:typea}, we study the case with R-symmetry breaking 
terms (A-type) which might not cause a metastability of SUSY breaking 
minimum, because corresponding SUSY vacua disappear when we set the 
vacuum energy at the SUSY breaking minimum vanishing. 
On the other hand, in section~\ref{sec:typeb}, we show that another 
class of R-symmetry breaking terms (B-type) makes SUSY breaking 
minimum metastable. 
Sec.~\ref{sec:conclusion rsugra} is summary of section \ref{sec:Generic arguments about supersymmetry breaking}. 
In Appendix~\ref{app:rmass}, we show some general features 
of R-axion masses, and find that the special SUSY 
solution exhibited in section~\ref{sec:rsugra} is at best a 
saddle point solution. 


\vspace*{2\baselineskip}

In section 3, we argue that conformal dynamics
can realize such a metastable SUSY 
breaking vacuum in global SUSY, based on \cite{Abe:2007ki}.
We start with a superpotential without R-symmetry.
However, we assume the conformal dynamics.
Because of that, certain couplings are exponentially 
suppressed.
Then, we could realize an R-symmetric superpotential 
or an approximately R-symmetric superpotential 
with tiny R-symmetry breaking terms.
It would lead to a stable or metastable SUSY breaking vacuum.
We study this scenario by using a simple model.
Also, we study 5D models, which have the same behavior.

The application of conformal dynamics provides several interesting aspects
in supersymmetric models. For example,
contact terms like $\int d^4\theta |X|^2|Q|^2$ 
are suppressed exponentially by conformal dynamics 
in the model that the chiral superfield $X$ 
belongs to the hidden conformal sector and 
the chiral superfield $Q$ belongs not to the 
conformal sector, but to the visible sector.
Such conformal suppression mechanism, i.e. conformal 
sequestering, is quite important to model building for SUSY
breaking \cite{Luty:2001jh,Dine:2004dv,Sundrum:2004un,Ibe:2005pj,
Schmaltz:2006qs,Murayama:2007ge}.
When $X$ contributes to SUSY breaking sizably, 
the above contact terms, in general, induce 
flavor-dependent soft SUSY breaking terms, soft sfermion masses 
and the so-called A-terms, 
and they lead to flavor changing neutral current (FCNC) processes, 
which are strongly constrained by current experiments.
However, conformal sequestering can suppress 
the above contact terms and flavor-dependent 
contributions to soft SUSY breaking terms.
Then, flavor-blind contributions such as 
anomaly mediation \cite{Randall:1998uk} would become dominant.
This situation could be also realized in our models.


\vspace*{2\baselineskip}

As we discuss in section 3, conformal dynamics realizes approximate R-symmetry. On the other hand, even if explicit soft SUSY breaking terms are included, R-symmetry tends to be recovered by conformal dynamics in almost cases. We discuss softly SUSY broken theories in section 4, and study how explicit R-symmetry breaking terms are suppressed by conformal dynamics. For example, gaugino mass is one of explicit R-symmetry breaking and soft SUSY breaking terms. It becomes exponentially suppressed as a gauge coupling approaches an infrared (IR) fixed point. A-terms, which are trilinear couplings of scalar fields, are also suppressed. We find that approximate R-symmetry can be also realized in softly broken SUSY theories by conformal dynamics.

Furthermore, it happens in more complicated models, such as the duality cascade we study in section \ref{sec:duality cascade}, that R-symmetry is recovered approximately at a low-energy scale.  
The duality cascade 
is a successive chain of the Seiberg dualities \cite{Seiberg:1994pq,Intriligator:1995au} from 
the ultraviolet (UV) region to the IR region and
reduces the rank of gauge groups one after another.
This leads to more complicated and 
interesting renormalization group (RG) flows of 
dual field theories. We discuss the duality cascade of softly SUSY broken theories in section \ref{sec:duality cascade}, based on Ref.\cite{Abe:2008sq}.
First, we review the duality cascade and show the unique RG flows, based on \cite{Klebanov:2000hb,Strassler:2005qs}.
After that, we study the RG flows of soft SUSY braking terms by using the spurion method \cite{Yamada:1994id,Hisano:1997ua,Jack:1997pa,Kobayashi:1998jq,
ArkaniHamed:1998kj,Terao:2001jw} and show that 
R-symmetry breaking and SUSY breaking terms are strongly suppressed as gauge couplings and yukawa couplings approach toward IR fixed points. 
However we find that B-term, which is a quadratic term, remains to be a finite value at a low energy scale.

Moreover, several models have been proposed to realize 
supersymmetric standard models (SSM) as well as 
their extensions at the bottom of the cascade 
\cite{Cascales:2005rj,Heckman:2007zp,Franco:2008jc}.
Those models are quite interesting and have opened 
possible candidates for high energy theories.
We consider the model with soft SUSY breaking terms and try to construct models
which possibly become realistic models at the bottom of the cascade. 
In our discussion, we assume that SUSY is softly broken at the beginning of 
the cascade.
Then, we study RG flows of SUSY breaking terms as well as 
supersymmetric couplings. Finally, we find some models to arrive at standard-model-like (SM-like) models and to cause gauge symmetry breaking corresponding to electro-weak (EW) symmetry breaking.

In section \ref{sec:rigid-SUSY}, we review briefly the RG flow of 
supersymmetric couplings in the duality cascade.
In section \ref{sec:SUSY-breaking cascade 2}, we study RG flows of SUSY breaking terms in the duality cascade.
In section \ref{sec:symm-br}, we study symmetry breaking due to the B-term
by using illustrative examples. In section \ref{sec:Illustrating model}, we give a simple example whose fields contents are similar to the minimal supersymmetric standard model (MSSM) or its extensions.
Section \ref{sec:conclusion dual} is conclusion in section 5. Section 6 is devoted to summary.
In Appendix \ref{app:superfield} and \ref{app:spurion}, we give a short introduction of the spurion method and the supergraph formalism \cite{Grisaru:1979}.

\section{Generic arguments about SUSY breaking}
\label{sec:Generic arguments about supersymmetry breaking}

We study generic arguments about vacuum structures based on \cite{Nelson:1993nf} and \cite{Abe:2007ax} in global and local SUSY. 
R-symmetry plays a key role in SUSY breaking. In section \ref{sec:global-SUSY}, we discuss vacuum structures of R-symmetric SUSY models in global SUSY.

\subsection{R-symmetry in global supersymmetric theory}
\label{sec:global-SUSY}

\subsubsection{The Nelson and Seiberg argument}
\label{sec:NS argument}
First, we review briefly the argument by Nelson and 
Seiberg~\cite{Nelson:1993nf} in R-symmetric global SUSY models. 
Let us consider the global SUSY model with 
$n$ chiral superfields $Q_I$ ($I=1,\ldots,n$) and 
their superpotential $W(Q_I)$.
In the case of global SUSY, F-flat conditions are 
determined by 
\begin{eqnarray}
W_{I} &=& 0, 
\label{eq:fflat}
\end{eqnarray}
where $W_{I}=\partial_{Q_I}W$.
Hereafter we use a similar notation for derivatives 
of functions $H(X)$ by fields $X$ as $H_X$.
The conditions (\ref{eq:fflat}) are $n$ complex equations for 
$n$ complex variables, and these can have a solution for generic 
superpotential. 

Now, we consider global SUSY models with a continuous global 
$U(1)_R$ symmetry and a nonvanishing superpotential.
Since the superpotential has the R-charge $2$, 
there exists at least one field with a nonvanishing R-charge.  
Suppose that the $n$-th field $Q_n$ is such a field with 
the nonvanishing R-charge, $q_{n} \ne 0$. 
Then, in the following field basis 
\begin{eqnarray}
\chi_i &=& \frac{Q_i}{Q_n^{q_{i}/q_{n}}}, 
\qquad (q_{\chi_i}=0), 
\nonumber \\
Y &=& Q_n, 
\qquad (q_Y=q_{n}\ne 0), 
\label{eq:rbasis1}
\end{eqnarray}
where $i=1,2,\ldots,n-1$, 
the superpotential can be written as 
\begin{eqnarray}
W_{NS} &=& Y^{2/q_Y}\zeta(\chi_i). 
\label{eq:rsp1}
\end{eqnarray}
Then the F-flat conditions (\ref{eq:fflat}) become 
\begin{eqnarray}
(2/q_Y)Y^{2/q_Y-1}\zeta(\chi_i) &=& 0, 
\label{eq:fflatx} \\
Y^{2/q_Y} \partial_{\chi_j} \zeta(\chi_i) &=& 0. 
\label{eq:fflatphi}
\end{eqnarray}
When we look for an R-symmetry breaking vacuum, 
$\langle Y \rangle \ne 0$, 
these conditions are equivalent to 
\begin{eqnarray}
\zeta(\chi_i) &=& 0, \qquad 
\partial_{\chi_j} \zeta(\chi_i) \ = \ 0. 
\label{eq:globalsusyvc}
\end{eqnarray}
These are $n$ complex equations for $n-1$ complex variables, 
that is, these are {\it over-constrained} conditions. 
These cannot be satisfied at the same time for a generic 
function $\zeta(\chi_i)$, and the SUSY can be broken. 
This is an observation by Nelson and Seiberg~\cite{Nelson:1993nf} 
that the existence of an R-symmetry is the necessary condition 
for a dynamical SUSY breaking, and is also the sufficient condition 
if the R-symmetry is spontaneously broken, $\langle Y \rangle \ne 0$.

However, R-symmetry is often unbroken, because 
the scalar potential, which 
is obtained from the superpotential (\ref{eq:rsp1}) and  
the K\"ahler potential $K(|Y|,\chi_i,\bar\chi_i) $, 
is found to have the global minimum at $Y=0$, unless 
the K\"ahler potential $K(|Y|,\chi_i,\bar\chi_i) $ is 
non-trivial.
Thus, SUSY is not broken dynamically with 
the NS superpotential (\ref{eq:rsp1}).

In the following section, we find out the condition for SUSY breaking and R-symmetry breaking concretely.
We discuss models with chiral superfields by $Q_I$ and their R-charges are denoted by $R[Q_I]=q_I$. Furthermore, we have $R[W]=2$ and $R[W_I]=2-q_I$. All of chiral superfields are classified by their R-charges into three classes, $X_a~(a=1, \dots , N_X)$, $\phi_{\al}~(\al=1, \dots ,N_{\phi})$, and $\Phi_{i}~(i=1, \dots ,N_{\Phi})$. R-charges of $X_a,~\phi_{\al}$, and $\Phi_{i}$ are given as $R[X_a]=2$, $R[\phi_{\al}]=0$, and $R[\Phi_{i}] \neq 0,2$, respectively. These are shown in the following table.

\begin{center}
\begin{tabular}{|c|c|c|}  \hline
chiral fields ($Q_I$) & R-charge ($R[Q_I]=q_I$) & The number ($N[Q_I]$) \\ \hline
$X_a$ & $R[X_a]=2$ & $N[X_a]=N_X$ \\ \hline
$\phi_{\al}$ & $R[\phi_{\al}]=0$ & $N[\phi_{\al}]=N_{\phi}$ \\ \hline
$\Phi_{i}$ & $R[\Phi_{i}] \neq 0,2$ & $N[\Phi_i]=N_{\Phi}$ \\ \hline
\end{tabular}
\label{notation}
\end{center}

\subsubsection{R-symmetric vacua}
We find that spontaneous R-symmetry breaking is the sufficient condition for SUSY breaking in the last subsection. 
However, this does not mean R-symmetric SUSY breaking vacua cannot exist. In this subsection, we look for R-symmetric vacua. 
If R-symmetric vacua exist, the vacuum expectation 
values (VEVs) of R-charged fields must vanish. This means that the following condition
\beq
<X_a>=<\Phi_i>=0
\eeq
must be satisfied and $W_I$ must also vanish except for $W_{X_a}$,
\beq
W_{\Phi_i}=W_{\phi_{\al}}=0.
\eeq
If the R-symmetric vacua are supersymmetric, F-flat conditions for $X_a$, which depend on only $\phi_i$, should be satisfied, i.e.,
\beq
W_{X_a}(\phi_{\al},X_a,\Phi_i) \Big|_{X_a=\Phi_i =0}=0.
\eeq
These are $N_X$ equations with $N_{\phi}$ variables. In the case that the number of equations $N_X$ is less than the number of variables $N_{\phi}$, the F-flat conditions can be solved generally. 
Based on the Nelson-Seiberg argument and the above result, the classification in the following table can be realized.\footnote{In subsection \ref{sec:runaway}, we discuss models with runaway directions where the relation $N_X > N_{\phi}$ is satisfied but SUSY is restored by the limit that a R-charged field $X_a$ or $\Phi$ goes to infinite.} 

\begin{center}
\begin{tabular}{|c|c|}  \hline
$N_X \leq N_{\phi} $ & $N_X > N_{\phi}$ \\ \hline
R-symmetric SUSY vacua exist. & SUSY is always broken. \\ \hline
\end{tabular}
\end{center}

\subsubsection{R-symmetry breaking vacua}
In this subsection, we look for the sufficient condition for R-symmetry breaking. 
 In this case, the VEV of at least one of $X_a,\Phi_i$ is nonzero, so we define $Y$ which is one of $X_a,\Phi_i$ with nonzero VEV. 
Under this assumption, all $W_I$ are described as 
\beq
W_I=Y^{\frac{2-q_I}{q}}f_I(Q_I/Y^{\frac{q_I}{q}}),
\eeq
where we define $q=R[Y]$ and each $f_I$ is a function which depends on $(N_X+N_{\Phi}+N_{\phi}-1)$ variables. 
We define $Y$ as $Q_1$ $(q=q_1)$, and define $z_J$ as $Q_J/Y^{ \frac{q_J}{q} }$ for $J=2, \dots ,N$ where $N$ satisfies $N=N_X+N_{\Phi}+N_{\phi}$.

In the following arguments, we assume that K\"ahler potential is canonical,
\beq
K=\sum_{I=1}^N |Q_I|^2.
\eeq
Then the scalar potential is written by
\beq
V=\sum_{I=1}^N |W_I|^2.
\eeq
Under the assumption and conditions, this scalar potential is described as
\beq
\begin{split}
V= &\sum_{I=1}^N |Y|^{\frac{2(2-q_I)}{q}}|f_I(z_J)|^2 \\
  = &\sum_{\al=1}^{N_{\phi}} |Y|^{\frac{4}{q}} |f_{\phi_{\al}}|^2+\sum_{i=1}^{N_{\Phi}} |Y|^{\frac{2(2-q_{\Phi_i})}{q}} |f_{\Phi_{i}}|^2+\sum_{a=1}^{N_{X}}  |f_{X_a}|^2.
\end{split}
\label{eq:redefinition}
\eeq
The scalar potential $V$ is monotonous about $|Y|$ or $1/|Y|$, as long as $\Phi_i$ do not include fields with $2-q_{\Phi_i}<0$, because all $f_I$ do not depend on $Y$. \footnote{ The phase direction of $Y$ is the Goldstone mode of $U(1)_R$ symmetry breaking.}
This leads classifications of R-symmetric models based on the assignments of fields.
We define $\tilde{Q}_I$ and $\Hat{Q}_I$, such that
$\tilde{Q}_I$ are the fields which couple with $X_a$, and 
$\Hat{Q}_I$ are the R-charged fields which couple with $X_a$.\footnote{$\tilde{Q}_I$ and $\Hat{Q}_I$ include $X_a$, and $\Hat{Q}_I$ do not include $\phi_{\al}$.}

For example, we consider the case that all R-charges of $\Phi_i$ are less than $2$.  
In this case, the potential is described as
\beq
V= \sum_{I=1}^N |Y|^{n_I} |f_I (z_J)|^2,
\eeq
where $f_I$ describe $f_{\phi_{\al}}$, $f_{\Phi_i}$, and $f_{X_a}$. 
In the case that $q$ is positive (negative), each $n_I$ is positive (negative) or vanishing.  
If R-symmetry breaking vacua exist, the stationary condition for $Y$ must be satisfied, 
\beq
\frac{\pa V}{ \pa |Y|}=\sum_{\al=1}^{N_{\phi}} \frac{4}{q} |Y|^{\frac{4}{q}-1} |f_{\phi_{\al}}|^2+\sum_{i=1}^{N_{\Phi}} \frac{4-2q_{\Phi_i}}{q}|Y|^{\frac{2(2-q_{\Phi_i})}{q}-1} |f_{\Phi_{i}}|^2=0.
\eeq 
We assume that $Y$ is non-vanishing, so that the unique solution is 
\beq
f_{\phi_{\al}}(z_J)=f_{\Phi_i}(z_J)=0 ~\textrm{for~all}~\phi_{\al},~\Phi_i.
\label{eq:R-breaking}
\eeq
This means that the F-flat conditions for all fields except for $X_a$ should be satisfied.
The F-flat conditions are $(N-N_X)$ equations,
\beq
W_{\phi_{\al}}=W_{\Phi_i}=0~\textrm{for~all}~\al,i.
\label{eq:F-flat 0}
\eeq
If we can find the solutions of these equations, the scalar potential along the slice $W_{\phi_{\al}}=W_{\Phi_i}=0$ is described as 
\beq
V(\tilde{z}_J)=\sum_{a=1}^{N_X}|f_{X_a}|^2,
\eeq
where $\tilde{z}_J$ is defined as $\tilde{z}_J \equiv \tilde{Q}_J/ Y^{\frac{q_{J}}{q}}$.
The fields $\tilde{z}_J$ must satisfy the $N[\tilde{z}]$ stationary conditions of this potential, where $N[\tilde{z}]$ stands for the number of $\tilde{z}_J$.
In order that all equations can be solved generally, the number of equations must not be larger than the number of the variables. 
The number of the equations is $(N-N_X+N[\tilde{z}])$ and there are $(N-1)$ complex variables. Eventually, we find the condition for R-symmetry breaking as $(N-N_X+N[\tilde{z}]) <  N$, i.e.,
\beq
0 < N_X -N[\tilde{z}].
\label{eq:R-breaking condition}
\eeq 
When this relation is satisfied, R-symmetry can be spontaneously broken and $(N_X-N[\tilde{z}])$ complex fields are flat directions.
The relation (\ref{eq:R-breaking condition}) corresponds to the condition that F-flat conditions for $X_a$ have no solution. This is because F-flat conditions for $X_a$ are $N_X$ equations with $N[\tilde{z}]$ variables, so the $N_X$ equations can not be solved if the relation (\ref{eq:R-breaking condition}) is satisfied. However, $U(1)_R$-invariant operators which consist of $\tilde{Q}_I$ only appear in $W_{X_a}$. Here we define $\omega _p$ $(p=1, \dots , N_{\omega })$ as $U(1)_R$-invariant independent operators which appear in $W_{X_a}$. For example, $\omega _p$ include $\phi_{\al}$ and operators which consist of $\Hat{Q}_I$, such as $(X_a/X_{b})~( a \neq b)$ on which $W_{X_a}$ depends. In other words, all $\Hat{Q}_I$ can not be fixed by $\pa_{\tilde{z}}V=0$. $N[\tilde{z}]$ is not larger than $N_{\omega }$. This fact limits (\ref{eq:R-breaking condition}) to 
\beq
0 < N_X -  N_{\omega }.
\label{eq:R-breaking condition 2} 
\eeq

On the other hand, under the condition $N_X \leq N_{\omega }$, R-symmetry breaking does not happen, so that there is an R-symmetric SUSY breaking vacuum. However, we consider models with only positive R-charge fields, so there is a possibility that $\Hat{Q_I}=0$ gives singular values to $W_{X_a}$.

To obtain (\ref{eq:R-breaking condition 2}), we limit the assignment of R-charges, but we find the same condition for satisfying both (\ref{eq:R-breaking}) and $\pa_{\tilde{z}_J}V=0$ in the model with superfields whose R-charges are more than 2. 
Moreover, the redefinition in (\ref{eq:redefinition}) leads to scalar potentials with both positive and negative power terms of $|Y|$. There would be another possibility that we would find a stationary point where $Y$ is also stabilized at a finite value. However, in the model with $\omega _p$ satisfying (\ref{eq:R-breaking condition 2}), a global minimum value of the scalar potential, $V_{min}$, is given by    
\beq
V_{min} =V_2( {\omega _p}_{min})=\sum_{a=1}^{N_X}|f_{X_a}({\omega _p}_{min})|^2,
\eeq
where $V_2(\omega _p)$ is defined as $\sum_{a=1}^{N_X}|f_{X_a}(\omega _p)|^2$ and $V_2( {\omega _p}_{min})$ is a minimum value of $V_2(\omega _p)$. 
${\omega _p}_{min}$ satisfy $\pa_{\tilde{z}_J}V=0$ as long as the equations (\ref{eq:F-flat 0}) with any $\omega _p$ are satisfied by the other fields.\footnote{ The limit $|Y|\rightarrow 0$ or $Y \rightarrow \infty$ becomes a solution for (\ref{eq:F-flat 0}). The functions $f_I(\tilde{z}_J)$ whose the coefficients $Y^{\frac{2-q_I}{q}}$ become infinite in the limit also need to vanish by the fields except for $\omega _p$.}
This is because the part of the scalar potential $V_2$ only depends on $\omega _p$, and $W_{\phi_{\al}}$ and $W_{\Phi_i}$ depend on not only $\omega _p$ but also the other fields. The F-flat conditions, $W_{\phi_{\al}}=W_{\Phi_{i}}=0$, can be satisfied by the fields which do not appear in $W_{X_a}$. 

On the other hand, models with $N_{\omega } \geq N_X$  have solutions for F-flat conditions for $X_a$ and eventually we find SUSY vacua in the limit $Y \rightarrow \infty$ or $Y \rightarrow 0$. In fact, we find runaway supersymmetric vacua in models with $N_{\omega } \geq N_X$ as we discuss in the following subsection.

\subsubsection{Runaway vacua}
\label{sec:runaway}
In this section, we study runaway directions in R-symmetric models generally.\footnote{See also Ref.\cite{Marques:2008va}. }
We classify $\Phi_i$ to $\Phi_{i_+}^+$, $\Phi_{i_-}^-$ and $\Phi_k$. The R-charges of $\Phi_{i_+}^+$ and $\Phi_{i_-}^-$ satisfy $R[\Phi_{i_+}^+]=q_{i_+} >2$ and $R[\Phi_{i_-}^-] <0$.
The fields $\Phi_k$ for $k=1, \dots , (N_{\Phi}-N[\Phi_{i_+}^+]-N[\Phi_{i_-}^-])$ describe the fields with R-charges $q_k$ ($0<q_k<2$).
Based on the above argument, the potential is described as 
\beq
\begin{split}
V=& \sum_a^{N_X} |f_{X_a}|^2+ \sum_{i_+} \left( \frac{1}{|Y|} \right)^{\frac{2|q_{i_+}-2|}{q}}|f_{\Phi_{i_+}}|^2+ \sum_{i_-} (|Y|)^{\frac{2|2-q_{i_-}|}{q}}|f_{\Phi_{i_-}}|^2 \\
&+ \sum_{k} |Y|^{\frac{2|2-q_{k}|}{q}}|f_{\Phi_k}|^2+ \sum_{\al} |Y|^{\frac{4}{q}}|f_{\phi_{\al}}|^2. 
\end{split}
\eeq
We assume that $q$ is positive. In this case, if we can find the direction that all $f_I(z_J)$ vanish except for $f_{X_a}$ and $f_{{\Phi_i}_+}$,
the scalar potential along such a direction becomes 
\beq
V= \sum_a^{N_X} |f_{X_a}|^2+ \sum_{i_+} \left( \frac{1}{|Y|} \right)^{\frac{2|q_{i_+}-2|}{q}}|f_{\Phi_+}|^2.
\eeq
This scalar potential in the limit $|Y| \rightarrow \infty  $ can be described as
\beq
V \longrightarrow \sum_a^{N_X} |f_{X_a}|^2.
\eeq

This means that $Y$ is a runaway direction. Especially, if $f_{X_a}=0$ for all $a$ can be satisfied,  
SUSY is restored by the limit $|Y| \rightarrow \infty$. 

On the other hand, when we assume that $Y$ is chosen as one of $\Phi^-_{i_-}$,
the potential on the slice $f_{\Phi^+_{i_+}}=0$ is given by 
\beq
\begin{split}
V= & \sum_a^{N_X} |f_{X_a}|^2+ \sum_{i_-} \left( \frac{1}{|Y|} \right)^{\frac{2|2-q_{i_-}|}{|q|}}|f_{\Phi_{i_-}}|^2 \\
&+ \sum_{k} \left( \frac{1}{|Y|} \right)^{\frac{2|2-q_{k}|}{|q|}}|f_{\Phi_k}|^2+ \sum_{\al} \left( \frac{1}{|Y|} \right)^{\frac{4}{|q|}}|f_{\phi_{\al}}|^2. 
\end{split} 
\eeq
In the limit $|Y| \rightarrow \infty$, the potential is
described as
\beq
V \longrightarrow \sum_a^{N_X} |f_{X_a}|^2.
\eeq
This limit also corresponds to $V \longrightarrow 0$ 
when $f_{X_a}=0$ for all $a$ are satisfied. The condition for the solution existing corresponds to $0 \geq (N_X-N_{\omega })$.   

Finally we conclude that there are some runaway directions in models with the fields whose R-charges are negative and/or more than 2. The scalar potential in the limit where at least one VEV of R-charged field is infinite, is given by $\sum_{a}|W_{X_a}|^2$.
Especially, the limit restores SUSY in models satisfying $N_X \leq  N_{\omega }$.
\footnote{ If we define $Y$ as a field with negative R-charge, the limit $Y \rightarrow \infty$ also gives the vanishing superpotential, $W \rightarrow 0$. If we consider supergravity effects, we would find that the limit $Y \rightarrow \infty$ restores SUSY in the models where  covariant derivatives $W_I+K_I W$ also go to $W_I$ in the limit. We need classify R-charge assignments to discuss the supergravity effect. This is our future work. }

\subsubsection{Example}

We show illustrating examples which describe the above generic arguments.
The fields $X_a~(a=1, \dots , n)$ have R-charge 2, and $\Phi$, $\ov{\Phi}$ and $\phi$ denote fields with R-charge $1$, $-1$ and $0$ fields respectively.
Based on the generic argument, we expect that the global minimum of the scalar potential exists along the slice  $W_{\phi}=W_{\Phi}=W_{\ov{\Phi}}=0$ although that may correspond to a runaway direction. It depends on $n$ whether the minimum preserves SUSY or not. 
We consider the renormalizable superpotential as follows,
\beq
W=\sum_{a=1}^n \left( f_a(\phi)X^a+\lambda_a X^a \Phi \ov{\Phi} \right) +\frac{1}{2} m(\phi) \Phi^2,
\label{eq:example(n,1,2)}
\eeq
where $m(\phi)$ is linear, and $f_a(\phi)$ are quadratic functions of $\phi$. 

The derivatives of $W$ are given by
\beq
\begin{split}
W_{X_a}= & f_a(\phi) + \lambda_a  \Phi \ov{\Phi}, \\
W_{\Phi}=& \sum_{a=1}^n \lambda_a X^a \ov{\Phi} + m(\phi) \Phi, \\
W_{\ov{\Phi}}=& \sum_{a=1}^n \lambda_a X^a \Phi, \\
W_{\phi}=& \sum_{a=1}^n \frac{\pa f_a(\phi)}{\pa \phi} X^a +\frac{1}{2} \frac{\pa m(\phi)}{\pa \phi} \Phi^2. 
\end{split}
\eeq

First, we consider the example with $n>3$, which corresponds to the case that a global minimum of the scalar potential is given by $V_2=|W_{X_a}|^2$ along the slice $W_{\phi}=W_{\Phi}=W_{\ov{\Phi}}=0$. The scalar potential $V_2(\tilde{Q})$ is obtained as
\beq
V_2(\Phi,\ov{\Phi},\phi) = \sum_{a=1}^n|f_a (\phi)+\frac{1}{2}\lambda_a \Phi \ov{\Phi}|^2,
\eeq
where the fields $\{ \tilde{Q} \}$ correspond to $\{ \Phi, \ov{\Phi}, \phi \}$. In this case, the F-flat conditions for all $X_i$, $W_{X_a}=0$, can not be solved. We look for the solutions for $\pa_{\tilde{Q}}V_2=0$. $V_2$ depends on two $U(1)_R$-invariant operators, $\phi(\equiv \omega _1)$ and $\Phi \ov{\Phi}(\equiv \omega _2)$. 

We find that one solution for $\pa_{\Phi}V_2=\pa_{\ov{\Phi}}V_2=0$ is $\Phi=\ov{\Phi}=0$ and the stationary condition for $\phi$, $\pa_{\phi}V_2=0$, is also satisfied as follows,
\beq
\begin{split}
\Phi=& \ov{\Phi}=0, \\
\sum_{a=1}^n \frac{\pa f_a(\phi)}{\pa \phi} &  \ov{f_a}(\phi) \Big|_{\phi=\phi^*} = 0.
\end{split}
\label{eq:R-breaking ex}
\eeq 
In this case, all $\tilde{Q}$ are fixed by the stationary condition of $V_2$.
When equation (\ref{eq:R-breaking ex}) is satisfied, the F-flat conditions, $W_{\phi}=W_{\Phi}=W_{\ov{\Phi}}=0$, are also satisfied by $X_a$,  
\beq
\sum_{a=1}^n \frac{\pa f_a(\phi)}{\pa \phi} X^a \Big|_{\phi=\phi^*}=0,
\eeq
where the scalar potential $V$ is estimated as $\sum_{a=1}^n |f_a|^2$, and  the VEVs of $(n-1)$ fields $X_a$ are flat directions. 

Furthermore, we find the other solution for $\pa_{\tilde{Q}}V_2=0$ as follows. 
The stationary conditions for $V_2$, $\pa_{\Phi}V_2=\pa_{\ov{\Phi}}V_2=\pa_{\phi}V_2=0$, are satisfied by $\omega _1(\equiv \phi)$ and $\omega _2(\equiv \Phi \ov{\Phi})$ satisfying the following equations 
\beq
\begin{split}
\Phi \ov{\Phi} = & -\frac{\sum_{a=1}^n \ov{\lambda}_a f_a (\phi) }{\sum_{a=1}^n |\lambda_a|^2}, \\
\sum_{a=1}^n \frac{\pa f_a(\phi)}{\pa \phi} & \ov{W}_{\ov{X}_a} = 0.
\end{split}
\label{eq:solution ex2}
\eeq
On the other hand, $\Phi$ must vanish in order that the F-flat conditions for $\Phi$ and $\ov{\Phi}$, $W_{\Phi}=W_{\ov{\Phi}}=0$ are satisfied, as long as $m(\phi)$ is non-vanishing.   
When $\omega _1$ and $\omega _2$ are fixed by (\ref{eq:solution ex2}), the limit $\Phi \rightarrow 0~(\ov{\Phi} \rightarrow \infty)$ gives the F-flat conditions and the fields $X_a$ satisfy the following equations,    
\beq
\begin{split}
\sum_{a=1}^n \lambda_a X^a=& 0, \\
\frac{\pa f_a(\phi)}{\pa \phi} X^a =& 0. 
\end{split}
\eeq   
In this runaway direction, $V$ is given by 
\beq
V \rightarrow \sum_{a=1}^n ( |f_a| ^2  - | \lambda _a |^2   |\Phi| ^2 |\ov{\Phi}| ^2), 
\eeq
where the VEVs of $(n-2)$ fields $X_a$ are flat directions.
Eventually, the either of these two solutions corresponds to the global minimum of the scalar potential $V$.
 
Second, we consider the model with $n=3$. In this case, the number of $X_a~(a=1,2,3)$ is equal to the number of $\tilde{Q}_I~(\phi,\Phi,\ov{\Phi})$ which couple with $X_a$. However, the vacuum structure is the same as in the case with $n>3$ because $V_2$ depends on only $U(1)_R$-invariant operators, $\phi$ and $\Phi \ov{\Phi}$. Based on the condition (\ref{eq:R-breaking condition 2}), the model with $n=3$ is classified as models satisfying $N_X > N_{\omega }=2$.

Now we consider the case with $n \leq 2$. Based on the generic argument, there is a runaway direction in this case. In fact, we discuss the model with $n=2$. When the fields, $\phi$, $\Phi \ov{\Phi}$, and $X_a$ are fixed by,
\beq
\begin{split}
\Phi \ov{\Phi} = & -\frac{f_1(\phi)}{\lambda_1}, \\
\sum_{a=1}^2 \lambda_a{X_a}  = & 0, \\
f_2(\phi) \lambda_1 - \lambda_2 f_1 (\phi) =& 0, \\
\sum_{a=1}^2 \frac{\pa f_a(\phi)}{\pa \phi} X^a =& 0, 
\end{split}
\eeq 
where the derivatives of $W$ are given by $W_{\ov{\Phi}}=W_{X_a}=0$, $W_{\phi}=(1/2) \pa_{\phi}m \Phi^2$ and $W_{\Phi}=m(\phi)\Phi$. When the field $\Phi$ vanishes, the field $\ov{\Phi}$ becomes infinite. 
The direction $\ov{\Phi} \rightarrow \infty$ corresponds to a runaway direction where SUSY is restored. 

In addition to the above runaway direction, the model with $n=2$ has a R-symmetric vacuum on the slice $\Phi=\ov{\Phi}=0$.

We give a comment about the case $N[\ov{\Phi}]=0$. The renormalizable superpotential is obtained by $\lambda_a=0$ in (\ref{eq:example(n,1,2)}) 
\beq
W=\sum_{a=1}^n f_a(\phi)X^a +\frac{1}{2} m(\phi) \Phi^2.
\eeq
In this case, the derivatives of $W$ are given by
\beq
\begin{split}
W_{X_a}= & f_a(\phi), \\
W_{\Phi}=& m(\phi) \Phi, \\
W_{\phi}=& \sum_{a=1}^n \frac{\pa f_a(\phi)}{\pa \phi} X^a +\frac{1}{2} \frac{\pa m(\phi)}{\pa \phi} \Phi^2. 
\end{split}
\eeq
We can not find a runaway direction, and the global minimum is obtained along the slice $W_{\Phi}=W_{\phi}=0$. The two F-flat conditions lead to the following solution,
\beq
\begin{split}
\Phi =& 0, \\
\sum_{a=1}^n \frac{\pa f_a(\phi)}{\pa \phi} X^a = & 0. 
\end{split}
\eeq
The field $\phi$ is fixed by $\pa_{\phi} V_2 =0$, so that the global minimum in the model with $n>1$ corresponds to SUSY breaking vacua with $(n-1)$ flat directions and the global minimum in the model with $n \leq 1$ describes SUSY vacua.

\subsubsection{Short summary}
We summarize the results we find in the previous subsections and classify models according to the number of fields with R-charge 2.

\vspace{1 \baselineskip}

$\bf{[1]~N_X \leq N_{\phi}}$

\vspace{0.5\baselineskip}

The F-flat conditions for $X_a$ can be solved generally, so R-symmetric and supersymmetric vacua can exist. If we find a SUSY breaking vacuum, it is metastable. When models have fields with negative R-charge and/or more than 2 R-charge, there are runaway directions. 

\vspace{1\baselineskip}

$\bf{[2]~N_{\phi} < N_X \leq  N_{\omega }}$\footnote{We assume that the condition, $N_{\phi} \leq N_{\omega}$, is always satisfied in our models.}

\vspace{0.5\baselineskip}

$N_{\omega }$ denotes the number of the $U(1)_R$-invariant independent operators of the fields which appear in the derivatives of $W$ by the fields $X_a$.
We can find R-symmetric SUSY breaking vacua in this model. 
However, R-symmetry can not be broken, as long as there is no field whose R-charge is negative or more than 2. On the other hand, there are runaway directions in the model with fields whose R-charges are either negative and/or more than 2. The limit that an R-charged field goes to infinite restores SUSY, so that R-symmetric SUSY breaking vacua are metastable. This does not conflict with the NS argument.

\vspace{1\baselineskip}

$\bf{[3]~N_{\omega } < N_X}$

\vspace{0.5\baselineskip}

There is no supersymmetric vacuum in this model with any assignment of R-charges. 
It is possible that R-symmetry is also broken generally. Eventually, there are SUSY breaking minima with $(N_X-N_{\omega })$ flat directions, where  the VEVs of R-charged fields correspond to the flat directions. Furthermore, the minimum of the partial potential $V_2=\sum_{a}|W_{X_a}|^2$ is a global minimum of the full potential $V$. 
The relation $N_{\omega } < N_X$ corresponds to the condition that there is no solution satisfying all F-flat conditions for $X_a$, so that this condition for R-symmetry breaking corresponds to the condition for SUSY breaking.

\vspace{2 \baselineskip}

We consider R-symmetric models to discuss SUSY breaking generally, and we find that we can realize SUSY breaking vacua, if we consider models which satisfy $N_X > N_{\omega }$. The O'Raifeartaigh model~\cite{O'Raifeartaigh:1975pr}, which we study in the next subsection, is well-known as one of those models.   

It is not a sufficient condition for spontaneous SUSY breaking that supersymmetric models have R-symmetry. However, recently it is proved that stable SUSY breaking vacua can not exist at tree level in models even without R-symmetry which have general polynomial superpotential and canonical K\"ahler potential \cite{Komargodski:2009jf,Ray:2006wk}. SUSY breaking vacua always have flat directions, as far as avoiding tachyonic modes.\footnote{The flat directions are also discussed generally in Ref. \cite{Komargodski:2009jf}.} This result and our argument indicate that it is appropriate that we concentrate on the O'Raifeartaigh model to discuss SUSY breaking generally, as we actually do later.

\subsubsection{The generalized O'Raifeartaigh model}

The O'Raifeartaigh model~\cite{O'Raifeartaigh:1975pr} 
is a good example of R-symmetric SUSY models, where SUSY 
is spontaneously broken. 
Its generalization is shown in Ref.~\cite{Intriligator:2007py} 
as the generalized OR model, which 
has the following superpotential,
 \begin{eqnarray}
W_{OR} &=& \sum_a g_a(\phi_i)\,X_a, 
\label{eq:orgsp}
\end{eqnarray}
where $a=1,2,\ldots,r$ and $i=1,2,\ldots,s$, and 
the numbers of fields are constrained as $r>s$. 
Their R-charges are assigned as $q_{X_a}=2$ and $q_{\phi_i}=0$, 
and $g_a(\phi_i)$ is a function of $\phi_i$.
Based on the discussion in the previous subsections,
this generalized OR model is a minimal model to cause SUSY breaking.
In this model, $F$-flat conditions for $X_a$ are just given by 
\begin{eqnarray}
\partial_{X_a} W &=& g_a(\phi_i) \ = \ 0.
\label{eq:orgfflat}
\end{eqnarray}
These are $r$ complex equations for $s$ complex variables, 
that is, these are {\it over-constrained} conditions 
for $r>s$. 
Therefore, there is no SUSY solution satisfying 
(\ref{eq:orgfflat}) for generic functions $g_a(\phi_i)$ 
with $r>s$. 
The superpotential of the generalized OR model (\ref{eq:orgsp})
is a specific form of the NS superpotential 
(\ref{eq:rsp1}).
In the generalized OR model, SUSY is always spontaneously broken 
independently of whether R-symmetry is spontaneously broken 
or not, or the fields $X_a$ develop nonvanishing vacuum expectation 
values or not.

The simplest OR model is the model with $r=1$ and $s=0$, 
and has the superpotential 
\begin{eqnarray}
 & & W_{(OR)_1}=fX_1 ,
\nonumber
\end{eqnarray}
where $f$ is a constant.
Obviously, SUSY is spontaneously broken in this model, 
because $W_{X_1}=f$.
The basic O'Raifeartaigh model corresponds 
to the model with $r=2$ and $s=1$, and 
$g_1(\phi)=f + \frac12 h\phi^2$ and $g_2(\phi)=m\phi$, 
and has the following superpotential, 
\begin{eqnarray}
W_{(OR)_{basic}} &=& (f+\frac{1}{2} h \phi^2) X_1 +m \phi X_2. 
\label{eq:oorsp}
\end{eqnarray} 
 The model has only a SUSY breaking pseudo-moduli space, 
\begin{eqnarray}
\phi &=& X_2 \ = \ 0, \qquad 
X_1 \ : \ \textrm{undetermined}, 
\label{eq:or1pms}
\end{eqnarray}
with $W_{X_1}=f$ as a global minimum of the potential. 
When integrating out heavy modes $X_2$ and $\phi$, we 
obtain $W_{(OR)_1}$ as an effective superpotential.
However, the flat direction along $X_1$ is lifted at the one-loop level 
by integrating out $\phi$, and the SUSY breaking 
vacuum in the quantum corrected OR model is given by
\begin{eqnarray}
\phi &=& X_2 \ = \ X_1 \ = \ 0. 
\label{eq:qor1vac}
\end{eqnarray}

These simple models suggest that the tadpole term of $X_a$ 
is important for SUSY breaking.
Indeed, we can show by simple discussion that 
non-vanishing terms of  $g_a(\phi_i)$ at $\phi_i=0$
are sources of SUSY breaking.
We assume that $g_a(\phi_i)$ are non-singular functions.
Then, we can always rewrite the superpotential (\ref{eq:orgsp}) as 
\begin{eqnarray}
W_{OR} &=& \sum_a f_a X_a +\sum_a \check{g}_a(\phi_i) X_a 
\nonumber \\ &=& 
\tilde{f} \tilde{X}_1 
+\sum_a \tilde{g}_a(\phi_i) \tilde{X}_a, \qquad 
(\tilde{g}_a(0) \ = \ 0), 
\label{eq:orgsprd}
\end{eqnarray}
where 
$f_a=g_a(0)$, $\check{g}_a(\phi_i)=g_a(\phi_i)-f_a$, 
$\tilde{X}_a=U_{ab}X_b$, 
$\tilde{g}_a(\phi_i)=\check{g}_a U^\dagger_{ab}$ 
and $U_{ab}$ is a constant unitary matrix defined by 
$f_a U^\dagger_{ab}=\tilde{f}_b=(\tilde{f},0,\ldots,0)$. 
In the following, we will frequently use this basis of 
fields and omit the tildes to simplify the notation. 
In this basis, the F-flat conditions for $X_a$, 
Eq.~(\ref{eq:orgfflat}), are written by 
\begin{eqnarray}
W_{X_a} &=& 
\ g_a(\phi_i)-\delta_{a1}f  
\ = \ 0 .
\label{eq:orgfflatsugrard}
\end{eqnarray}
Together with 
$W_{\phi_i}=\sum_a X_a \partial_{\phi_i}g_a(\phi_i)=0$, 
we find that, if $f=0$, there is a solution $X_a=\phi_i=0$ 
and SUSY is not broken. 
Then it is obvious in the field basis (\ref{eq:orgsprd}) 
that a nonvanishing $f$ is the source of dynamical 
SUSY breaking in the generalized OR model.

In the generalized OR model with the above field basis, 
the field $X_1$ plays a special role, 
while each of $X_a$ ($a \neq 1$) has the qualitatively same 
character as others $X_b$ ($b \neq 1$).
Thus, the simple model with $r=2$ and $s=1$, 
and the superpotential,
\begin{eqnarray}
W_{(OR)_2}=(f+g_1(\phi))X_1+g_2(\phi) X_2,
\nonumber
\end{eqnarray}
shows qualitatively generic aspects of 
the generalized OR model.
Its scalar potential is written as 
\begin{eqnarray}
V&=& |f+g_1(\phi)|^2+|g_2(\phi)|^2+|W_{\phi}|^2,
\nonumber 
\end{eqnarray}
and stationary conditions are obtained as 
\begin{eqnarray}
V_{X_1}&=&\overline {W_{\phi}}g'_1(\phi) =0, \nonumber \\
V_{X_2}&=&\overline {W_{\phi}}g'_2(\phi) =0, \nonumber \\
V_{\phi}&=&\overline {W_{\phi}}W_{\phi \phi}
+(\overline f+ \overline {g_1(\phi)})g'_1(\phi) 
+ \overline {g_2(\phi)}g'_1(\phi) =0, 
\nonumber 
\end{eqnarray}
where $g'_a(\phi)=dg_a(\phi)/d\phi$ and $W_\phi = \sum_a X_a g'_a(\phi)$.
Unless $W_{\phi}$ does not vanish, we would have 
{\it over-constrained} conditions, i.e., 
$g'_1(\phi) = g'_2(\phi) =0$ for generic functions.
Thus, in general, the solution of the above 
stationary conditions corresponds to 
\begin{eqnarray}
 & & W_{\phi} = X_1 g'_1(\phi) + X_2g'_2(\phi) =0 ,
\nonumber \\
 & & (\overline f+ \overline {g_1(\phi)})g'_1(\phi) 
+ \overline {g_2(\phi)}g'_1(\phi) =0. 
\label{eq:or2-min}
\end{eqnarray}
The latter is the condition to fix $\phi$.
For a fixed value of $\phi$, a ratio between 
$X_1$ and $X_2$ is fixed by the former condition, 
but the linear combination 
\begin{eqnarray}
X_1g'_2(\phi)-X_2g'_1(\phi), 
\label{eq:flatdirgen}
\end{eqnarray}
remains undetermined.
That is the pseudo-flat direction, and 
would be lifted by loop effects.
Similarly we can discuss models with 
several fields $X_a$ and $\phi_i$ ($r>s$).

\subsection{Explicit R-symmetry breaking and 
metastable vacua}
\label{sec:rsugra explicit R-breaking in global SUSY}
In order to have Majorana gaugino masses in addition to soft scalar 
masses, the R-symmetry must be broken spontaneously or explicitly 
at the SUSY breaking minimum we are living. 
On the other hand, as shown in the previous section, 
the NS argument requires an exact R-symmetry 
for the dynamical SUSY breaking. Then, an appearance of 
an unwanted massless Goldstone mode, an R-axion, is inevitable 
in such R-symmetry breaking minimum. Does this mean the dynamical 
SUSY breaking is phenomenologically disfavored ? 

Recently, it has been argued by Intriligator, Seiberg and 
Shih~\cite{Intriligator:2007py} that our world must reside in a 
metastable state, in order to avoid the above conflict between 
gaugino masses and the massless R-axion. The argument is as follows. 
Consider a theory with an approximate R-symmetry which has a small 
R-symmetry breaking parameter $\epsilon$. 
In the limit $\epsilon \to 0$, the R-symmetry 
becomes exact, and the theory possesses a SUSY breaking 
ground state due to the NS argument. For a nonzero 
but tiny parameter $\epsilon$, this SUSY breaking minimum 
still remains as a local minimum of the potential, although there 
appear SUSY ground states somewhere in the field space 
due to explicit R-symmetry breaking effects. As long as 
the parameter $\epsilon$ is small enough, the separation between 
the SUSY breaking minimum and the supersymmetric vacua 
is large, and the former can be a long-lived metastable vacuum. 
These facts were exhibited by ISS based on the O'Raifeartaigh model 
as a simple example of dynamical SUSY breaking model 
with R-symmetry. 
Indeed, such O'Raifeartaigh-type model can be realized 
in some region of the moduli space of SUSY Yang-Mills 
theories~\cite{Intriligator:2006dd}. 

Here following the discussion by ISS we study generic 
aspects of explicit R-symmetry breaking terms, 
and SUSY preserving vacua.
We also classify  explicit R-symmetry breaking terms 
in global SUSY models.
In addition, we discuss metastability.

The simplest R-symmetry breaking term is the 
constant term $W_{R\!\!\!\!/} = c$, but 
the constant term does not play any role in global SUSY theory.
Thus, we do not discuss about adding the constant term in this section.
It is obvious that when we add any R-symmetry breaking term 
$W_{R\!\!\!\!/}(Y,\chi)$ to the NS superpotential 
(\ref{eq:rsp1}), that can relax {\it over-constrained} 
conditions and F-flat conditions can have SUSY 
solutions.

The generalized OR model has richer structure 
in explicit R-symmetry breaking terms.
To see such structure, we consider the 
generalized OR model with 
three types of typical R-symmetry breaking terms, 
i) a function including only 
$\phi_i$ fields $W_{R\!\!\!\!/}=w(\phi)$, 
ii) a function including only $X_a$ ($a \neq 1$), 
$W_{R\!\!\!\!/}=w(X_a)$,
and iii) a function including only $X_1$, 
$W_{R\!\!\!\!/}=w(X_1)$.
The first type of R-symmetry breaking terms 
$W_{R\!\!\!\!/}=w(\phi)$ do not change 
F-flat conditions for $X_a$, i.e., 
$\partial_{X_a} W = f\delta_{a1}+g_a(\phi_i)=0$.
Hence, there is no SUSY solution.
 
For the second type of R-symmetry breaking terms 
$W_{R\!\!\!\!/}=w(X_a)$ ($a\neq 1$), F-flat conditions 
are obtained as 
\begin{eqnarray}
W_{X_1} &=& f+g_1(\phi_i) \ = \ 0, 
\nonumber \\
W_{X_a} &=& g_a(\phi_i)+w_{X_a}(W_a) \ = \ 0 
\qquad {\rm for ~~} a\neq 1, \nonumber \\
W_{\phi_i} &=& \sum_a X_a 
\partial_{\phi_i} g_a(\phi_i) \ = \ 0.
\nonumber
\end{eqnarray}
Thus, if $w_{X_a}(W_a) \neq 0$ for all of 
$X_a$, {\it over-constrained} conditions 
can be relaxed and a SUSY solution 
can be found.
If all of $\phi_i$ vanish, we have $g_1(\phi_i)=0$ 
and the condition $W_{X_1}=0$ can not be 
satisfied.
Hence, the SUSY minimum, which 
appears by adding $W_{R\!\!\!\!/}=w(X_a)$ ($a\neq 1$), 
corresponds to the point, where some of $\phi_i$ 
develop nonvanishing vacuum expectation values.

For the third type of R-symmetry breaking terms 
$W_{R\!\!\!\!/}=w(X_1)$, F-flat conditions are 
obtained as 
\begin{eqnarray}
W_{X_1} &=& f+g_1(\phi_i) +\partial_{X_1} w(X_1) \ = \ 0, 
\nonumber \\
W_{X_a} &=& g_a(\phi_i) \ = \ 0 
\qquad {\rm for ~~} a\neq 1, \nonumber \\
W_{\phi_i} &=& \sum_a X_a 
\partial_{\phi_i} g_a(\phi_i) \ = \ 0.
\nonumber
\end{eqnarray}
If $r=s+1$, the {\it over-constrained} conditions 
can be relaxed.
In this case, the point $\phi_i=0$ for all of $i$ 
can be a solution for $W_{X_a}=0 $ for $a \neq 1$.
Furthermore, the conditions,
\begin{eqnarray}
 & & f+\partial_{X_1} w(X_1) \ = \ 0, \qquad 
\sum_a X_a \partial_{\phi_i} g_a(\phi_i) \ = \ 0,
\nonumber
\end{eqnarray}
should be satisfied.

When R-symmetry breaking terms 
include $X_1$ and $X_a$ ($a \neq 1$), 
{\it over-constrained} conditions 
can be relaxed and a solution 
for F-flat conditions would correspond 
to $\phi_i \neq 0$ for some of $\phi_i$.

The SUSY breaking minimum is found 
in the generalized OR model without explicit 
R-symmetry breaking terms, as discussed 
in the previous subsection.
As discussed above, 
SUSY vacua can appear,
when we add the definite form of 
explicit R-symmetry breaking terms 
to the generalized OR model.
Thus, the previous SUSY breaking minimum 
is a metastable vacuum, 
if such R-symmetry breaking effects are 
small around the SUSY breaking minimum 
and the SUSY breaking vacuum itself is 
not destabilized by such 
R-symmetry breaking terms.
 
As an illustrating example, we consider 
the basic OR model (\ref{eq:oorsp}) 
with explicit R-symmetry breaking terms.
ISS introduced an explicit R-symmetry breaking term in the 
superpotential\footnote{See also Ref.~\cite{Intriligator:2007cp}.}, 
$W=W_{(OR)_{basic}}+W_{R\!\!\!\!/}$, where 
\begin{eqnarray}
W_{R\!\!\!\!/} &=& \frac{1}{2}\epsilon m X_2^2. 
\label{eq:rbrkex1}
\end{eqnarray}
In this case, there appears a SUSY minimum, 
\begin{eqnarray}
\phi &=& \sqrt{-\frac{2f}{h}}, \qquad 
X_2 \ = \ -\frac{1}{\epsilon} \phi, \qquad 
X_1 \ = \ \frac{m}{\epsilon h}, 
\nonumber
\end{eqnarray}
which is far away from the (local) SUSY breaking 
minimum (\ref{eq:qor1vac}) 
for a sufficiently small $\epsilon \ll 1$. 
In addition, the SUSY breaking minimum is 
not destabilized by the above R-symmetry breaking 
term (\ref{eq:rbrkex1}).
Then the original 
SUSY breaking vacuum (\ref{eq:qor1vac}) becomes metastable 
which can be parametrically long-lived for $\epsilon \ll 1$.

Instead, if we consider the following R-breaking term~\cite{Dine:2006gm} 
\begin{eqnarray}
W_{R\!\!\!\!/} &=& \frac{1}{2}\epsilon m X_1^2, 
\label{eq:rbrkex2}
\end{eqnarray}
the newly appeared SUSY point is found as 
\begin{eqnarray}
\phi &=& X_2 \ = \ 0, \qquad 
X_1 \ = \ -\frac{f}{\epsilon m}. 
\nonumber
\end{eqnarray}
In this case, the pseudo-moduli space (\ref{eq:or1pms}) 
disappears at the tree level. 
However, the SUSY 
breaking point (\ref{eq:qor1vac}) remains as a local 
minimum due to the one-loop mass for $X_1$, but becomes 
metastable. Then the situation is similar to the above example. 
We easily find that any R-breaking terms 
which consist of only $\phi$ do not restore SUSY.

Now, let us study whether the 
SUSY breaking minimum, which is found without 
R-symmetry breaking terms, is destabilized by adding 
R-symmetry breaking terms.
We consider the generalized OR model with $(r=2,s=1)$, i.e., 
$W_{(OR)_2}$,  
whose stationary conditions (\ref{eq:or2-min}) are studied in the 
previous subsection.
Their solutions are denoted by $X_a=X^{(0)}_a$ and 
$\phi=\phi^{(0)}$.
First, we add a small R-symmetry breaking term, 
$W_{R\!\!\!\!/} = \epsilon w(X_2)$, which 
depends only on $X_2$.
Then, the scalar potential is written as 
\begin{eqnarray}
 & & V=|f+g'_1(\phi)|^2+|g_2(\phi)+\epsilon w'(X_2)|^2 +
|W_\phi|^2,
\nonumber
\end{eqnarray}
where $W_\phi=X_1g'_1(\phi)+X_2g'_2(\phi)$.
In addition, we assume that the stationary conditions of $V$ 
are satisfied by $X_a=X^{(0)}_a+\delta X_a$ and 
$\phi=\phi^{(0)} + \delta \phi$, and that all of $\delta X_a$ and 
$\delta \phi$ are of ${\cal O} (\epsilon)$.
For example, the stationary condition along $\phi$, 
$V_\phi =0$, gives the following condition,
\begin{eqnarray}
 & & \left( \sum_a |g'_a(\phi^{(0)})|^2 +
\sum_a(\overline f\delta_{a1}+\overline{g_a(\phi^{(0)})})g''_a(\phi^{(0)}))
\right) \delta \phi 
+  \epsilon g'_2(\phi^{(0)}) \ \overline {w'(X_2^{0})} =0,
\nonumber
\end{eqnarray}
where we have used the stationary conditions (\ref{eq:or2-min})  
at $X_a=X_a^{(0)}$ and $\phi = \phi^{(0)}$.
This is the equation to determine $\delta \phi$.
The stationary condition along $X_1$, $V_{X_1}=0$, reduces to 
\begin{eqnarray}
 & & g'_1(\phi^{(0)}) \ \overline{\delta W_\phi} =0,
\nonumber
\end{eqnarray}
where 
\begin{eqnarray}
 & & \delta W_\phi = \sum_a g'_a(\phi^{(0)})\delta X_a 
+ \sum_aX_a^{(0)}g''_a(\phi)\delta \phi .
\nonumber
\end{eqnarray}
Thus, this shows a relation among $\delta X_a$ and $\delta \phi$ 
unless $g'_1(\phi^{(0)})=0$.
On the other hand, the stationary condition along 
$X_2$, $V_{X_2}=0$, leads to the following equation,
\begin{eqnarray}
 & & \epsilon w''(X_2^{(0)}) \ \overline{g_2(\phi^{(0)})}=0.
\nonumber
\end{eqnarray}
This is not an equation among $\delta X_a$ and $\delta \phi$, 
but implies that the stationary condition is destabilized 
unless $w''(X_2^{(0)}) \ \overline{g_1(\phi^{(0)})}  =0$.
In the above basic O'Raifeartaigh model, we have 
$g_1(\phi^{(0)})=0$.
Thus, the SUSY breaking minimum is not destabilized by 
adding the mass term of $X_2$, $w(X_2) =\frac12 mX_2$, i.e., 
$w''(X_2) \neq 0$ at $X_2=0$.

Now, let us add an R-symmetry breaking term, 
$W_{R\!\!\!\!/} = \epsilon w(X_1)$, which 
depends only on $X_1$.
Similarly, we can examine stationary conditions of 
the scalar potential, 
\begin{eqnarray}
 & & V=|f+g'_1(\phi)+\epsilon w'(X_1)|^2+|g_2(\phi)|^2 +
|W_\phi|^2.
\nonumber
\end{eqnarray}
The stationary conditions along $X_2$ and $\phi$ 
give an equation to determine $\delta \phi$ and 
a relation among $\delta X_a$ and $\delta \phi$.
However, the stationary condition along $X_1$, 
$V_{X_1}=0$, leads to 
\begin{eqnarray}
 & & w''(X_1^{(0)})  
\left( \overline{f}+\overline{g_1(\phi^{(0)})}\right)  =0.
\nonumber
\end{eqnarray}
If this condition is not satisfied, 
the stationary condition at the 
SUSY breaking vacuum is destabilized.
Indeed, the basic O'Raifeartaigh model has 
$f+g_1(\phi)=f$ at $\phi =0$.
Thus, when we add the mass term of $X_1$, 
$w(X_1)=\frac12 mX_1^2$, i.e., $w'' \neq 0$, 
the SUSY breaking minimum becomes destabilized 
at the tree level as shown above.
Note that this kind of destabilization would be 
related to the existence of the flat direction 
(\ref{eq:flatdirgen}) in the OR model with global SUSY.\footnote{
Such flat direction would be lifted by supergravity effects.}

The above discussion shows that 
adding generic R-symmetry breaking terms 
can destabilize the SUSY breaking minimum, which 
is found in the model without such explicit 
R-symmetry breaking terms.
In order to realize metastability of 
the original SUSY breaking minimum, 
we need a certain type of R-symmetry breaking terms. 
Alternatively, loop-effects would be helpful not to 
destabilize the original SUSY breaking minimum 
by R-symmetry breaking terms.

\subsection{R-symmetry in supergravity}
\label{sec:rsugra}

In the previous section, based on the argument by ISS, 
we have shown that a certain type of explicit R-symmetry breaking 
terms can restore SUSY, and the original SUSY 
breaking vacuum can become metastable when 
a certain (but not generic) class of explicit R-symmetry breaking terms 
are added and/or loop effects stabilize 
the original SUSY breaking minimum. 
The metastable minimum 
can be parametrically long-lived if the coefficient of the 
R-breaking term is sufficiently small with which the SUSY 
ground state is far from the metastable state in the field space. 

This argument has been performed in a decoupling limit of gravity. 
As we find in the above discussion, however, we have to treat a large 
distance between some separated minima in the field space. This may 
imply that large vacuum values of some fields might be involved in 
the analysis, where supergravity effects could become sizable. 
Moreover, in global SUSY, the SUSY breaking minima 
always have a positive vacuum energy with the magnitude of the 
SUSY breaking scale, which never satisfies the observation 
that the vacuum energy almost vanishes. 
In such a sense, we would be forced to consider supergravity.

Note that, even in supergravity, it is often a hard task to tune 
the vacuum energy at the stationary points of the scalar potential 
to be almost vanishing. This might require a large R-symmetry breaking 
effect specialized to supergravity, i.e., a constant term in the 
superpotential~\cite{Bagger:1994hh}. The existence of such a special 
R-symmetry breaking term could also affect the ISS argument of 
metastability. 
Loop effects have contributions to the vacuum energy.
Here we assume that such loop effects are subdominant, 
and we tune our parameters such that we realize $V \approx 0$ 
at the tree level.
Hereafter we use the unit with $M_{Pl}=1$, where 
$M_{Pl}$ denotes the reduced Planck scale.

\subsubsection{Nelson-Seiberg argument}

In this subsection, we study the NS argument 
within the framework of supergravity theory.
In the case of supergravity, F-flat conditions (\ref{eq:fflat}) are 
modified as 
\begin{eqnarray}
D_IW &\equiv& W_I+K_I W \ = \ 0, 
\nonumber
\end{eqnarray}
where $K$ denotes the K\"ahler potential, $K(|Y|,\chi_i,\bar \chi_i)$. 
In the field basis (\ref{eq:rbasis1}) with the superpotential 
(\ref{eq:rsp1}), these are written as 
\begin{eqnarray}
D_{\chi_i}W &=& Y^{2/q_Y}(\zeta_i+K_i \zeta) \ = \ 0, 
\nonumber \\
D_Y W &=& (2/q_Y+YK_Y)Y^{2/q_Y-1}\zeta \ = \ 0.
\nonumber
\end{eqnarray}
Then, we find the following two candidates of 
R-breaking SUSY solutions 
in supergravity, 
\begin{eqnarray}
\zeta_i &=& 0, \qquad 
\zeta \ = \ 0, 
\label{eq:susyvc1}
\end{eqnarray}
and 
\begin{eqnarray}
D_{\chi_i} \zeta &=& \zeta_i+K_i \zeta \ = \ 0, \qquad 
2/q_Y+YK_Y \ = \ 0. 
\label{eq:susyvc2}
\end{eqnarray}

The first conditions (\ref{eq:susyvc1}) contain $n$ complex equations 
for $n-1$ complex variables, and the situation is the same as the case 
of global SUSY (\ref{eq:globalsusyvc}), that is, the solution 
does not exist for a generic function $\zeta$. 
On the other hand, the second conditions (\ref{eq:susyvc2}) are $n$ 
complex equations for $n$ complex variables which can have a solution. 
This corresponds to a SUSY stationary point specialized to 
R-symmetric supergravity.

In this subsection, we analyze the special SUSY stationary 
solution (\ref{eq:susyvc2}) which appears due to purely the 
supergravity effect and does not obey the NS condition. 
Then, in the following we assume that there is a solution for 
\begin{eqnarray}
2/q_Y+YK_Y &=& 0. 
\label{eq:susyvc2:2}
\end{eqnarray}

For instance, if the K\"ahler potential is given by 
\begin{eqnarray}
K &=& \sum_{n_Y=1}c_{n_Y}|Y|^{2n_Y}
+\hat{K}(\chi_i,\bar\chi_i), 
\label{eq:kxpol}
\end{eqnarray}
the condition (\ref{eq:susyvc2:2}) becomes 
\begin{eqnarray}
2/q_Y+\sum_{n_Y=1}n_Y c_{n_Y}|Y|^{2n_Y} &=& 0. 
\nonumber
\end{eqnarray}
Then, we need at least one negative value of $\{ c_{n_Y},q_Y \}$ 
to have a solution. In the simplest minimal case with $c_{n_Y>1}=0$ 
(and then $K_{Y\bar{Y}}=c_1>0$), a negative charge, $q_Y<0$,  
is required. 

A nontrivial point of this solution is that this SUSY 
stationary point is always tachyonic as we can see from the 
arguments in Appendix~\ref{app:rmass}. In addition, we can find a 
SUSY breaking minima along the direction $D_{\chi_i} \zeta=0$ 
(the first condition in Eq.~(\ref{eq:susyvc2})), 
if we assume that $\chi_i$ receives a heavy SUSY mass 
$m^2_{\chi_i} \gg m_{3/2}^2$ by the condition $D_{\chi_i} W=0$. 
This is a reasonable assumption because $\chi_i$ has a vanishing 
R-charge and $\zeta(\chi_i)$ in $W$ is assumed to be a generic function. 

The scalar potential along $D_{\chi_i} \zeta=0$ is found to be 
\begin{eqnarray}
v(Y) &=& V \Big|_{D_{\chi_i} f=0} \ = \ 
e^K \Big( K_{Y\bar{Y}}^{-1}|2/q_Y+K_Y Y|^2
-3|Y|^2 \Big) |Y|^{2(2/q_Y-1)}|\zeta|^2. 
\nonumber
\end{eqnarray}
Again, for the minimal K\"ahler potential (\ref{eq:kxpol}) 
with $c_1=1$ and $c_{n_Y>1}=0$, the stationary condition 
\begin{eqnarray}
\partial_Y v(Y) &=& 
e^{\hat{K}( \langle \chi_i \rangle, \langle \bar\chi_i \rangle)} 
e^{|Y|^2} |Y|^{2/q_Y-2} (2/q_Y+|Y|^2)
\nonumber \\ && \times 
\Big( |Y|^4+2(2/q_Y-1)|Y|^2+(2/q_Y)^2-2/q_Y \Big) 
\ = \ 0, 
\nonumber
\end{eqnarray}
leads to solutions 
\begin{eqnarray}
|Y|^2 &=& -2/q_Y, 
\label{eq:susysp}
\end{eqnarray}
and 
\begin{eqnarray}
|Y|^2 &=& 1-2/q_Y \pm \sqrt{1-2/q_Y}. 
\label{eq:susubm}
\end{eqnarray}
The first solution (\ref{eq:susysp}) corresponds to the 
SUSY saddle point and the second solutions 
(\ref{eq:susubm}) are SUSY breaking minima. 
We can find this kind of SUSY breaking minima 
in a similar way for more generic K\"ahler potential. 

We can study the same system in a different view point. 
We redefine the field $Y$ as 
\begin{eqnarray}
T &=& -\frac{2}{aq_Y} \ln Y, 
\label{eq:def:t}
\end{eqnarray}
where $a$ is a real constant. 
In this basis, the K\"ahler potential and the superpotential 
(\ref{eq:rsp1}) is written as 
\begin{eqnarray}
K &=& K(T+\bar{T},\chi_i,\bar\chi_i), 
\nonumber \\
W &=& e^{-aT} \zeta(\chi_i). 
\label{eq:rsp3}
\end{eqnarray}
This type of K\"ahler and superpotential appear in the 
four-dimensional effective theory derived from superstring theory, 
where $T$ may be a modulus field associated to some 
compactified dimensions. In such a case, the K\"ahler 
potential is typically given by 
\begin{eqnarray}
K &=& -n_T \ln (T+\bar{T})
+\hat{K}(\chi_i,\bar\chi_i), 
\nonumber
\end{eqnarray}
where $n_T$ is a fractional number, and the $T$-dependence 
of the superpotential (\ref{eq:rsp3}) may originate from 
nonperturbative effects such as string/D-brane instanton effects 
and gaugino condensation effects, where the corresponding gauge coupling 
is determined by the vacuum value of $T$. 
In this case, the scalar potential along $D_{\chi_i} \zeta=0$ 
is given by 
\begin{eqnarray}
v(T) &=& V \Big|_{D_{\chi_i} f=0} \ = \ 
e^K \Big( K_{T\bar{T}}^{-1}(K_T-a)^2-3 \Big) |e^{-aT} \zeta |^2, 
\nonumber
\end{eqnarray}
and then the stationary condition
\begin{eqnarray}
\partial_t v(t) &=& 
-e^{\hat{K}( \langle \chi_i \rangle, \langle \bar\chi_i \rangle)} 
e^{-at}t^{-n_T-1}
\nonumber \\ && \times 
(at+n_T)
\Big( (a^2/n_T) t^2
+2a(1-1/n_T)t+n_T-3 \Big) \ = \ 0, 
\nonumber
\end{eqnarray}
results in a SUSY saddle point $t=-n/a$ and 
SUSY breaking minima 
\begin{eqnarray}
t &=& -(n_T/a)(1-1/n_T) 
\pm (n_T|a|/a^2)\sqrt{5/n_T+1/n_T^2}, 
\nonumber
\end{eqnarray}
where $t=T+\bar{T}$.

In the literature, there are examples of the models 
which have this kind of vacuum structure of the potential. 
Typical superstring models have several moduli $T_I$ with the K\"ahler 
potential $K=\ln \prod_I (T_I+\bar{T}_I)^{-n_{T_I}}$. 
The superpotential induced by some nonperturbative effects 
is given by 
\begin{eqnarray}
W &=& \sum_n A_n e^{\sum_I a_n^I T_I}, 
\nonumber
\end{eqnarray}
where $A_n$ and $a_n^I$ are constants. If the number of the 
moduli is the same as or larger than the number of the 
nonperturbative terms appearing in the 
superpotential~\cite{Barreiro:1999hp}, 
we can define an R-symmetry. A particular linear combination of 
$T_I$'s corresponds to $T$ in Eq.~(\ref{eq:rsp3}) which is 
determined by the condition that all the remaining 
combinations corresponding $\chi_i$'s receive a heavy 
mass by the SUSY condition $D_{\chi_i}W=0$. 
This is possible for certain values of $a_n^I$. 
For the two moduli with double nonperturbative terms, i.e., 
racetrack models,  
a detailed analysis was carried out in Ref.~\cite{Abe:2005pi}.

We stress that the analysis of the SUSY breaking 
minimum as well as the SUSY saddle point in this 
subsection is based on the assumption that all the other fields 
$\chi_i$ than $Y$ or $T$ are stabilized by $D_{\chi_i}W=0$, 
that is, by the SUSY masses~\cite{deAlwis:2005tf}. 
We comment that these stationary solutions have a nonvanishing 
and negative vacuum energy. We need to uplift the SUSY 
breaking minimum to a Minkowski vacuum in order to identify 
this minimum as the one we are living. For such purpose, 
we need another sector which provides the uplifting energy 
and is well sequestered in order not to spoil the original 
structure of dynamical SUSY breaking. 
Such sector can be realized by a dynamically generated 
F-term~\cite{Lebedev:2006qq,Abe:2006xp} for which the discussions 
in the following sections would be important. 

In summary, there is a possibility of special SUSY 
stationary solution in R-symmetric supergravity with a 
generic superpotential. However, it is always a saddle point 
at best and we find SUSY breaking minima with lower 
vacuum energy. This may imply that the NS argument for a dynamical 
SUSY breaking is qualitatively correct also in this 
case, although there is a SUSY solution. 
Furthermore,  the NS argument 
still holds in supergravity as long as the 
K\"ahler potential satisfies $2/q_Y+YK_Y \ne 0$ for any value of $Y$ 
in the field basis (\ref{eq:rbasis1}). 
For instance, in typical models with $q_Y>0$ 
and $K=|Y|^2$, we always find $2/q_Y+YK_Y > 0$.

\subsubsection{Generalized O'Raifeartaigh model in supergravity}
Now we consider the generalized OR model (\ref{eq:orgsp}) in supergravity. 
The F-flat conditions (\ref{eq:orgfflat}) for $X_a$ become  
\begin{eqnarray}
D_{X_a} W &=& 
\partial_{X_a} W + (\partial_{X_a} K)\,W 
\nonumber \\ &=& 
\sum_b M_{ab}(X_c,\phi_i)\,(g_b(\phi_i) + \delta_{b1}f) 
\ = \ 0, 
\label{eq:orgfflatsugra}
\end{eqnarray}
where 
\begin{eqnarray}
M_{ab}(X_c,\phi_i) &=& 
\delta_{ab}+K_{X_a}X_b. 
\nonumber
\end{eqnarray}
We define its determinant as 
\begin{eqnarray}
\Delta &\equiv& \det M_{ab} 
\ = \ 1+\sum_a K_{X_a} X_a. 
\label{eq:det1kx}
\end{eqnarray}

If there is no solution for $\Delta=0$, the matrix $M_{ab}$ 
has an inverse matrix and consequently the F-flat conditions 
(\ref{eq:orgfflatsugra}) are reduced to the same ones as 
Eq.~(\ref{eq:orgfflat}) in the global SUSY, 
\begin{eqnarray}
g_a(\phi_i) + \delta_{a1}f&=& 0, 
\nonumber
\end{eqnarray}
which does not allow a solution for $r>s$ in general. 
However, in the limit $f \rightarrow 0$ in the tilde basis 
(\ref{eq:orgsprd}), these equations are satisfied at $\phi_i=0$.
Thus, the constant $f$ represents the typical size of 
SUSY breaking effects and $g_a(\phi_i)$ as the global 
SUSY case.
We comment that the situation changes if there exists a 
solution of $\Delta=0$. Actually, the condition $\Delta=0$ 
is an analogue of the second condition in Eq.~(\ref{eq:susyvc2}). 
Then, we can carry out a similar analysis as in 
the previous subsection also for this OR model.
That is straightforward and is 
omitted here. Note that the condition $\Delta=0$ is never satisfied 
for a minimal K\"ahler potential, 
\begin{eqnarray}
K &=& \sum_a |X_a|^2 +\sum_i|\phi_i|^2. 
\label{eq:minkahler}
\end{eqnarray}
In the following, we just assume that there is no solution 
for $\Delta=0$. 

We comment that, even in supergravity, the scalar potential 
is positive,  $V >  0$, in the generalized OR model 
(\ref{eq:orgsprd}) with the minimal K\"ahler potential 
(\ref{eq:minkahler}).
In this case, the scalar potential is written as 
\begin{eqnarray}
V &=& e^{K}\left[ (\bar{g}_a +\delta_{a1}\bar f) \big\{ \delta_{ab} 
+(|X_c|^2 -1) \bar{X}_a X_b \big\} (g_b + \delta_{b1}f)
+|X_a D_{\phi_i} g_a|^2  \right] .
\nonumber 
\end{eqnarray}
For any vacuum values of $X_a$, we can always rotate their basis as 
\begin{eqnarray}
& & U_{ab}\,X_b =\hat{X}_a 
=(0,\ldots,0,\hat{X}_c,0,\ldots,0),
\nonumber
\end{eqnarray}
 by a unitary matrix $U(X_a)$, 
and in this basis we can write 
\begin{eqnarray}
e^{-K}V &=& 
\big\{ (|\hat{X}_c|^2-1/2)^2+3/4 \big\} |\hat{g}_c|^2
+ \sum_{a \neq c} |\hat{g}_a|^2
+\sum_{i}|\hat{X}_c D_{\phi_i} \hat{g}_c|^2 
\ > \ 0, 
\nonumber
\end{eqnarray}
where $\hat{g}_a=(U^\dagger)_{ab}\,(g_b +\delta_{b1}f)$. 
Note that $\hat{g}_a$ are now 
$X_a$-dependent functions. 
As discussed above, the conditions, $\hat g_a(\phi)=0$,  
can not be satisfied at the same time.
Thus, the vacuum energy must be positive, $V > 0$.
Since typical magnitudes of $\hat g_a(\phi)$ would be 
of ${\cal O}(f)$, we would estimate $V \sim f^2$.
To realize the almost vanishing vacuum energy $V \approx 0$ 
at this SUSY breaking minimum, we need 
a negative and sizable contribution to 
the vacuum energy, which can be generated by 
R-symmetry breaking effects, e.g., the constant term 
in the superpotential.

We would find the features like this in the models whose superpotentials do not have quadratic terms of $X_a$, $m_{ab}X^aX^b$,  
\beq
W=X_a g^a(\phi_i, \Phi_I) +\Hat{w}(\phi_i,\Phi_I),
\eeq
where the R-charges of $\Phi_I$ are not zero or 2. This is because $\Phi_I=0$ for all $I$ could be a solution of the F-flat conditions for $\Phi_I$. 
This leads the same situation as the above because of $\Hat{w}(\phi_i,\Phi_I)=0$. However, this argument is formed on the slice, $\Phi_I=0$, and it is not easy to discuss whether vacua with negative vacuum energy exist in the directions, $\Phi_I \neq 0$.

\subsection{Explicit R-symmetry breaking in supergravity}
\label{sec:r-breaking-sugra}

Here we study explicit R-symmetry breaking terms 
in supergravity and 
examine whether SUSY solutions can be found
by adding explicit R-symmetry breaking terms 
to the NS model and the generalized OR model.
In the previous section, we have pointed out that 
there is a SUSY 
stationary point when the condition (\ref{eq:susyvc2:2}) 
or the condition $\Delta =0$ is satisfied.
In the following sections, we consider the models, where 
such conditions are not satisfied, and 
SUSY is broken in the NS and generalized OR models 
even within the framework of supergravity like global 
SUSY theory.

First we consider the NS model with 
explicit R-symmetry breaking terms 
$W_{R\!\!\!\!/}=w(Y,\chi_i)$. 
The total superpotential is written as,
\begin{eqnarray}
 & & W=Y^{2/q_Y}\zeta(\chi_i) + w(Y,\chi_i).
\nonumber 
\end{eqnarray}
In this case, F-flat conditions of supergravity theory, 
$D_YW=D_{\chi_i} W=0$, do not lead to {\it over-constrained} 
conditions for any non-vanishing function $w(Y,\chi_i)$.
It is remarkable that within the framework of 
supergravity theory the constant term 
$W_{R\!\!\!\!/} = c$ breaks 
R-symmetry and even such term is enough to 
relax the  {\it over-constrained} conditions.

\subsubsection{Generalized O'Raifeartaigh model}
Let us study more explicitly the generalized OR model 
with explicit R-symmetry breaking terms 
$W_{R\!\!\!\!/}=w(X_a,\phi_i)$.
The total superpotential is written as,
\begin{eqnarray}
 & & W=fX_1 +\sum_{a=1}^rg_a(\phi_i)X_a+w(X_a,\phi_i).
\nonumber
\end{eqnarray}
First, we consider the case with 
the constant R-symmetry breaking term, 
$W_{R\!\!\!\!/}=c$.
In this case, F-flat conditions are 
written explicitly as 
\begin{eqnarray}
D_{X_a}W &=& f\delta_{a1}+g_a(\phi_i) + K_{X_a}\left( 
fX_1 +\sum_{a=1}^rg_a(\phi_i)X_a+c \right) \ = \ 0, 
\nonumber \\
D_{\phi_i}W &=& \sum_a X_a \partial_i g_a(\phi_i) 
+ K_{\phi_i}\left( 
fX_1 +\sum_{a=1}^rg_a(\phi_i)X_a+c \right) \ = \ 0.
\nonumber 
\end{eqnarray}
The former conditions are not always {\it over-constrained} 
for $c \neq 0$.
Furthermore, the vacuum expectation value of $W$ and 
at least $(r-s)$ vacuum values of $K_{X_a}$ are required to be 
non-vanishing.
Otherwise, the former conditions become {\it over-constrained}  
for generic functions $g_a(\phi_i)$.
Furthermore, when  $K_{X_a}$ for $a\neq 1$ does 
not vanish, all vacuum values of $\phi_i$ can not 
vanish to satisfy $D_{X_a}W=g_a(\phi_i)+K_{X_a}W=0$.
Thus, a SUSY solution can be found by 
adding $W_{R\!\!\!\!/}=c$.
This solution corresponds to 
the AdS SUSY minimum, because 
non-vanishing $\langle W \rangle$ is required 
and the scalar potential at this point 
is evaluated as $V=-3e^K|W|^2 < 0$.
The values of the constant $c$ and $\langle W \rangle$ 
must be sizable, because this AdS SUSY point disappears 
in the limit that $c\rightarrow 0$ or 
$ \langle W \rangle \rightarrow 0$.
Magnitudes of  $c$ and $\langle W \rangle$ are 
expected to be comparable with $f$ when 
$K_{X_a}={\cal O}(1)$.
Hence, we can find the new type of 
SUSY solution, which can not be found 
in global SUSY theory. 
However, that requires large  values of $c$ and $\langle W \rangle$, 
 and may have sizable effects on the previous SUSY breaking 
minimum, which is found in the generalized OR model without 
R-symmetry breaking terms.

Similarly, we can discuss the case that 
R-symmetry breaking terms include only $\phi_i$ 
fields, i.e.,  $W_{R\!\!\!\!/}=w(\phi_i)$.
In this case, F-flat conditions along $X_a$, $D_{X_a}W=0 $, 
are written as 
\begin{eqnarray}
D_{X_a}W &=& f\delta_{a1}+g_a(\phi_i) + K_{X_a}\left( 
fX_1 +\sum_{a=1}^rg_a(\phi_i)X_a+ w(\phi_i)\right) \ = \ 0 .
\nonumber 
\end{eqnarray}
Thus, the situation is quite similar to the case 
with $W_{R\!\!\!\!/}=c$.
To have a SUSY solution, it is required that 
$\langle W \rangle$, $\langle w(\phi_i) \rangle $ 
and at least $(r-s)$ vacuum values of $K_{X_a}$ must 
be non-vanishing.
Sizes of $\langle W \rangle$ and $\langle w(\phi_i) \rangle$ 
are expected to be comparable with $f$.

Finally, we consider the case that R-symmetry breaking terms include 
$X_a$ fields, $W_{R\!\!\!\!/}=w(X_a,\phi_i)$.
In this case, F-flat conditions along $X_a$, $D_{X_a}W=0 $, 
are written as 
\begin{eqnarray}
D_{X_a}W &=& f\delta_{a1}+g_a(\phi_i) + \partial_{X_a}w(X_a,\phi_i)
+ K_{X_a}W =0.
\nonumber 
\end{eqnarray}
When $K_{X_a}W$ is sufficiently small, 
the above F-flat conditions correspond to 
F-flat conditions in global SUSY theory.
In such a case, we have a SUSY solution when 
$w(X_a,\phi_i)$ depend on at least $(r-s)$ $X_a$'s.
Otherwise, the global SUSY solution can not 
be found, but a SUSY solution with 
$\langle w(X_a,\phi_i) \rangle \neq 0$ 
and $\langle W \rangle \neq 0$ can be found 
within the framework of supergravity theory.
Such situation is similar to the case with $W_{R\!\!\!\!/}=c$.

We have discussed that the NS model and generalized OR 
model with R-symmetry breaking terms have 
SUSY solutions with $\langle W \rangle \neq 0$ 
in supergravity theory.
If the SUSY breaking minimum, which is found 
without R-symmetry breaking terms, is not destabilized 
by the presence of R-symmetry breaking terms, 
the previous SUSY breaking minimum 
would correspond to a SUSY breaking metastable vacuum.
However, a sizable vacuum value of superpotential 
is required unless $\partial_{X_a}w(X_a,\phi_i) \neq 0$ 
for at least $(r-s)$ $X_a$ fields.
Such large superpotential (even if that is a constant term) 
would affect the stability of 
the previous SUSY breaking minimum.

Furthermore, we have another reason to have 
a large size of $\langle W \rangle$ at 
the previous SUSY breaking minimum.
At the previous SUSY breaking minimum, the vacuum 
energy is estimated as 
$V \sim |f|^2 > 0$ for $\langle W \rangle =0$.
To realize the almost vanishing vacuum energy, $V\approx 0$, 
we need a non-vanishing value of $\langle W \rangle$, 
which are comparable with $f$.
In this case, supergravity effects at the 
previous SUSY breaking minimum are not negligible.
This purpose to realize  $V\approx 0$ has 
the implication even for the case that 
R-symmetry breaking terms include more than $(r-s)$ 
$X_a$ fields.
In this case, we can find a (global) SUSY solution even for 
$\langle W \rangle =0$.
However, realization of $V\approx 0$ requires 
a sizable vacuum value of $\langle W \rangle $, 
although values $\langle W \rangle $ at the 
SUSY breaking minimum and SUSY preserving minimum 
are not the same.
Hence, it is quite non-trivial whether 
one can realize a metastable SUSY breaking vacuum 
with  $V\approx 0$ in supergravity theory, 
which has a SUSY minimum.
We will study this possibility concretely  
by using simple classes of the generalized OR models 
in the following sections.
We will concentrate ourselves to the minimal K\"ahler potential 
(\ref{eq:minkahler}) in most cases of the following 
discussions.

\subsubsection{Classification of R-breaking terms in supergravity}

In this subsection and the following sections, 
we consider minutely the previous discussions 
about the explicit R-symmetry breaking in the supergravity 
framework by examining concrete examples. 
We introduce the explicit R-symmetry breaking terms 
$W_{R\!\!\!\!/}$ into the above supergravity OR model, 
\begin{eqnarray}
W_{R\!\!\!\!/} &=& c(\phi_i)
+\frac{1}{2}\sum_{a,b}
m\,\epsilon_{ab}(\phi_i) X_a X_b 
+\cdots, 
\label{eq:rbrkexg}
\end{eqnarray}
where $c(\phi_i)$ and $\epsilon_{ab}(\phi_i)$ 
are generic functions of $\phi_i$ including $\phi$-independent 
constants, and the ellipsis denotes the higher order terms in $X_a$. 
Note that, as mentioned before, only the $\epsilon_{ab}(\phi_i)$ terms 
are relevant to the recovery of SUSY vacua in the case of global SUSY. 
Now we have the total superpotential, $W=W_{OR}+W_{R\!\!\!\!/}$.
The F-flat conditions (\ref{eq:orgfflatsugrard}) are modified as 
\beq
D_{X_a}W = 
\sum_b M_{ab} \bigg( g_b(\phi_i) +\delta_{a1}f_1 
+\sum_{c,d}  M^{-1}_{bc} \epsilon_{cd}(\phi_i) X_d 
+\Delta^{-1} K_{X_b} W_{R\!\!\!\!/} \bigg)  =  0. 
\label{eq:orgfflatsugrarb}
\eeq
Here we find that all the terms in $W_{R\!\!\!\!/}$ 
including $c(\phi_i)$ are accompanied by $X_a$ in the 
above F-flat conditions and then have a possibility 
for restoring SUSY, contrary to the case of 
global SUSY explained in the previous subsection. 

Most notably, just a constant superpotential 
\begin{eqnarray}
W_{R\!\!\!\!/} &=& c, 
\label{eq:constantsp}
\end{eqnarray}
i.e., $c(\phi_i)=c$ and $\epsilon_{ab}(\phi_i)=0$, can restore 
SUSY. In this case with the minimal K\"ahler potential 
(\ref{eq:minkahler}), we find a solution for 
Eq.~(\ref{eq:orgfflatsugrarb}) as 
\begin{eqnarray}
\bar{X}_a &=& -c^{-1} \Delta\,g_a(\phi_i), 
\label{eq:orgsugragsol}
\end{eqnarray}
where $\Delta=1+\sum_a|X_a|^2$ defined in Eq.~(\ref{eq:det1kx}) 
is real and positive. 
{}From Eq.~(\ref{eq:orgsugragsol}), $X_a$ can be written in terms 
of $\phi_i$, and then $\Delta$ is given by 
\begin{eqnarray}
\Delta &=& \frac{|c|^2 \pm |c| 
\sqrt{|c|^2-4 \sum_a |g_a(\phi_i)|}}{2 \sum_a |g_a(\phi_i)|}, 
\nonumber
\end{eqnarray}
which should be a real number. 
Therefore, in order for the SUSY solution (\ref{eq:orgsugragsol}) to be 
valid, the constant superpotential $c$ must satisfy the condition 
\begin{eqnarray}
4 \sum_a |g_a( \langle \phi_i \rangle )|^2 \le |c|^2, 
\label{eq:orgsugrassc}
\end{eqnarray}
where $\langle \phi_i \rangle$ are solutions of $D_{\phi_i}W=0$ 
under the condition (\ref{eq:orgsugragsol}). 

Because $X_1$ is distinguished in the superpotential 
(\ref{eq:orgsprd}), we divide the generic R-breaking terms 
(\ref{eq:rbrkexg}) into two parts: 
\begin{eqnarray}
W_{R\!\!\!\!/} &=& 
W^{(A)}_{R\!\!\!\!/}+W^{(B)}_{R\!\!\!\!/}, 
\nonumber
\end{eqnarray}
where 
\begin{eqnarray}
W^{(A)}_{R\!\!\!\!/}(X_{a \ne 1};\,\phi_i) &=& 
c(\phi_i)+\frac{1}{2}\sum_{a,b \ne 1}
m\,\epsilon_{ab}(\phi_i) X_a X_b 
+\cdots, 
\label{eq:arb} \\
W^{(B)}_{R\!\!\!\!/}(X_1;\,X_{a \ne 1},\phi_i) &=& 
\sum_{a \ne 1} m\,\epsilon_{a1}(\phi_i) X_a X_1 
+\frac{1}{2} m\,\epsilon_{11}(\phi_i) X_1^2 
+\cdots. 
\label{eq:brb}
\end{eqnarray}
The ellipses denote the higher order terms 
in terms of $X_{a \ne 1}$ in $W^{(A)}_{R\!\!\!\!/}$, 
and those of $X_1$ and $X_{a \ne 1}$ in $W^{(B)}_{R\!\!\!\!/}$. 
Without loss of generality, we can assume that $\epsilon_{11}(0)$ 
is real and positive among $\epsilon_{ab}(0)$, which is referred 
as $\epsilon$ in Sec.~\ref{sec:typeb}.

\subsection{Type-A breaking: Polonyi-like models}
\label{sec:typea}
In this section, we study the effect of R-breaking terms 
(\ref{eq:arb}) which we call the A-type breaking, that is, the total superpotential is written by
$$
W \ = \ W_{OR}+W^{(A)}_{R\!\!\!\!/}. 
$$ 
Because this type of breaking terms does not contain $X_1$, 
we find the Polonyi model~\cite{Polonyi:1977pj} 
\begin{eqnarray}
W \big|_{X_{a \ne 1}=0,\phi_i=0} 
&=& W_{\rm Polonyi} \ \equiv \ fX_1+c, 
\label{eq:polonyi}
\end{eqnarray}
in the hypersurface $X_{a \ne 1}=0$, $\phi_i=0$ of the 
scalar potential, where $c=c(0)$. 
This hypersurface would be a stationary 
plane in the $X_{a \ne 1}$- and the $\phi_i$-directions 
if $\partial_{\phi_i} g_{a \ne 1}(0)$ are sufficiently 
large, which correspond to SUSY masses for 
$X_{a \ne 1}$ and $\phi_i$ on that plane. 

Moreover, if $m_1^i$ and/or $h_1^{ij}$ in 
\begin{eqnarray}
g_1(\phi_i) &=& m_1^i\phi_i +h_1^{ij} \phi_i \phi_j +\cdots, 
\label{eq:hx1phi2}
\end{eqnarray}
are nonvanishing, the Polonyi model in this hypersurface 
can be affected/modified by a tree-level SUSY mass 
and/or a one-loop SUSY breaking mass for $X_1$. 
Then, we further classify the A-type breaking models into 
two cases, $g_1(\phi_i)=0$ and $g_1(\phi_i) \ne 0$.

\subsubsection{Decoupled case: $g_1(\phi_i)=0$}
\label{sec:typea-decoupled}
In the case with $g_1(\phi_i)=0$, the superpotential of 
the A-type breaking models is written as 
\begin{eqnarray}
W &=& fX_1 +\sum_{a \ne 1} g_a(\phi_i)X_a 
+c(\phi_i)
+\frac{1}{2} \sum_{a,b \ne 1}m 
\epsilon_{ab}(\phi_i) X_a X_b +\cdots 
\nonumber \\ &=& 
c+fX_1+\frac{1}{2}\mu_{AB}\Phi_A \Phi_B +\cdots, 
\nonumber
\end{eqnarray}
where $\Phi_A=(X_{a \ne 1},\phi_i)$ with the index $A=(a \ne 1,i)$. 
The SUSY mass matrix $\mu_{AB}$ is given by 
the R-breaking components, 
$\mu_{a \ne 1,b \ne 1}=m\epsilon_{ab}(0)$, 
$\mu_{ij}= \partial_{\phi_i} \partial_{\phi_j} c(0)$ and 
the R-symmetric components, 
$\mu_{a \ne 1,i}=2 \partial_{\phi_i} g_a(0)$. 
After the unitary rotation which makes $\mu_{AB}$ diagonal, 
the above superpotential takes the form of 
\begin{eqnarray}
W &=& 
c+fX_1+\frac{1}{2}\mu_{A} \Phi_A^2 +\cdots, 
\label{eq:modelad}
\end{eqnarray}
where $\mu_A$ represents the eigenvalues of $\mu_{AB}$. 
Because of the SUSY mass $\mu_A$, the field 
$\Phi_A$ would be integrated out without affecting the 
low energy dynamics of $X_1$, because $X_1$ is completely 
decoupled in the present case\footnote{We may have to 
assume that the K\"ahler mixing is also zero or negligible 
between $X_1$ and the others.}. 

Then, the effective action for $X \equiv X_1$ is just determined 
by the Polonyi superpotential (\ref{eq:polonyi}), where the 
phase of $c$ and $f$ can be eliminated by the $U(1)_R$ 
rotation and the rephasing of $X_1$. Assuming the minimal 
K\"ahler potential (\ref{eq:minkahler}) for simplicity, 
the effective scalar potential is minimized by a real vacuum 
value $X=\bar{X}=x$ satisfying the stationary condition 
\begin{eqnarray}
V_X &=& e^GG_X(G_{XX}+G_X^2-2) \ = \ 0, 
\nonumber
\end{eqnarray}
where $G=K+\ln |W|^2$ and
\begin{eqnarray}
G_{XX}+G_X^2-2 &=& fW^{-1}(x^3 +f^{-1}c x^2 -2f^{-1}c), 
\label{eq:vxsb} \\
G_X &=& fW^{-1}(x^2 +f^{-1}c x +1). 
\label{eq:vxss}
\end{eqnarray}
The F-flat condition for $X$ corresponds to $G_X=0$, and 
the SUSY breaking stationary point is determined 
by the condition $G_{XX}+G_X^2-2=0$. 

As we declared, we persist in obtaining a vanishing vacuum 
energy at the SUSY breaking minimum. Then in 
addition to the stationary condition $G_{XX}+G_X^2-2=0$, 
we set $V=e^G(G^{X \bar{X}}|G_X|^2-3)=0$. 
In this case, we have to take a definite value 
of the constant $c$ and find two solutions 
\begin{eqnarray}
(x,\,f^{-1}c) &=& 
(\sqrt{3}-1,\,2-\sqrt{3}), 
\label{eq:plsol1}
\end{eqnarray}
and 
\begin{eqnarray}
(x,\,f^{-1}c) &=& 
(-\sqrt{3}-1,\,2+\sqrt{3}). 
\label{eq:plsol2}
\end{eqnarray}
The mass eigenvalues of $({\rm Re}X,\,{\rm Im}X)$ are computed 
as $(2\sqrt{3}f^2,\,(4-2\sqrt{3})f^2)$ for the first solution 
(\ref{eq:plsol1}), and $(-2\sqrt{3}f^2,\,(4+2\sqrt{3})f^2)$ for 
the second one (\ref{eq:plsol2}), at this SUSY breaking 
Minkowski stationary point where $W=f$. Then, only the first 
solution (\ref{eq:plsol1}) can be a minimum of the potential, 
while the second one (\ref{eq:plsol2}) is a saddle point. 
We comment that $\phi_i$ and $X_{a \ne 1}$ directions would not 
possess tachyonic masses at these points for sufficiently large 
SUSY mass $\mu_{A}$ compared with the SUSY 
breaking mass $f$. Therefore, the candidate for our present universe, 
where the SUSY is broken with (almost) vanishing vacuum energy, 
is the first solution (\ref{eq:plsol1}).

In addition to a SUSY breaking solution satisfying 
$G_{XX}+G_X^2-2=0$, we have a SUSY solution 
$G_X=0$ due to the R-breaking effect $c \ne 0$, that is, 
\begin{eqnarray}
x_\pm &=& \frac{1}{2}(-f^{-1}c \pm \sqrt{(f^{-1}c)^2-4}), 
\label{eq:plsusy}
\end{eqnarray}
if the R-breaking constant $c$ satisfies 
\begin{eqnarray}
|f^{-1}c| \ge 2. 
\label{eq:plsusycd}
\end{eqnarray}
Note that this condition (\ref{eq:plsusycd}) corresponds to 
Eq.~(\ref{eq:orgsugrassc}) in the previous general argument 
for the generalized OR model. 
The mass eigenvalues of $({\rm Re}X,\,{\rm Im}X)$ are computed as 
$W_\pm^2 (x_\pm^2-2)(x_\pm^2+1)$ and $W_\pm^2 (x_\pm^2-1)(x_\pm^2+2)$ 
at this SUSY AdS stationary point where 
\begin{eqnarray}
|W_\pm| &=& |fx_\pm +c| 
\ = \ \frac{1}{2}
\left| f(f^{-1}c \pm \sqrt{(f^{-1}c)^2-4}) \right| \ > \ 0, 
\nonumber
\end{eqnarray}
and then we obtain 
\begin{eqnarray}
V &=& -3e^G \ = \ -3e^{x_\pm^2}|W_\pm|^2 \ < \ 0. 
\nonumber
\end{eqnarray}
Remark that, in the vanishing (one of) R-breaking limit, 
$c \to 0$, the condition (\ref{eq:plsusycd}) is not satisfied, 
and the SUSY solution (\ref{eq:plsusy}) disappears. 
In the other words, this SUSY solution is a 
consequence of the R-breaking constant term $c$ in the 
superpotential. Due to the appearance of this SUSY 
solution, there is a possibility that the SUSY breaking 
point determined by $G_{XX}+G_X^2-2=0$ becomes a metastable vacuum 
as in the case of global SUSY explained previously. 

However, this is not the case. Interestingly, if we tune the 
R-breaking constant superpotential $c$ as $f^{-1}c=2-\sqrt{3}$ 
so that the solution (\ref{eq:plsol1}) with the vanishing vacuum 
energy is realized, the condition 
(\ref{eq:plsusycd}) is not satisfied and the SUSY 
stationary solution (\ref{eq:plsusy}) disappears. In such a sense, 
the constant R-breaking term $c$ does not lead to a metastability 
of SUSY breaking Minkowski minimum (\ref{eq:plsol1}).

Next, we consider the SUSY stationary solutions 
outside the Polonyi slice $X_{a \ne 1}=0$, $\phi_i=0$. 
For the superpotential (\ref{eq:modelad}), 
the F-flat directions are determined by 
\begin{eqnarray}
D_{\Phi_A}W &=& K_A W +\mu_A \Phi_A +\cdots \ = \ 0, 
\nonumber \\
D_{X_1}W &=& f +K_{X_1}W \ = \ 0, 
\nonumber
\end{eqnarray}
which can be satisfied by distinguishing a single field 
$\Phi_B \ne 0$ for  $^\exists B$ as 
\begin{eqnarray}
W &=& -K_B^{-1} \Phi_B (\mu_B +\cdots) 
\ = \ -K_{X_1}^{-1} f \qquad (\textrm{for} \ ^\exists B), 
\label{eq:ospsusysol} \\
\Phi_A &=& K_A \ = \ 0 \qquad 
(\textrm{for} \ A \ne B), 
\nonumber
\end{eqnarray}
where the ellipsis represents the higher order terms of $\Phi_B$. 
The first line gives two complex equations 
for two complex variables $X_1$ and $\Phi_{^\exists B}$, 
which have a solution in general. 

For example, if the K\"ahler potential is minimal 
(\ref{eq:minkahler}), all the parameters in the 
superpotential are real and there is no higher order terms 
of $\Phi_B$ (no ellipses in the above expressions), 
then the solution for Eq.~(\ref{eq:ospsusysol}) is found as 
\begin{eqnarray}
\Phi_B^2 &=& 
-2 \left( \frac{c}{\mu_B}
+\frac{f^2}{\mu_B^2} +1 \right) \ > \ 0, \qquad 
\Phi_{A \ne B} \ = \ 0. 
\label{eq:newsol}
\end{eqnarray}
For this value of $\Phi_B$, the remaining condition 
$D_{X_1}W=0$ is satisfied by 
$$
X_1=f/\mu_B. 
$$
Note that the number of these SUSY points 
is $n_X+n_\phi-1$ because the solution (\ref{eq:newsol}) 
is valid for every choice of $B=(b \ne 1,j)$. 
In order for the solution (\ref{eq:newsol}) to be valid, 
the parameter $\mu_B$ must satisfy 
\begin{eqnarray}
\mu_B^2 +c\mu_B +f^2 \ \le \ 0. 
\nonumber
\end{eqnarray}
This leads to the same condition (\ref{eq:plsusycd}) 
for the R-breaking constant term $c$ as in the 
Polonyi-type SUSY solution. 

In summary, the A-type breaking terms (\ref{eq:arb}) 
can restore SUSY in the generalized OR model (\ref{eq:orgsp}) 
or equivalently (\ref{eq:orgsprd}) in general. 
However, if we tune the R-breaking constant term in the 
superpotential so that the SUSY breaking minimum has a 
vanishing vacuum energy, i.e., (\ref{eq:plsol1}), the SUSY 
solutions (\ref{eq:plsusy}) and (\ref{eq:newsol}) disappear. 
Therefore, in this sense, the A-type R-symmetry breaking terms 
do not lead to a metastability of the SUSY breaking (Minkowski) 
vacuum aside from a possibility of the existence of more complicated 
SUSY solutions than (\ref{eq:newsol}).

\subsubsection{Generic case: $g_1(\phi_i) \ne 0$}
\label{sec:atypegen}
Now we turn on a nonvanishing $g_1(\phi_i)$ as in 
Eq.~(\ref{eq:hx1phi2}). With this term, the tree-level 
(field dependent) mass matrices in the $\phi_i=0$ plane 
contain the following contributions, 
\begin{eqnarray}
V_{X_1 \bar{X_1}} \big|_{\phi_l=0} 
&=& |m_1^i|^2+\cdots, 
\nonumber \\
V_{\phi_i \bar\phi_{\bar{j}}} \big|_{\phi_l=0}
&=& m_1^i \bar{m}_1^{\bar{j}}
+4h_1^{ik} \bar{h}_1^{\bar{j}\bar{k}}|X_1|^2 +\cdots, 
\nonumber \\
V_{\phi_i \phi_j} \big|_{\phi_l=0} 
&=& h_1^{ij}\bar{f} +\cdots, 
\nonumber \\
V_{X_1 \phi_i} \big|_{\phi_l=0}
&=& 2h_1^{ij}\bar{m}_1^{\bar{j}}X_1 +\cdots, 
\label{eq:g1mass}
\end{eqnarray}
where the ellipses represent the original terms 
involving $X_{a \ne 1}$, those coming from $c(\phi_i)$, 
and the supergravity corrections. Here the doubled 
indices are summed up. 
The K\"ahler covariant derivatives of the superpotential 
in the hypersurface $\phi_i=0$, $X_{a \ne 1}=0$ are given by 
$$
D_{X_1}W \big|_{\phi_l=0} \ = \ f +K_{X_1}W, \qquad 
D_{X_{a \ne 0}}W \big|_{\phi_l=0} \ = \ 0, \qquad 
D_{\phi_i}W \big|_{\phi_l=0} \ = \ m_1^i X_1. 
$$
{}From the third equation, we find that $\phi_i$ can not 
be integrated out prior to $X_1$ by the F-flat condition 
$D_{\phi_i}W=0$ unlike before. This is because, with the 
nonvanishing $m_1^i$, the source field $X_1$ for SUSY 
breaking shares a common SUSY mass with $\phi_i$ 
as shown in Eq.~(\ref{eq:g1mass}). 

In this case, the purely $X_1$-direction is no longer special 
in the scalar potential. We have to treat $X_1$ and $\phi_i$ 
at the same time. The analysis is quite complicated, and then 
we consider the case with $m_1^i=0$ in the following, where 
$g_1(\phi_i)$ starts from the quadratic term in $\phi_i$, and 
the $\phi_i$ can be integrated by their F-flat conditions 
$D_{\phi_i}W=0$ resulting $\phi_i=0$. We will comment about 
the case with $m_1^i \ne 0$ in Sec.~\ref{ssec:typebgen} together 
with more general R-breaking terms. 
The components of the mass matrices (\ref{eq:g1mass}) are 
now reduced to 
$$
V_{\phi_i \bar\phi_{\bar{j}}} \big|_{\phi_l=0}
\ = \ 
4h_1^{ik} \bar{h}_1^{\bar{j}\bar{k}}|X_1|^2 +\cdots, 
\qquad 
V_{\phi_i \phi_j} \big|_{\phi_l=0}
\ = \ 
h_1^{ij}\bar{f} +\cdots. 
$$
{}From the second equation, we observe that some linear 
combinations of ${\rm Re}\,\phi_i$ and ${\rm Im}\,\phi_j$ 
become tachyonic in the $\phi_i=0$ plane if $|h_1^{ij}f|$ 
dominate the SUSY mass for $\phi_i$. 
The $X_1$-dependence in the first one indicates that a 
SUSY breaking mass of $X_1$ is generated at the 
one-loop level, which is proportional to $h_1^{ij}$. 

Therefore, the effective potential after integrating out 
$\phi_i$ and $X_{a \ne 1}$ is given by 
\begin{equation}
V = V^{(0)}+V^{(1)},~ 
V^{(0)}  =  e^G (G^{X \bar{X}}|G_X|^2-3),~ 
V^{(1)}  =  m_X^2 |X|^2, 
\label{eq:qcorsugra}
\end{equation}
where $X \equiv X_1$, $G=K+ \ln |W|^2$, and the effective 
superpotential $W=W_{\rm Polonyi}$ is shown in 
Eq.~(\ref{eq:polonyi}). The one-loop mass $m_X$ is 
determined by $h_1^{ij}$ as well as $f$, which would be 
considered as an independent parameter in the effective action. 
The stationary condition $V_X=0$ results in~\cite{Abe:2006xp} 
$$
X \ \simeq \ 2 fc/m_X^2, 
$$
for $c \sim f \sim m_X \ll 1$ in the unit with $M_{Pl}=1$, and 
the vanishing vacuum energy at this minimum requires 
$$
c \ = \ f/\sqrt{3} +{\cal O}(f^3/m_X^2). 
$$
The SUSY is broken at this Minkowski minimum 
with $D_XW=f+{\cal O}(f^2)$ and $W=f/\sqrt{3}+{\cal O}(f^2)$.

\subsection{Adding type-B breaking: Metastable universe}
\label{sec:typeb}
In the previous section, we have analyzed the generalized OR model 
with the explicit R-symmetry breaking terms (\ref{eq:brb}) 
which do not involve the source field $X_1$ for the 
dynamical SUSY breaking. 

In this section, we study more general case with the 
R-breaking terms (\ref{eq:arb}) including $X_1$, i.e.,  
$$
W \ = \ W_{OR}+W^{(A)}_{R\!\!\!\!/}+W^{(B)}_{R\!\!\!\!/}.
$$
In the type-B breaking terms (\ref{eq:arb}), the first term 
with $\epsilon_{a \ne 1,1}(0)$ gives the common SUSY 
mass for $X_1$ and $X_{a \ne 1}$ in the $\phi_i=0$ plane. 
Then the situation is similar to the case with a nonvanishing 
$m_1^i$ in Eq.~(\ref{eq:hx1phi2}), that is, we can not integrate 
out $X_{a \ne 1}$ prior to $X_1$, and we will include this case 
also in Sec.~\ref{ssec:typebgen}. 

By setting $\epsilon_{a \ne 1,1}(0)=0$, the superpotential 
in the hypersurface $\phi_i=X_{a \ne 1}=0$ is given by 
\begin{eqnarray}
W &=& fX+\frac{1}{2}m \epsilon X^2+c
+\cdots, 
\label{eq:qdcpl}
\end{eqnarray}
where $X \equiv X_1$, $\epsilon=\epsilon_{11}(0)$ 
and the ellipsis stands for the higher order terms in $X$.

\subsubsection{Decoupled case: $g_1(\phi_i) =0$}
As in the previous section, we first consider the case 
with $g_1(\phi_i) =0$, where $X_1$ is decoupled from the others in 
the superpotential. In this case the hypersurface $\phi_i=X_{a \ne 1}=0$  
would be stable in the $\phi_i$-, $X_{a \ne 1}$-direction 
as in Sec.~\ref{sec:typea-decoupled}. The effective theory in this slice is 
described by the superpotential (\ref{eq:qdcpl}). 

With the minimal K\"ahler potential (\ref{eq:minkahler}), 
real parameters $f$, $c$, $m$ and no higher order terms (ellipsis) 
in the superpotential (\ref{eq:qdcpl}) for simplicity, 
the SUSY breaking and SUSY stationary 
conditions are respectively given by Eqs.~(\ref{eq:vxsb}) and 
(\ref{eq:vxss}). In the limit $\epsilon \to 0$ of Eq.~(\ref{eq:qdcpl}), 
the SUSY breaking solution is given by Eq.~(\ref{eq:plsol1}). 
Then we can find the deviation of $X$ from this point assuming 
$\epsilon \ll 1$ and $m \sim c^{1/3} \sim f^{1/2}$. We find a 
SUSY breaking minimum with a vanishing vacuum energy at 
\begin{eqnarray}
X_{SB} &=& X_0 +\delta X, \qquad 
X_0 \ = \ \sqrt{3}-1, \qquad 
\delta X \ = \ 
-\frac{\epsilon m}{2f} +{\cal O}(\epsilon^2), 
\label{eq:btypesbmin}
\end{eqnarray}
where the constant superpotential term $c$ is tuned as 
\begin{eqnarray}
c &=& 
(2-\sqrt{3})f 
+(2 \sqrt{3}-3) \epsilon m 
+{\cal O}(\epsilon^2).
\label{eq:cbtype}
\end{eqnarray}

On the other hand, a SUSY solution, 
\begin{eqnarray}
X_{SUSY} &\simeq& -\frac{2f}{\epsilon m}, 
\label{eq:btypessmin}
\end{eqnarray}
arises as a consequence of the B-type R-breaking term represented by 
the parameter $\epsilon$, although the vacuum energy is set to be 
vanishing at the SUSY breaking minimum. This is unlike the 
case of SUSY solutions (\ref{eq:plsusy}) and (\ref{eq:newsol}) 
caused by the introduction of A-type R-breaking terms (\ref{eq:arb}). 
The shift of SUSY breaking minimum $\delta X$ in Eq.~(\ref{eq:btypesbmin}) 
is rewritten as 
$$
\delta X/X_0 \ \simeq \ 
\frac{1}{\sqrt{3}-1}\,\frac{1}{X_{SUSY}},
$$ 
and we find 
$$
|X_{SUSY}| \ > \ \frac{1}{\sqrt{3}-1} \ \sim \ {\cal O}(1),
$$
in order for the shift $\delta X$ to reside in a perturbative region, 
$|\delta X/X_0| < 1$. 

This means that the vacuum value of $|X|$ at the 
newly appeared SUSY vacuum must be larger than the Planck 
scale $M_{Pl}=1$, where the supergravity calculation might not be valid. 
It would be possible that the potential is lifted for $|X|>1$ by 
the effect of quantum gravity, the above SUSY vacuum is 
washed out and the SUSY breaking minimum remains as a global 
minimum. If the supergravity approximation is valid even for $|X|>1$ 
by any reason, we obtain a constraint on the R-breaking parameter 
$\epsilon$ as 
$$\epsilon \ < \ 2(\sqrt{3}-1)|f/m|,$$ 
from the above condition. 

\begin{figure}[t]
\centerline{\epsfig{file=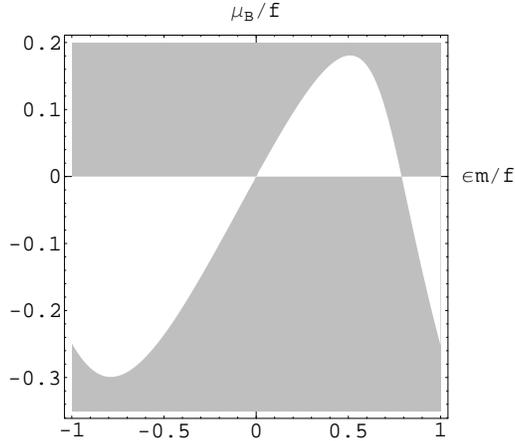,width=0.5\linewidth}}
\caption{Parameter region (white) of $\mu_B$ and $\epsilon$
allowing the SUSY solution (\ref{eq:newsolgen}). 
All the parameters are assumed to be real and the constant 
term $c$ is fixed by the vanishing vacuum energy condition 
(\ref{eq:cbtype}) at the SUSY breaking minimum 
(\ref{eq:btypesbmin}). In the shaded region, the SUSY 
solution (\ref{eq:newsolgen}) is not allowed and the 
SUSY breaking solution (\ref{eq:btypesbmin}) 
does not become metastable due to the R-breaking effect 
parameterized by $\epsilon$. We find no allowed region 
in the limit $\epsilon \to 0$ which corresponds to the 
solution (\ref{eq:newsol}).}
\label{fig:1}
\end{figure}

We also find a SUSY minimum outside the hyperplane 
$\phi_i=X_{a \ne 1}=0$, which is a generalization 
of Eq.~(\ref{eq:newsol}), given by 
\begin{eqnarray}
\Phi_B^2 &=& 
-\frac{2}{\mu_B}\, 
\left\{ \mu_B +c 
+\frac{f^2}{\mu_B-\epsilon m} 
\left( 1+\frac{\epsilon m}{2(\mu_B-\epsilon m)} \right) 
\right\} \ \ge \ 0, 
\nonumber \\
\Phi_{A \ne B} &=& K_{A \ne B} \ = \ 0, \qquad 
X \ = \ \frac{f}{\mu_B-\epsilon m}, 
\label{eq:newsolgen}
\end{eqnarray}
where we assumed the minimal K\"ahler potential 
(\ref{eq:minkahler}), and the absence of the higher 
order terms of $X$ in the superpotential for concreteness. 
In the limit $\epsilon \to 0$, this solution is 
reduced to (\ref{eq:newsol}). 
In contrast to (\ref{eq:newsol}), the above solution 
(\ref{eq:newsolgen}) does not disappear in all of the parameter 
region, even after the vacuum energy at the SUSY 
breaking minimum is set to zero as in Eq.~(\ref{eq:btypesbmin}). 
Such parameter region of $\mu_B$ and $\epsilon$ allowing the 
SUSY solution is shown in Fig.~\ref{fig:1}. 
In the shaded region, the SUSY solution 
(\ref{eq:newsolgen}) is not allowed and the SUSY 
breaking solution does not become metastable due to the 
R-breaking effect represented by $\epsilon$. 
Note that we find no allowed region along the $\epsilon = 0$ 
axis, which corresponds to the case of the solution 
(\ref{eq:newsol}).

\subsubsection{Generic case: $g_1(\phi_i) \ne 0$}
\label{ssec:typebgen}
Finally we introduce nonvanishing $g_1(\phi)$. As in subsection~\ref{sec:atypegen}, 
we first consider the case with $m_1^i=0$ in Eq.~(\ref{eq:hx1phi2}). 
In this case we can still integrate $\phi_i$ and $X_{a \ne 1}$ by 
use of $D_{\phi_i}W=D_{X_{a \ne 1}}W=0$ 
resulting in $\phi_i=X_{a \ne 1}=0$. 

The remnant of these heavy fields would be the one-loop mass 
$m_X$ for $X_1=X$ in Eq.~(\ref{eq:qcorsugra}). The effective 
scalar potential is in the same form as Eq.~(\ref{eq:qcorsugra}) 
but the effective superpotential $W$ in $G=K+ \ln |W|^2$ is now 
replaced by Eq.~(\ref{eq:qdcpl}). 
For $\epsilon \ll c \sim f \sim m_X \ll 1$ in the unit with $M_{Pl}=1$, 
we can obtain a SUSY breaking Minkowski minimum 
\begin{eqnarray}
X_{SB} &=& \frac{2fc}{m_X^2}\,\big( 1+{\cal O}(\epsilon^2) \big), 
\label{eq:btypesbminmx}
\end{eqnarray}
where the R-breaking constant 
$$
c \ = \ f/\sqrt{3}+{\cal O}(f^3/m_X^2;\,\epsilon^2), 
$$
is determined by the vanishing vacuum energy condition. 

The SUSY ground state in the hyperplane 
$\phi_i=X_{a \ne 1}=0$ which originates from the R-breaking 
parameter $\epsilon$ is the same as Eq.~(\ref{eq:btypessmin}), 
and the above breaking minimum becomes metastable. 
Unlike (\ref{eq:btypesbmin}), the SUSY breaking minimum 
(\ref{eq:btypesbminmx}) is not affected by the R-breaking term 
at ${\cal O}(\epsilon)$ due to the one-loop mass $m_X$, that is, 
the SUSY minimum (\ref{eq:btypessmin}) is independent 
of the SUSY breaking minimum (\ref{eq:btypesbminmx}) 
at this order. There might exist SUSY points 
analogous to Eq.~(\ref{eq:btypessmin}) outside the hypersurface 
$\phi_i=X_{a \ne 1}=0$ also in this case, but the solution 
would be more complicated due to the nonvanishing $h_1^{ij}$ 
in Eq.~(\ref{eq:hx1phi2}). 

Finally we comment about the case with $m_1^i \ne 0$ in 
Eq.~(\ref{eq:hx1phi2}). In this case, as mentioned in 
Sec.~\ref{sec:atypegen}, the field $X_1$ has a SUSY 
mass with the same magnitude as those of $\phi_i$'s as shown in 
Eq.~(\ref{eq:g1mass}). Then the field $X_1$ in the field basis 
(\ref{eq:orgsprd}) is no longer special. In this generalized OR model 
with most general R-breaking terms, the total superpotential 
would be written as 
\begin{eqnarray}
W &=& fX_1 +\sum_{a=1} g_a(\phi_i)X_a 
+c(\phi_i)
+\frac{1}{2} \sum_{a,b=1} m 
\epsilon_{ab}(\phi_i) X_a X_b +\cdots 
\nonumber \\ &=& 
c+fX_1+\frac{1}{2}\mu_{IJ}\Phi_I \Phi_J +\cdots, 
\nonumber
\end{eqnarray}
where $\Phi_I=(X_a,\phi_i)$, $I=(a,i)$ and the ellipses 
denote the higher order terms in $\Phi_I$. 
The SUSY mass matrix $\mu_{IJ}$ is given by 
the R-breaking components, 
$\mu_{ab}=m\epsilon_{ab}(0)$, 
$\mu_{ij}= \partial_{\phi_i} \partial_{\phi_j} c(0)$ and 
the R-symmetric components, 
$\mu_{ai}=2 \partial_{\phi_i} g_a(0)$. 
Note that $\mu_{1i}=2\partial_{\phi_i} g_1(0)=2m_1^i$. 
After the unitary rotation which makes $\mu_{IJ}$ diagonal, 
the above superpotential takes the form of 
$$
W \ = \ 
c+fU_{1I}\Phi_I+\frac{1}{2}\mu_{I} \Phi_I^2 +\cdots, 
$$
where $U_{IJ}$ is the rotation matrix and 
$\mu_I$ represents the eigenvalues of $\mu_{IJ}$. 
The F-flat conditions, $D_IW=W_I+K_IW=0$, allow a solution 
in general and SUSY would not be broken for 
$m_1^i \sim f$.

\subsection{Conclusion in section 2}
\label{sec:conclusion rsugra}

In section \ref{sec:global-SUSY}, we considered $N=1$ global supersymmetric models 
with a continuous global $U(1)_R$ symmetry. We discussed the features of models with SUSY breaking vacua and runaway directions. For example, models with fields whose R-charges are negative and/or more than 2 R-charge have runaway directions. Furthermore, models which satisfy $N_X > N_{\omega }$ have no solution for F-flat conditions, so that the condition $N_X > N_{\omega }$ is a sufficient condition for SUSY breaking. The generalized O'Raifeartaigh model satisfies the condition.   

In section \ref{sec:rsugra explicit R-breaking in global SUSY}, we studied the effect of explicit R-symmetry breaking terms in detail. 
In global supersymmetric models, based on the argument by ISS, 
we have shown that a specific type of explicit R-symmetry breaking 
terms can restore SUSY, and the original SUSY breaking vacuum 
can become metastable when a certain (but not generic) class 
of explicit R-symmetry breaking terms are added and/or loop 
effects stabilize the original SUSY breaking minimum. 

We also considered $N=1$ local supersymmetric models 
with a continuous global $U(1)_R$ symmetry in section \ref{sec:rsugra}.
We have executed similar analyses in R-symmetric supergravity models. 
First we examined the general argument by NS in supergravity and 
found that it also holds with local SUSY except for the  
nontrivial case where the K\"ahler potential allows solution for 
the second condition in Eq.~(\ref{eq:susyvc2}). We presented 
concrete examples of this exception.
These models lead to AdS SUSY 
stationary solutions and associated SUSY breaking vacua with 
lower vacuum energy.
We found the general argument that this 
class of SUSY solutions corresponds to at best a saddle point, 
referring to Appendix~\ref{app:rmass}. 

In section \ref{sec:r-breaking-sugra}, \ref{sec:typea}, and \ref{sec:typeb}, we studied the generalized OR model in supergravity with 
explicit R-symmetry breaking terms. We analyzed the structure of 
newly appeared SUSY stationary points as a consequence of the 
R-breaking effect and classified them. We have shown that these 
SUSY solutions disappear for type-A breaking terms (\ref{eq:arb}), 
when we tune the R-breaking constant term in the superpotential 
such that the original SUSY breaking minimum has 
a vanishing vacuum energy. In this sense, the introduction of 
explicit R-breaking terms do not always lead to a metastability 
of the SUSY breaking vacuum. 
On the other hand, the introduction of type-B breaking terms 
(\ref{eq:brb}) could cause a metastability of SUSY Minkowski 
minimum. We examined a parameter region which yields metastable 
vacuum in some concrete examples.

\section{Metastable supersymmetry breaking vacua from 
conformal dynamics }

In section 2, we have argued that an approximately R-symmetric superpotential 
with tiny R-symmetry breaking terms is favored to avoid conflicts with experimental results.
In this section, we suggest the models to realize the tiny R-symmetry breaking terms and cause metastable SUSY breaking vacua effectively.

We start with a superpotential without R-symmetry.
Based on the Nelson-Seiberg argument \cite{Nelson:1993nf}, SUSY would not be broken in this situation. 
However, we assume the conformal dynamics.
Because of that, certain couplings are exponentially 
suppressed at a low-energy scale.
Then, we could realize an R-symmetric superpotential 
or an approximate R-symmetric superpotential 
with tiny R-symmetry breaking terms.
It would lead to a stable or metastable SUSY breaking vacuum.
We study this scenario by using a simple model. Also, we study 5D models, which have the same behavior.

As the other good point in our model, 
contact terms between the hidden conformal sector and the visible sector are
suppressed exponentially by conformal dynamics.
As we discuss in the introduction, such conformal suppression mechanism, i.e. conformal 
sequestering, is quite important to model building for SUSY
breaking \cite{Luty:2001jh,Dine:2004dv,Sundrum:2004un,Ibe:2005pj,
Schmaltz:2006qs,Murayama:2007ge}.
The suppression can lead the situation that flavor-blind contributions such as 
anomaly mediation \cite{Randall:1998uk} would become dominant.

This section is organized as follows.
In section \ref{sec:4D conformal model}, we give a 4D simple model 
to realize our conformal scenario.
In section \ref{sec:5D model}, we study 5D models, which have 
the same behavior.
Section \ref{sec:short summary in 3} is short summary of our models.


\subsection{4D conformal model}
\label{sec:4D conformal model}

Our model is the $SU(N)$ gauge theory with $N_f$ flavors 
of chiral matter fields $\phi_i$ and $\tilde \phi_i$, which are 
fundamental and anti-fundamental representations of 
$SU(N)$.
The flavor number satisfies $3N \geq  N_f \geq \frac32 N$, 
and that corresponds to the conformal window 
\cite{Seiberg:1994pq,Intriligator:1995au},  
that is, this theory has an IR fixed point \cite{Banks:1981nn}.
The  Novikov-Chifman-Veinstein-Zaharov (NSVZ) beta-function of physical gauge coupling 
$\alpha=g^2/8\pi^2$ is  
\begin{equation}
\beta^{\rm NSVZ}_\alpha= - \frac{\alpha^2}{1-N\alpha} (3N-N_f+N_f \gamma_\phi),
\label{eq:NSVZ}
\end{equation}
where $\gamma_\phi$ is the anomalous dimension of 
$\phi_i$ and $\tilde \phi_i$ \cite{Novikov:1983uc, ArkaniHamed:1997ut}.
Since the IR fixed point corresponds to $\beta^{\rm NSVZ}_\alpha=0$, 
around that point the matter fields $\phi_i$ and $\tilde \phi_i$ have 
anomalous dimensions $\gamma_\phi= -(3N-N_f)/N_f$, which 
are negative.

In addition to the fields $\phi_i$ and $\tilde \phi_i$, 
we introduce singlet fields $\Phi_{ij}$ for $i,j=1,\cdots,N_f$.
The gauge invariance allows the following superpotential 
at the renormalizable level,
\begin{equation}
W=h\phi_i\Phi_{ij}\tilde \phi_j + f {\rm Tr}_{ij }\Phi_{ij} + 
\frac{m}{2} {\rm Tr}_{ik} \Phi_{ij}\Phi_{jk} + \frac{\lambda}{3} 
{\rm Tr}_{i\ell}\Phi_{ij}\Phi_{jk}\Phi_{k\ell}.
\label{W-1}
\end{equation}
Here we have preserved the $SU(N_f)$ flavor symmetry.
Even if the $SU(N_f)$ flavor symmetry is broken, e.g 
by replacing $f {\rm Tr}_{ij }\Phi_{ij}$ by $f_{ij}\Phi_{ij}$, 
the following discussions would be valid.
For simplicity, we assume that all of couplings, 
$h$, $f$, $m$, $\lambda$, are real, although the following 
discussions are available for the model with 
complex parameters, $h$, $f$, $m$ and $\lambda$.
We can add the mass terms of $\phi_i$ and $\tilde \phi_j$ 
to the above superpotential.
We will comment on such terms later, but 
at the first stage we study the superpotential 
without the mass terms of $\phi_i$ and $\tilde \phi_j$.

If $m =\lambda =0$, the above superpotential 
corresponds to the superpotential of the 
Intriligator-Seiberg-Shih (ISS2) model 
\cite{Intriligator:2006dd}.\footnote{The flavor number does not satisfy $3N \geq  N_f \geq \frac32 N$ which corresponds to the conformal window in the ISS2 model\cite{Intriligator:2006dd}. However, we call the model we introduce here the ISS2 model.} The ISS2 model corresponds to the generalized OR model we discuss in the section 2: $\Phi_{ij}$ corresponds to the R-charge 2 field described as $X_a$, and $\phi_i$, $\tilde{\phi}_i$ corresponds to the R-charge 2 field described as $\phi$ in the generalized OR model. As we discuss later, the ISS2 model always causes SUSY breaking without explicit R-symmetry breaking terms.\footnote{The gaugino condensation contribution can be included in explicit R-symmetry breaking terms in \cite{Intriligator:2006dd}.}

We consider that our theory is an effective theory with 
the cutoff $\Lambda$.
We assume that dimensionless parameters $h$ and $\lambda$ 
are of $O(1)$ and dimensionful parameters 
$f$ and $m$ satisfy $f \approx m^2$ and $m \ll \Lambda$.
We denote physical couplings as 
$\hat h=(Z_\phi Z_{\tilde \phi} Z_\Phi)^{-1/2}h$, 
$\hat f_{ij} = (Z_\Phi)^{-1/2} f_{ij}$, 
$\hat m = (Z_\Phi)^{-1} m$ and $\hat \lambda = (Z_\Phi)^{-3/2} \lambda$, 
where $Z_\phi, Z_{\tilde \phi}, Z_\Phi$ are wavefunction renormalization 
constants for $\phi,\tilde \phi, \Phi$, respectively.

The F-flat conditions are obtained as 
\begin{eqnarray}
\partial_{\Phi_{ij}}W &=& h\phi_i \tilde \phi_j + f\delta_{ij} 
+ m \Phi_{ij} + \lambda \Phi_{jk}\Phi_{ki}
=0, \\
\partial_{\phi_i} W &=& h \Phi_{ij}\tilde \phi_j =0, \\
\partial_{\tilde \phi_j} W &=& h \phi_i \Phi_{ij} =0.
\end{eqnarray}
These equations have a supersymmetric solution for generic values of 
parameters, $h, m, \lambda$.
To see such a supersymmetric solution, 
following \cite{Intriligator:2006dd} we decompose $\phi, \tilde \phi$ 
and $\Phi$ as 
\begin{equation}
\Phi = \left(
\begin{array}{cc}
Y & Z^T \\
\tilde Z & X
\end{array}
\right), 
\qquad 
\phi = \left(
\begin{array}{c}
\chi \\
\rho
\end{array}
\right),
\qquad 
\tilde \phi^T = \left(
\begin{array}{c}
\tilde \chi \\
\tilde \rho
\end{array}
\right),
\end{equation}
where $Y$, $\chi$ and $\tilde \chi$ are $N\times N$ matrices, 
$X$ is an $(N_F-N)\times (N_F-N)$ matrix, 
$Z$, $\tilde Z$, $\rho$ and $\tilde \rho$ are 
$(N_F-N)\times N$ matrices.
Let us consider the slice with $Z=\tilde Z = \rho =0$ 
in the field space, where the first derivatives of $W$ 
reduce to 
\begin{equation}
W_{\Phi_{ij}} = \left(
\begin{array}{cc}
f\delta_{ij} + h \chi_i \tilde \chi_j + mY_{ji} + \lambda 
Y_{jk}Y_{ki} & 0 \\
0 & f\delta_{ij} + mX_{ji} + \lambda X_{jk}X_{ki} 
\end{array}
\right),
\end{equation}
\begin{equation}
W_{\phi_i}^T= \left(
\begin{array}{c}
hY_{ij}\tilde \chi_j \\
0
\end{array}
\right), \qquad 
W_{\tilde \phi_j} = \left(
\begin{array}{c}
h\chi_i Y_{ij} \\
0
\end{array}
\right).
\end{equation}
Here, we have used the same indices for 
$\Phi_{ij}$, $\phi_i$, $\tilde \phi_j$ and their submatrices.
Thus, the fields $X_{ij}$ and the others are decoupled 
in the F-flat conditions, $W_{\Phi_{ij}}=W_{\phi_i}=W_{\tilde
  \phi_j}=0$.
The F-flat condition $W_{\Phi_{ij}}=0$ for $X_{ij}$ has a 
solution as $X_{ij}=x_s\delta_{ij}$ with 
\begin{equation}
x_s=\frac{-m \pm \sqrt{m^2-4f\lambda}}{2\lambda} .
\end{equation}
The F-flat conditions $W_{\Phi_{ij}}=W_{\phi_i}=W_{\tilde
  \phi_j}=0$ for $Y_{ij}, \chi_i$ and $\tilde \chi_j$ have 
the following solution,
\begin{equation}
f\delta_{ij} + h \chi_i \tilde \chi_j = 0, \qquad  Y_{ij}=0.
\label{sol-Y}
\end{equation}
In addition, the D-flat conditions correspond to 
$|\chi_i| = |\tilde \chi_i|$.

There is another solution,  
$\chi_i=\tilde \chi_j=0$ and $Y_{ij}=x_s\delta_{ij}$.
However, only the above solution (\ref{sol-Y}) survives at the 
IR region, as $\hat m$ and $\hat \lambda$ become to vanish as 
we will see later.
Thus, we concentrate to the solution (\ref{sol-Y}).
At any rate, the superpotential (\ref{W-1}) does not have R-symmetry, 
and there is a supersymmetric minimum.

The above aspect is the behavior of this model around 
the energy scale $\Lambda$.
Now let us study the behavior around the IR region.
We assume that the gauge coupling is around the IR fixed point, 
i.e. $\beta_\alpha \approx 0$, and that $\phi_i$ and $\tilde \phi_i$ 
have negative anomalous dimensions $\gamma_\phi$.
In addition, we assume that the physical Yukawa coupling $\hat h$ 
is driven toward IR fixed points.
The beta-function of $\hat h$ is obtained as 
\begin{equation}
\beta_{\hat h} = \hat h (\gamma_\phi +\gamma_{\tilde \phi} + \gamma_\Phi).
\end{equation}
The condition of the fixed point leads to 
$2 \gamma_\phi + \gamma_\Phi =0$.
Since $\gamma_\phi < 0$, we obtain a positive anomalous dimension 
for $\Phi_{ij}$.
Then, physical couplings, $\hat f$, $\hat m$ and 
$\hat \lambda$, are suppressed exponentially 
toward the IR direction as 
\begin{eqnarray}
\hat f(\mu) &=& \left( \frac{\mu}{\Lambda} \right)^{\gamma_\Phi}
\hat f(\Lambda), \qquad   
\hat m(\mu) = \left( \frac{\mu}{\Lambda} \right)^{2 \gamma_\Phi}
\hat m(\Lambda), \nonumber \\   
\hat \lambda(\mu) &=& \left( \frac{\mu}{\Lambda} \right)^{3\gamma_\Phi}
\hat \lambda(\Lambda).
\end{eqnarray}
Thus, the mass parameter $\hat m$ and 3-point coupling $\hat \lambda$ 
are suppressed faster than $\hat f$.
If we neglect $\hat m$ and $\hat \lambda$ but not 
$\hat f$, the above superpotential becomes 
the superpotential of the ISS2 model, and 
there is a SUSY breaking minimum around $\Phi_{ij}=0$ because of 
the rank condition.

Let us see more explicitly.
We concentrate ourselves to the potential of 
the fields $X_{ij}$, because $X_{ij}$ 
contribute to SUSY breaking in the ISS2 model.
Furthermore, we consider their overall direction, 
i.e. $X_{ij}=x\delta_{ij}$, and we use the 
canonically normalized basis, $\hat x$.
Then, the above superpotential (\ref{W-1}) leads to the 
following scalar potential,
\begin{equation}
V_{\rm SUSY} = (N_f -N)|\hat f + \hat m \hat x +\hat \lambda \hat x^2|^2 .
\end{equation}
In addition, around $\hat x=0$, SUSY is broken and 
that generates one-loop effective potential of $\hat x$.
Around $\hat x=0$, the mass term $m_x^2|\hat x|^2$ 
in the one-loop effective potential would be important.
Hence, we analyze the potential, $V= V_{\rm SUSY} + m_x^2|\hat x|^2$, 
and we use $m_x^2$, which has been calculated in
\cite{Intriligator:2006dd}, i.e.
\begin{equation}
m_x^2=\frac{\hat h^3 \hat f}{8 \pi^2}N(N_f - N) (\log 4 -1 ).
\end{equation}
Note that $m_x^2$ is suppressed toward the IR region like $\hat f$.
We consider only the real part of $\hat x$.
The stationary condition $\partial_{\hat x} V=$ is written as 
\begin{equation}
(\hat f + \hat m \hat x +\hat \lambda \hat x^2)(\hat m + 2\hat \lambda
\hat x) + m_x^2\hat x=0.
\label{st-cod}
\end{equation}
At a high energy scale corresponding to $Z_\Phi=O(1)$, 
we have $|\hat f|, |\hat m|^2 \gg m^2_x$, because 
$m_x^2$ is smaller than $\hat f$ by a loop factor.
The potential and the stationary condition are 
controlled by $|\hat f|, |\hat m|^2$, $\hat \lambda$, 
but not $m_x$.
Thus, there is no (SUSY breaking) minimum around $x=0$, 
but we have a supersymmetric minimum 
\begin{equation}
\hat x_s =\frac{-\hat m \pm 
\sqrt{\hat m^2-4\hat f\hat \lambda}}{2\hat \lambda} .
\label{susy-min}
\end{equation}
However, toward the IR direction, 
$\hat m^2$ becomes suppressed faster than $m_x^2$.
Then, the couplings $\hat f$ and $m_x^2$ are important 
in the potential.
Around $\hat x=0$, the stationary condition (\ref{st-cod}) becomes 
\begin{equation}
\hat f \hat m + m_x^2\hat x + \cdots = 0,
\end{equation}
that is, the stationary condition is satisfied 
with 
\begin{equation}
\hat x_{sb} \approx - \frac{\hat f \hat m}{m_x^2}.
\end{equation}
At this point, SUSY is broken, 
and this point becomes close to $\hat x_{sb} =0$  toward the IR.
Around $\hat x=0$, the size of mass is estimated by 
$m_x$, because the other terms are suppressed.
Hence, the SUSY breaking metastable vacuum 
corresponding to $\hat x \sim 0$ 
appears at the IR energy scale, where $\hat m^2 \ll m_x^2$.
Moreover, the previous SUSY vacuum (\ref{susy-min}) 
moves to a point far away from the origin $\hat x=0$, 
because it behaves like
\begin{equation}
\hat x_s =\frac{-\hat m \pm \sqrt{\hat m^2-4\hat f\hat \lambda}}{2\hat \lambda} 
\sim \left( \frac{\Lambda}{\mu} \right) ^{\gamma_\Phi}.
\end{equation}

Both breaking scales of the $SU(N)$ gauge symmetry and supersymmetry 
at the metastable SUSY breaking point $\hat x =0$ are 
determined by $O(\hat f(\mu))$.
Thus, such an energy scale is estimated as 
$\mu_{IR}^2 \sim \hat f(\mu_{IR})$, i.e. 
\begin{equation}
\mu_{IR} \sim \left( \frac{\hat f(\Lambda)}{\Lambda^{\gamma_\Phi}} 
\right)^{1/(2-\gamma_\Phi)},
\end{equation}
and at this energy scale conformal renormalization group flow 
is terminated.

So far, we have assumed that the mass term of 
$\phi_i$ and $\tilde \phi_i$, $m_\phi \phi_i \tilde \phi_i$ vanishes.
Here, we comment on the case with such terms. 
The physical mass $\hat m_\phi$ becomes enhanced as 
\begin{equation}
\hat m_\phi(\mu) = \left( \frac{\mu}{\Lambda} \right)^{2 \gamma_\phi} 
 \hat m_\phi(\Lambda),
\end{equation}
because of the negative anomalous dimension $\gamma_\phi$.
At $\mu \sim \hat m_\phi(\mu)$, the matter fields $\phi_i$ $\tilde \phi_i$
decouple and this theory removes away from the conformal window.
Thus, if $\hat m_\phi(\mu) > \mu_{IR}$, the conformal renormalization 
group flow is terminated at $\mu_D \sim  \hat m_\phi(\mu_D)= 
(\mu_D/\Lambda)^{2\gamma_\phi}\hat m_\phi(\Lambda)$.

We have studied the scenario that 
conformal dynamics leads to metastable SUSY breaking 
vacua.
As an illustrating example of our idea, 
we have used the simple model.
Our scenario could be realized by other models.

\subsection{5D model}
\label{sec:5D model}

There would be an AdS dual to our conformal scenario.
Indeed, we can construct simply various models 
within the framework of 5D orbifold theory.
Renormalization group flows in the 4D theory correspond to 
exponential profiles of zero modes like $e^{-c_i Ry}$, where 
$R$ is the radius of the fifth dimension,\footnote{
We assume that the radion is stabilized.} 
$y$ is the coordinate for the extra dimension, i.e. $y=[0,\pi]$ and 
$c_i$ is a constant.
The parameter $c_i$ corresponds to anomalous dimension in the 4D theory,
and each field would have a different constant $c_i$.
In 4D theory, values of anomalous dimensions are constrained 
by concrete 4D conformal dynamics.
However, constants $c_i$ do not have such strong constraints, 
although they would correspond to some charges.
Hence, 5D models would have a rich structure 
and one could make model building rather simply.
Here we show a simple 5D model.
We consider the 5D theory, whose 5-th dimension is 
compactified on $S^1/Z_2$. 
Two fixed points on $S^1/Z_2$ correspond to $y=0$ and $y=\pi$.
We introduce three bulk fields $X$, $\phi_1$, $\phi_2$.
They correspond to chiral multiplets of 
bulk hyper-multiplets and zero modes of their partners in hyper-multiplets 
$X^c$, $\phi_1^c$, $\phi_2^c$
are projected out by the $Z_2$ orbifold projection.
We assume that zero mode profiles of $X$, $\phi_1$ and $\phi_2$ 
behave along the $y$ direction as $e^{-c_X R y}$, $e^{-c_1 Ry}$ and 
$e^{-c_2 Ry}$, respectively.
We integrate $y$ and obtain their kinetic term coefficients $Y_i$ of 
4D effective theory, that is, the field corresponding to 
the zero mode profile  $e^{-c_i Ry}$ has the following 
kinetic term coefficient \cite{Gherghetta:2000qt,Marti:2001iw}
\begin{equation}
Y_i =\frac{1}{c_i}\left( 1 - e^{-2c_i\pi R} \right).
\end{equation}
In the limit $c_i \rightarrow 0$, $Y_i$ becomes $2\pi R$.
Their superpotential is not allowed in the bulk, 
but is allowed on the boundary.

Suppose that the following superpotential is allowed 
only on the $y=\pi$ boundary,
\begin{equation}
\int dy \delta (y-\pi) W^{(\pi)},
\end{equation}
\begin{eqnarray}
W^{(\pi)} & = & fe^{-c_X Ry}X + me^{-2c_X Ry}X^2 + 
h  e^{-3c_X Ry}X^3  \nonumber \\
 & & +m_{12} e^{-(c_1+c_2) Ry}\phi_1 \phi_2
+ m_2 e^{-2c_2 Ry}\phi_2^2  \\
 & & +\sum_{i,j} h_{ij} e^{-(c_X+c_i+c_j) Ry} X \phi_i \phi_j.
\nonumber
\end{eqnarray}
Here we have assumed extra $Z_2$ symmetry, under which 
$X$ has the even $Z_2$ charge and $\phi_1$ 
and $\phi_2$ have the odd $Z_2$ charge.
That allows the mass term $m_{11}\phi^2$, but 
we have assumed it vanishes by the same reason as 
why we did not add the mass term $m_{ij} \phi_i \tilde \phi_j$ 
in the superpotential (\ref{W-1}).
We assume that $f \approx m^2 \approx m_{12}^2 \approx m_2^2$ 
and $h,h_{ij} =O(1)$.
We take  
\begin{equation}
c_1 = 0, \qquad c_2=c_X,
\end{equation}
and $c_X >0$ with $c_X\pi R =O(1)$ and 
$e^{-c_X\pi R }\ll 1$.
The 4D superpotential $\hat W$ becomes 
\begin{equation}
\hat W = e^{-c_X\pi R}(fX+m_{12}\phi_1\phi_2 + h_{11}X\phi_1^2)
 + e^{-2c_X\pi R}\Delta W.
\label{W-4D}
\end{equation}
When we neglect $\Delta W$, the superpotential $W$ 
corresponds to the O'Raifeartaigh model \cite{O'Raifeartaigh:1975pr}, 
that is, SUSY is broken.
Such a minimum is metastable and there is a SUSY minimum, 
when we take into account $\Delta W$ \cite{Abe:2007ax}.
The O'Raifeartaigh model with the following superpotential,
\begin{equation}
W_{0} = \hat f X + \hat m_{12} \phi_1 \phi_2 
+ \hat h_{11} X \phi_1^2,
\end{equation}
leads to the SUSY breaking minimum of scalar potential 
$V=|\hat f|^2$ at $\phi_1=\phi_2=0$ and arbitrary $X$, 
that is, it has the pseudo-flat direction.
One-loop effects lift up this pseudo-flat direction, 
and the field $X$ has the mass $m_X$, 
\begin{equation}
m_X^2 = O\left( \frac{1}{4\pi^2} 
\frac{\hat f^2 \hat h^4_{11}}{\hat m_{12}^2}\right),
\end{equation}
around $X=0$.
In the case with $h_{11}=O(1)$ in the superpotential (\ref{W-4D}), 
we would have a rather small mass $m_X$ by the suppression factor 
$e^{-c_X \pi R}$.
To have larger mass $m_X$, we can assume the following superpotential 
$W^{(0)}$ at $y=0$ as 
\begin{equation}
\int dy \delta(y) W^{(0)}= \int dy \delta(y) h^{(0)}_{11}X\phi_1^2.
\end{equation}
In this case, the 4D superpotential becomes
\begin{equation}
\hat W = (h^{(0)}_{11}-h_{11}e^{-C_X \pi R})X\phi_1^2  + 
e^{-c_X\pi R}(fX+m_{12}\phi_1\phi_2 )
 + e^{-2c_X\pi R}\Delta W.
\end{equation}
This leads to the metastable SUSY breaking minimum around $X=0$ 
and the field $X$ can have a larger mass around $X=0$ than the 
previous model, because the coupling $h^{(0)}_{11}$ has no 
suppression factor like $e^{-c_X\pi R}$.
The SUSY breaking source $F^X$ is quasi-localized around $y=0$.

We can construct more various models for approximately 
R-symmetric superpotential with metastable SUSY breaking 
vacua in 5D theory.

\subsection{Short summary}
\label{sec:short summary in 3}
We have studied the scenario that conformal dynamics leads to 
approximately R-symmetric superpotential with a metastable 
SUSY breaking vacuum.
We have shown a simple model to realize our scenario.
We can make 5D models with the same behavior.
Since in our 4D scenario, metastable SUSY breaking vacua 
are realized by conformal dynamics, 
such a SUSY breaking source would be sequestered from the 
visible sector by conformal dynamics. We indicate that conformal dynamics leads the suppressions of cross terms between the visible sector and the hidden sector $\Phi_{ij}$. Our model, which have conformal dynamics in the hidden sector, also leads to the situation that anomaly mediation is dominant, so that the suppression to avoid large FCNC processes is realized.


In our scenario, at a high energy scale, there would be 
only SUSY minimum and at low energy metastable 
SUSY breaking vacuum would appear.
To realize the initial condition such that 
a metastable SUSY breaking is favored at a high energy scale,
finite temperature effects would be important, 
because finite temperature effects might 
favor a metastable SUSY breaking vacuum \cite{Abel:2006cr}.

\vspace*{2\baselineskip}

We suggest that conformal dynamics leads to approximately R-symmetric superpotential at low energy. In fact, conformal dynamics can realize this situation even in softly SUSY breaking theories. For example, gaugino mass, that is one of soft SUSY breaking terms and also a explicit R-symmetry breaking term, is strongly suppressed according to gauge coupling approaching toward an IR fixed point. We show how explicit R-symmetry breaking terms are suppressed in softly SUSY broken theories in section \ref{sec:SUSY-breaking}.

We introduce the RG flow of SUSY breaking parameters in softly SUSY breaking theories in section \ref{sec:SUSY-breaking}.
We discuss models with conformal dynamics and soft SUSY breaking terms, and we do not specify the origin of SUSY breaking there.  
Finally, we find the effects of SUSY breaking and explicit R-symmetry breaking are suppressed strongly by conformal dynamics.
The suppression can be also realized in more complicated models, such as the Klebanov-Strassler (KS) model \cite{Klebanov:2000hb} we show in section \ref{sec:duality cascade},
so that this result leads to the other scenario that the visible sector, which includes SSM, has conformal dynamics. 
In section \ref{sec:duality cascade}, we introduce the model with conformal dynamics, which is expected to realize SSM at a low-energy scale.

\section{RG flow of soft SUSY breaking terms in conformal dynamics}
\label{sec:SUSY-breaking}

Here, we study the RG flow of SUSY breaking terms 
in softly broken supersymmetric theories and show that approximate R-symmetry is realized.
In this study, all soft SUSY breaking terms are free parameters, and a gauge coupling and yukawa couplings have IR fixed points.  
The soft SUSY breaking terms, such as gaugino masses and A-terms, also break R-symmetry explicitly.  

It is convenient to use the spurion method 
\cite{Yamada:1994id,Hisano:1997ua,Jack:1997pa,Kobayashi:1998jq,
ArkaniHamed:1998kj,Terao:2001jw} to derive 
RG equations of soft SUSY breaking terms from 
those for supersymmetric couplings. In the appendix \ref{app:superfield} and \ref{app:spurion}, we discuss the justification of the spurion method based on superfield perturbation and symmetry.

\subsection{A gauge theory with conformal fixed point in a softly SUSY broken theory}
\label{sec:3-1}

Now we consider a $SU(N)$ gauge theory with $N_f$ flavors of fundamental and anti-fundamental matter fields $(Q,~\ov{Q})$ and vanishing superpotential $W=0$. $N_f$ satisfies $\frac{3}{2} N \leq N_f \leq 3N$, so that the gauge coupling has an IR fixed point as we discuss in section \ref{sec:4D conformal model}. We find that explicit R-symmetry breaking terms in the superpotential goes to zero by the conformal dynamics in section \ref{sec:4D conformal model}. Here our model does not have superpotential, so a explicit R-symmetry breaking term is only gaugino mass, which is one of soft SUSY breaking terms.      
The soft SUSY breaking term is given by as follows, 

\beq
\begin{split}
L_{soft}^{SQCD} =& -\int d^4 \theta (m^2_{ij} \theta^2 \ov{\theta}^2) Q^{i\dagger}Q^j -\int d^4 \theta (\ov{m}^2_{ij} \theta^2 \ov{\theta}^2) \ov{Q}^{i\dagger}\ov{Q}^j \\
& - \int d^2 \theta (M_{1/2} \theta^2)\frac{1}{g^2} Tr(W^{\al} W_{\al}) +h.c, 
\end{split} 
\label{eq:soft-SQCD}
\eeq
where $i$ denotes the flavor index $(i=1, \dots ,N_f)$, $M_{1/2}$, $m^2_{ij}$ and $\ov{m}^2_{ij}$ denote soft SUSY breaking terms: gaugino mass, squared scalar masses respectively. 

We use the spurion method to analyze the RG flows of the soft SUSY breaking terms, so we give a brief review on the spurion method. The more precise argument about the spurion method is given in the appendix \ref{app:spurion}. 
We also discuss the dual gauge theory, whose gauge symmetry is $SU(N_f-N)$ with $N_f$ flavor pairs and superpotential is given by the yukawa couplings, in the next section. 

\vspace*{2\baselineskip}

Let us consider a generic gauge theory with a gauge coupling $g$, 
a gaugino mass $M_{1/2}$, yukawa couplings $y_{ijk}$, corresponding 
A-terms $a_{ijk}$ and soft scalar masses $m_i$.
\beq
\begin{split}
L_{soft} = &-\int d^4 \theta (m^2_{ij} \theta^2 \ov{\theta}^2) \Phi^{i\dagger}\Phi^j - \int d^2 \theta (M_{1/2} \theta^2)\frac{1}{g^2} Tr(W^{\al} W_{\al}) \\ 
& - \int d^2 \theta \frac{1}{6} (h_{ijk} \theta^2)\Phi^i \Phi^j \Phi^k + h.c.
\end{split}
\label{eq:soft}
\eeq

We define the following superfield couplings 
\begin{eqnarray}
\tilde \alpha &=& \alpha \left( 1 + M_{1/2}\theta^2 + \ov{M}_{1/2} \bar \theta^2 
+ (2 |M_{1/2}|^2 + \Delta_g) \theta^2 \bar \theta^2 \right), \\
\tilde y_{ijk} &=& y_{ijk} - a_{ijk} \theta^2 + 
\frac{1}{2}(m_i^2+m_j^2+m_k^2)y_{ijk}\theta^2 \bar \theta^2,
\end{eqnarray}
where $\al$ is defined as $\al \equiv g^2/(8 \pi^2)$ and $\Delta_g$ is written as \cite{Kobayashi:1998jq}\footnote{We show the function $F(\al)$ in appendix \ref{app:spurion}.}
\begin{equation}
\Delta_g = - \frac{F(\alpha)}{\alpha} 
\left[\sum_iT_i m_i^2 - T_G|M_{1/2}|^2
\right].
\end{equation}
Then, beta-functions of superfields $\tilde \alpha$ and
$\tilde y_{ijk}$ ($\bar{\tilde y}_{ijk}$) including soft SUSY breaking terms are 
obtained from those of $\alpha$ and $y_{ijk}$ (${\ov{y}}_{ijk}$),
$\beta_\alpha(\alpha,y_{ijk},\bar{y}_{ijk})$ and $\beta_{y_{ijk}}(\alpha,y_{ijk},\bar{y}_{ijk})$ 
by replacing $\alpha$ and $y_{ijk}$ ($\bar{y}_{ijk}$) by $\tilde \alpha$,
$\tilde y_{ijk}$ ($\bar{\tilde{y}}_{ijk}$), i.e.,
\begin{equation}
\label{eq:spurion-RG}
\mu \frac{d \tilde \alpha}{d \mu} = \beta_\alpha(\tilde \alpha, \tilde
y_{ijk}, \bar{\tilde{y}}_{ijk}), \qquad 
\mu \frac{d \tilde y_{ijk}}{d \mu} = \beta_{y_{ijk}}(\tilde \alpha, \tilde
y_{ijk}, \bar{\tilde{y}}_{ijk}).
\end{equation}
That implies that the beta-function of the gaugino mass $M_{1/2}$ is 
obtained as 
\begin{equation}
\mu \frac{dM_{1/2}}{d\mu} = \left( M_{1/2}\alpha 
\frac{\partial}{\partial \alpha}  - a_{ijk}  
\frac{\partial}{\partial y_{ijk}}  \right) \left(  
\frac{\beta_\alpha}{\alpha}\right)
\equiv
D_1 \left(  
\frac{\beta_\alpha}{\alpha}\right).
\label{eq:RG-gaugino}
\end{equation}

The RG equation for the soft scalar mass $m_i$ of a
chiral superfield $\phi_i$ is also easily obtained as
\begin{equation}
\mu \frac{d m^2_i}{d \mu} =
\left. 
\gamma_i(\tilde{\alpha}, \tilde{y}_{ijk}, \bar{\tilde{y}}_{ijk})
\right|_{\theta^2 \bar{\theta}^2}.
\end{equation}
These equations are found to be consistent with
the equations for the $\theta^2 \bar{\theta}^2$
components of Eqs. (\ref{eq:spurion-RG}).
Explicitly, the RG equations are written down as
\begin{eqnarray}
\mu \frac{d m^2_i}{d \mu} &=& D_2 \gamma_i \ ,\\
D_2 &=& D_1 \bar{D}_1 
+ (|M_{1/2}|^2 + \Delta_g)\alpha \frac{\partial}{\partial \alpha} 
\nonumber \\
& &
+ \frac{1}{2}(m^2_i + m^2_j + m^2_k)
\left(
y_{ijk }\frac{\partial}{\partial y_{ijk}}
+
\bar{y}_{ijk }\frac{\partial}{\partial \bar{y}_{ijk}}
\right) .
\label{eq:RG-scalar mass}
\end{eqnarray}

\vspace*{2\baselineskip}

Let us turn back to the RG flows of the soft SUSY breaking terms in SQCD given by (\ref{eq:soft-SQCD}).
We find an IR attractive fixed-point of the gauge coupling, by solving the equation that the beta-function given by (\ref{eq:NSVZ}) is equal to zero. 
We consider the perturbation around the fixed point as 
$\alpha=\alpha^* + \delta \alpha$, where $\delta \alpha \ll 1$.
The beta-function of $\delta \alpha$ around the fixed point
is written as 
\begin{equation}\label{eq:RG-da}
\mu \frac{d \delta \alpha}{d \mu} = 
\left( \frac{\partial \beta_{\alpha}}
{\partial {\alpha}}  
\right)_{\alpha=\alpha^*} \delta \alpha
\equiv
\Gamma \delta \alpha.
\end{equation}
Because this fixed point is the IR attractive, that leads to 
$\Gamma > 0$.
Then, the spurion method leads immediately 
to the RG flow of the gaugino mass, that is,   
the gaugino mass is renormalized as
\begin{equation}\label{eq:M-RG-CFT}
M_{1/2}(\mu) = M_{1/2}(\mu_0) 
\left( \frac{\mu}{\mu_0}\right)^{\Gamma}.
\end{equation}
This is because 
the spurion method tells that 
\beq
\delta \tilde{\alpha} = \alpha^* M_{1/2}\theta^2
- F(\alpha^*)
\sum_i T_i m^2_i \theta^2 \bar{\theta}^2 
\eeq
also decrease exponentially toward the IR direction.

Thus the gaugino mass $M_{1/2}$ is found to be
exponentially suppressed around the IR fixed point.
Eventually we find that the explicit R-symmetry breaking term, gaugino mass, is also suppressed by conformal dynamics, and we can also show that the sum $\sum_i T_i m^2_i$ 
is exponentially suppressed in this theory.

Furthermore, it is straightforward to extend this discussion 
to the theory with a gauge coupling and yukawa couplings 
and to show that the gaugino mass $M_{1/2}$ and the A-term $a_{ijk}$ 
as well as the sums $\sum_i T_i m^2_i$ and 
$m^2_i + m^2_j + m^2_k$ are exponentially suppressed.

\subsection{The dual gauge theory in a softly SUSY broken theory}

We have shown that SQCD with $W=0$, gaugino mass and soft scalar masses recovers approximate R-symmetry at low energy. Here we discuss soft SUSY breaking terms in the dual gauge theory. Based on the Seiberg's argument \cite{Seiberg:1994pq,Intriligator:1995au}, the gauge symmetry and the superpotential are given by $SU(\tilde{N})=SU(N_f-N)$ with $N_f$ flavor pairs $(q,~\ov{q})$ and $W=y_{ij} q_i M_{ij} \ov{q}_j$. The soft SUSY breaking terms are described as 
 
\beq
\begin{split}
L_{soft}^d = &-\int d^4 \theta (m^2_{ij} \theta^2 \ov{\theta}^2) q^{i\dagger}q^j -\int d^4 \theta (\ov{m}^2_{ij} \theta^2 \ov{\theta}^2) \ov{q}^{i\dagger}\ov{q}^j \\
&- \int d^2 \theta (\tilde{M}_{1/2} \theta^2)\frac{1}{g'^2} Tr(W^{\al} W_{\al}) - \int d^2 \theta \frac{1}{6} (a_{ij} \theta^2)\ov{q}^i M^{ij} q^j + h.c.
\end{split}
\label{eq:soft-d}
\eeq

We define $\tilde{\al}$ and $\tilde{y}_{ij}$ in the dual gauge theory as follows, 

\begin{eqnarray}
\tilde{\al}' &=& \alpha' \left( 1 + \tilde{M}_{1/2}\theta^2 + \ov{\tilde{M}}_{1/2} \bar \theta^2 
+ (2 |\tilde{M}_{1/2}|^2 + \Delta_g) \theta^2 \bar \theta^2 \right), \\
\tilde y_{ij} &=& y_{ij} - a_{ij} \theta^2 + 
\frac{1}{2}(m_i^2+m_j^2+m_{ij}^2)y_{ij}\theta^2 \bar \theta^2,
\end{eqnarray}
where where $\al'$ is defined as $\al' \equiv g'^2/(8 \pi^2)$ and $\Delta_g$ is written as\footnote{We show the function $F_d(\al')$ in appendix \ref{app:spurion}.} 
\begin{equation}
\Delta_g = - \frac{F_d(\al')}{\al'} 
\left[\sum_iT_i m_i^2 - T_G|\tilde{M}_{1/2}|^2
\right].
\end{equation}

We obtain the RG equations of $\tilde{M}_{1/2}$ by replacing $M_{1/2}$ and $y_{ijk}$ in (\ref{eq:RG-gaugino}) by $\tilde{M}_{1/2}$ and $y_{ij}$. The RG equations of $a_{ij}$ are also obtained as follows based on the last section and appendix \ref{app:spurion}, 

\beq
\mu \frac{d a_{ij}}{d\mu} = \frac{1}{2} (\gamma_i+\gamma_j+\gamma_{ij})a_{ij}-(D_1\gamma_i+ D_1\gamma_j+ D_1\gamma_{ij})y_{ij}. 
\eeq

In the dual-side, the gauge coupling and the yukawa couplings have IR-fixed points, and in fact this behavior corresponds to our model in section \ref{sec:4D conformal model}. 

It is also found that these RG equations lead to
very interesting properties of the soft SUSY 
breaking parameters at the vicinity of an 
IR attractive fixed point 
\cite{Terao:2001jw,Karch:1998qa,Kobayashi:2001kz} as the last section.
Deviations of the gauge coupling  and
the yukawa coupling  from their fixed
point values, $\delta \alpha' = \alpha' - \alpha'^*$ and 
$\delta y_{ij} = y_{ij} - y^*_{ij}$, decrease exponentially.
Then the spurion method tells that both of
\begin{eqnarray}
\delta \tilde{\alpha}' &=& \alpha'^* \tilde{M}_{1/2}\theta^2 +\alpha'^* \ov{\tilde{M}}_{1/2} \ov{\theta}^2
- F(\alpha'^*)
\sum_i T_i m^2_i \theta^2 \bar{\theta}^2, \\
\delta \tilde{y}_{ij} &=&
-a_{ij} \theta^2
+\frac{1}{2}(m^2_i + m^2_j + m^2_{ij})y_{ij}^*  \theta^2 \bar{\theta}^2 ,
\end{eqnarray}
also decrease exponentially toward the IR direction.
Therefore, the gaugino mass $M_{1/2}$ and the A-term $a_{ij}$ 
are found to be suppressed\footnote{That implies that the ratio 
$a_{ij}/y_{ij}$ is also suppressed exponentially, 
because the yukawa coupling $y_{ij}$ has a fixed point.} 
and the soft scalar masses satisfy the
IR sum rules given by $\sum_i T_i m^2_i=0$ and 
$m^2_i + m^2_j + m^2_{ij}=0$.

We may also understand this as follows.
When we use the one-loop anomalous dimensions, 
we can show that at the fixed point the gauge coupling and 
yukawa coupling are related as
$y^*=Cg'^*$, where $C$ is a constant determined by 
group-theoretical factors \cite{Pendleton:1980as}.
At the fixed point, this relation is realized as 
the relation between superfield couplings as 
$|\tilde y|^2/(8 \pi^2) = C^2 \tilde \alpha'$, 
and their $\theta^2 \bar \theta^2$-terms lead to 
\cite{Kawamura:1997cw,Kobayashi:2000di}
\begin{equation}\label{eq:sum-rule}
m^2_q+m^2_{\bar q}+m^2_M = |\tilde{M}_{1/2}|^2.
\end{equation}
Since the gaugino mass $\tilde{M}_{1/2}$ is exponentially 
damping toward the conformal fixed point, the sum 
$m^2_q+m^2_{\bar q}+m^2_M$ is also exponentially damping 
as mentioned above.

Finally, we find that conformal dynamics realizes approximate R-symmetry at a low energy scale because of the suppressions of gaugino mass and A-term even in softly SUSY broken theories. However, a few parameters, such as B-term which is a quadratic term of scalar, are not controlled by the dynamics. As we discuss later, these parameters cause gauge symmetry breaking and disturb the RG flow of the gauge coupling.  

In section \ref{sec:duality cascade}, we discuss more complicated model with conformal dynamics, which have two gauge couplings and nonzero superpotential. In that case, we also find the feature that approximate R-symmetry is realized at low energy, and the B-term frightens the gauge symmetries. We discuss not only the aspect, but also the other applications of conformal dynamics for phenomenology.

\section{The Duality cascade of softly broken 
supersymmetric theories}
\label{sec:duality cascade}

As we discuss in section \ref{sec:SUSY-breaking}, conformal dynamics restores R-symmetry. We find that this suppression can be realized in more complicated model with SUSY breaking parameters. We discuss the duality cascade with soft SUSY breaking terms based on \cite{Abe:2008sq}. 

The duality cascade is caused by conformal dynamics, and has interesting RG flow of gauge couplings and yukawa coupling \cite{Klebanov:2000hb,Strassler:2005qs}.
After a brief review on the duality cascade of rigid SUSY theory, we discuss RG flows of soft SUSY breaking terms in the duality cascade, and show that the dynamics also leads the suppression of soft SUSY breaking terms and leads to approximate R-symmetry. However, B-term, which is a quadratic scalar term, stay in a finite value at low energy.

Conformal dynamics plays important roles in 
various aspects of (supersymmetric) field theories 
and particle phenomenology. For example,
in Ref. \cite{Karch:1998qa,Kobayashi:2001kz,Nelson:2000sn,Kobayashi:2001is}, conformal dynamics in the visible sector leads not only realistic hierarchies of quark and lepton masses, but also the alignment of A-terms to avoid large FCNC processes. At the same time, sfermion masses are exponentially 
suppressed toward the IR fixed point.\footnote{
A similar dynamics would be useful to control a 
large radiative correction on Higgs soft masses \cite{Kobayashi:2004pu}.}
In the later of this section , we study the model with conformal dynamics in the visible sector. This model is based on the duality cascade, and has unique RG flows. Finally, we suggest models with conformal dynamics in the visible sector, which have gauge symmetries and matters of SSM. Then we find that the models cause gauge symmetry breaking corresponding to EW symmetry breaking.

\vspace*{2\baselineskip}

We show the dynamics of the duality cascade in the section \ref{sec:rigid-SUSY}. 
Conformal fixed points and conformal field theories (CFTs)
are essential in Seiberg duality \cite{Seiberg:1994pq,Intriligator:1995au}.
That leads to more complicated and 
interesting RG flows of 
dual field theories, that is, 
the duality cascade \cite{Klebanov:2000hb,Strassler:2005qs},
which is a successive chain of the dualities from 
UV region to the IR region and
reduces the rank of gauge groups one after another.
Furthermore, the AdS/CFT (gravity/gauge) correspondence
\cite{Maldacena:1997re}
suggests that the cascading theories would be realized
in supergravity theory with a warped background, that is, 
the Klebanov-Strassler warped throat.
In the supergravity description, the energy scale of the field theory
corresponds to the distance from a tip of the throat.
The duality cascade process means that the charges of D-branes
disappear as the probe brane gets closer to the tip.
The investigation of the duality cascade from the string/supergravity 
viewpoint is a highly non-trivial check for the gravity/gauge correspondence.

In section \ref{sec:SUSY-breaking cascade 2}, we study more about the duality cascade in the model with soft SUSY breaking terms, based on \cite{Abe:2008sq}. 
As we comment in the introduction, several models have been proposed to realize 
supersymmetric standard models as well as 
their extensions at the bottom of the cascade 
\cite{Cascales:2005rj,Heckman:2007zp,Franco:2008jc} recently.
To explain how we obtain the standard model like theories 
with fewer ranks from the infinitely many string vacua, which would generally have gauge groups with large ranks,
those models are quite interesting and have opened 
possible candidates for high energy theories.
Those models are exactly supersymmetric.
At any rate, supersymmetry is broken in Nature 
even if supersymmetric theory is realized at high energy.
Thus, if the cascading theories are relevant to the particle physics 
at the weak scale, 
supersymmetry should be broken at a certain stage, e.g. 
at the top or bottom of the cascade (high or low energy) or 
between them (intermediate energy).
Here we assume that SUSY is softly broken at the beginning of 
the cascade.
Then, we study RG flows of SUSY breaking terms as well as 
supersymmetric couplings.

This section is organized as follows.
In section \ref{sec:rigid-SUSY}, we review briefly the RG flow of 
supersymmetric couplings in the duality cascade.
In section \ref{sec:SUSY-breaking cascade 2}, we study RG flows of SUSY breaking terms in the duality cascade.
In section \ref{sec:symm-br}, we study symmetry breaking due to the B-term
by using illustrative examples. In section \ref{sec:Illustrating model}, we give a simple example whose fields contents are similar to the MSSM or its extensions.
Section \ref{sec:conclusion dual} is conclusion and discussion of section \ref{sec:duality cascade}.

\subsection{RG flow in duality cascade of 
rigid supersymmetric theories}
\label{sec:rigid-SUSY}

Here, we give a brief review on 
the RG flow in the duality cascade of rigid supersymmetric 
theories \cite{Klebanov:2000hb,Strassler:2005qs}.
We consider the gauge group $SU(kN) \times SU((k-1)N)$ 
and we denote their gauge couplings, $g_k$ and $g_{k-1}$.
Also, our model has two chiral multiplets $Q_r$ $(r=1,2)$ 
in the bifundamental representation of $SU(kN) \times SU((k-1)N)$, 
i.e. the fundamental representation for $SU(kN)$ and 
the anti-fundamental representation for $SU((k-1)N)$, 
and two chiral multiplets $\bar Q_s$ $(s=1,2)$ 
in the anti-bifundamental representation.
Then we introduce the following superpotential,
\begin{equation}
\label{eq:W-4}
W=h~{\rm tr} \det_{r,s} (Q_r \bar Q_s)=h\left[ 
(Q_1)^\alpha_a(\bar Q_1)^a_\beta (Q_2)^\beta_b (\bar Q_2)^b_\alpha - 
(Q_1)^\alpha_a(\bar Q_2)^a_\beta (Q_2)^\beta_b (\bar Q_1)^b_\alpha
\right],
\end{equation}
where the indices $\alpha$ and $\beta$ are group indices for 
$SU(kN)$ and the indices $a$ and $b$ are group indices for 
$SU((k-1)N)$.

Now, we study the RG flow of gauge couplings $g_k$ and 
$g_{k-1}$ and the quartic coupling $h$ and their 
fixed points.
The fields $Q_r$ and $\bar Q_s$ have the same anomalous dimension, 
which we denote by $\gamma_Q$.
In the NSVZ scheme \cite{Novikov:1983uc}, 
beta-function of the gauge coupling $g$ in generic gauge theory 
is  written as 
\begin{equation}
\mu \frac{d \alpha}{d\mu}=\beta_\alpha = -\tilde{F}(\alpha)[3T_G - \sum_i T_i(1-\gamma_i)],
\end{equation} 
where $\alpha=g^2/(8\pi^2)$ and 
\begin{equation}
\tilde{F}(\alpha) =  \frac{\alpha^2}{1-T_G \alpha}.
\end{equation} 
Here,  $T_i$ and $\gamma_i$ denote 
Dynkin indices and anomalous dimensions 
of the chiral matter fields, 
while $T_G$ denotes the Dynkin index of 
the adjoint representation.
For example, we have $T_G=N$ for the $SU(N)$ gauge group and 
$T_i=1/2$ for the fundamental representation of the $SU(N)$ gauge group.
Using this scheme, beta-functions of the gauge couplings $g_k$ and 
$g_{k-1}$ are written as  
\begin{eqnarray}
\beta_{\alpha_k} &=& -\tilde{F}(\alpha_k)N[k+2+2(k-1)\gamma_Q], \\
\beta_{\alpha_{k-1}} &=& -\tilde{F}(\alpha_{k-1})N[k-3+2k\gamma_Q].
\end{eqnarray}
In addition, we can write the beta-function of $\eta = h \mu$ as 
\begin{equation}
\beta_\eta = \eta (1+2\gamma_Q).
\end{equation}

Suppose that both $SU(kN)$ and $SU((k-1)N)$ sectors are within the conformal 
window \cite{Seiberg:1994pq}, 
i.e. $3k/2 \leq 2(k-1) \leq 3k$ and $3(k-1)/2 \leq 2k \leq 3(k-1)$.
Then, we have two fixed points 
\cite{Banks:1981nn,Seiberg:1994pq,Intriligator:1995au}, 
\begin{equation}
\label{eq:fp-k-1}
{\rm A}: ~~~ k-3+2k\gamma_Q = 0, \qquad \alpha_k=\eta=0,
\end{equation}
and 
\begin{equation}
\label{eq:fp-k}
{\rm B}: ~~~ k+2+2(k-1)\gamma_Q = 0, \qquad \alpha_{k-1}=\eta=0.
\end{equation}
The anomalous dimension $\gamma_Q$ is a function of the couplings.
We represent a value of the gauge coupling $g_{k-1}$ 
($g_k$) at the first (second) fixed point by 
$g_{k-1}^*$ ($g_{k}^*$).

At the vicinity of the first fixed point A given by
(\ref{eq:fp-k-1}) with $g_{k-1} \approx g_{k-1}^*$ and 
$0 < \alpha_k, \eta \ll 1$ (region I), 
it is found that $\beta_{\alpha_{k-1}} \approx 0$, 
$\beta_{\alpha_k} <0$ and $\beta_\eta >0$, that is, 
$\alpha_k$ increases and $\eta$ decreases toward the IR direction.
Thus, the theory would flow to the other fixed point 
B given by (\ref{eq:fp-k}) toward the IR direction.
On the other hand, 
around the fixed point B with 
$g_{k} \approx g_{k}^*$ and $0 < \alpha_{k -1}, \eta \ll 1$
(region II), 
it is found that $\beta_{\alpha_{k}} \approx 0$, 
$\beta_{\alpha_{k -1}} >0$ and $\beta_\eta <0$.
Hence, the quartic operator 
$h~{\rm tr} \det_{r,s} (Q_r \bar Q_s)$ is relevant and 
the coupling $\eta$ increases toward the IR, while 
$\alpha_{k-1}$ shrinks.

We could examine the RG flows of the gauge couplings
$\alpha_k$ and $\alpha_{k-1}$, if we admit using
the anomalous dimension $\gamma_Q$ obtained in the
1-loop level. 
For a sufficiently large $N$, the anomalous dimension $\gamma_Q$
is given as
\begin{equation}
\gamma_Q = -N(k\alpha_k + (k-1)\alpha_{k-1}) .
\end{equation}
In Fig.~2, the RG flows in the coupling space
$(\alpha_k, \alpha_{k-1})$ obtained in the NSVZ
scheme are shown in the case of $k=5$. 
Here we rescale the couplings as $N\alpha \to \alpha$.
The points A $(0, 0.05)$ and B $(0.175, 0)$ represent
the fixed points. The renormalized trajectory (R.T.) 
connecting these fixed points is shown by the bold line.

The flows in the region I are subject to
the conformal dynamics around the UV fixed point A, while
the flows in the region II are subject to that around the
IR fixed point B.
The convergence in the region I is not strong, since the
fixed point coupling $\alpha_{k-1}^*$ is not so strong
in the case of $k=5$.
It is seen in Fig.~2 that the R.T. bends on the way and 
the behavior of the R.T. changes quickly there.
Thus the RG property on the R.T. in Fig.~2 may be
characterized well as that in the region I or II.

\begin{figure}[hbt]
\begin{center}
\includegraphics[width=0.65\textwidth ,clip]{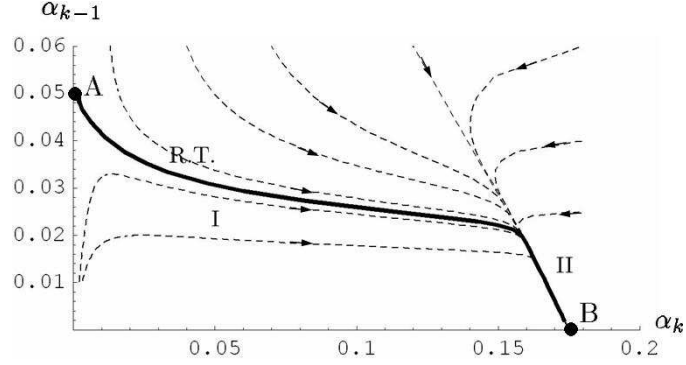}
\caption{RG flows in the coupling space 
$(\alpha_k, \alpha_{k-1})$ in the case of $k=5$. 
The points A and B represents
the UV and IR fixed points respectively. 
The renormalized trajectory connecting
these fixed points is shown by the bold line.
}
\end{center}
\end{figure}

The theory around the fixed point B 
is strongly coupled and would be well-described by 
its Seiberg dual \cite{Seiberg:1994pq,Intriligator:1995au}, 
which has the gauge group 
$SU((k-1)N) \times SU((k-2)N)$ and two bifundamental 
chiral multiplets $q_r$ and two anti-bifundamental chiral 
multiplets $\bar q_s$ and another kind of 
chiral multiplets $M_{rs}$, 
which correspond to $Q_r \bar Q_s$ and belong to  
the adjoint representation for $SU((k-1)N)$ 
\footnote{In~the~followings,~we~will~ignore~the~irrelevant~mesons~$M^0_{rs}$~which~are~singlet~for~$SU((k-1)N)$.}
and 
singlet for $SU((k-2)N)$.
This dual theory has the following superpotential,
\begin{equation}\label{eq:W-dual}
W=y~{\rm tr}~\bar q_r M_{rs} q_s + m~{\rm tr} \det_{r,s} M_{rs}.
\end{equation}
The second term is the mass term of $M_{rs}$, which corresponds to 
$h~{\rm tr}~\det_{r,s}(Q_r \bar Q_s)$.
The mass $m$ would be related with the coupling $h$ as 
\begin{equation}
h(\Lambda_k) \Lambda_k \sim \frac{m(\Lambda_k)}{\Lambda_k},
\end{equation}
where $\Lambda_k$ is a typical energy scale of 
$SU(kN)$ gauge theory such as the energy scale, where 
the theory enters the conformal regime, i.e. 
$g_k(\Lambda_k) \approx g^*_k$.
The $\beta$-function of $\alpha_{k-2}$ is written 
in a way similar to those of $\alpha_{k-1}$ and $\alpha_{k}$.
In addition, the $\beta$-function of $y$ is written as 
\begin{equation}
\beta_y = \frac{y}{2}(\gamma_M + 2\gamma_q),
\end{equation}
where $\gamma_M$ is the anomalous dimension of $M_{rs}$.
The dual theory has a non-trivial fixed point, $g_{k-2}=g^*_{k-2}$ 
and $y=y^*$ when $g_{k-1}=0$, where $g^*_{k-2}, y^* \neq 0$.
At the fixed point, it is satisfied that $\gamma_M = - 2\gamma_q$, 
that is $M_{rs}$ has the same conformal dimension as 
$Q_r \bar Q_s$.
Thus, at the vicinity of the fixed point, $g_{k-2}=g^*_{k-2}$, 
$y=y^*$ and $g_{k-1}=0$, both operators, 
$h~{\rm tr}~\det_{r,s}(Q_r \bar Q_s)$ and 
$m~{\rm tr} \det_{r,s} M_{rs}$ are relevant, 
and the mass $m/\mu$ and coupling $h\mu$ increase toward the 
IR direction.
Because the fields $M_{rs}$ become heavy, we integrate out them 
and the effective superpotential results 
in \cite{Strassler:2005qs}\footnote{See also \cite{Kobayashi:2000di}.}
\begin{equation}
W = \tilde h~{\rm tr} \det_{rs} q_r \bar q_s,
\end{equation}
where $\tilde h = -y^2/m$.
The operator $\tilde h~{\rm tr} \det_{rs} q_r \bar q_s$ 
is irrelevant and the coupling $\tilde h$ decreases 
toward the IR direction.
Thus, the low energy effective theory is the same as the 
starting theory except replacing the gauge group 
$SU(kN) \times SU((k-1)N)$ by $SU((k-1)N) \times SU((k-2)N)$.
This RG flow would continue and the low-energy effective theory 
would become  the $SU((k-n)N) \times SU((k-n-1)N)$ gauge theory 
with the quartic superpotential 
$W = \tilde h~{\rm tr} \det_{rs} q_r \bar q_s$ until 
the theory becomes outside of the conformal window.
The RG flow toward the IR is illustrated as 
\begin{eqnarray}
\begin{array}{ccc}
(g_k\approx 0, g_{k-1} \approx g_{k-1}^*,\eta \approx 0) &
&   \\
\downarrow & & \\
(g_k\approx g_k^*, g_{k-1} \approx 0,\eta \approx 0) & 
\leftrightarrow & 
(g_{k -2}\approx g_{k -2}^*, g_{k-1} \approx 0,y \approx y^*, m/\mu
\approx 0) \\ \downarrow & {\rm dual} & \downarrow \\
(g_k\approx g_k^*, g_{k-1} \approx 0,\eta \gg 1) & \leftrightarrow & 
(g_{k -2}\approx g_{k -2}^*, g_{k-1} \approx 0,y \approx y^*, 
m/\mu \gg 1 ) \\
& & ~~~~~~~~~~~~~~~~~~~~~~~~~~ \downarrow {\rm integrating~out }~M_{rs} \\ 
& & (g_{k -2}\approx g_{k -2}^*, g_{k-1} \approx 0, \tilde \eta
\approx 0).
\\
\end{array} \nonumber
\end{eqnarray}

At the end of cascade we would obtain the 
$SU(2N) \times SU(N)$ gauge theory. 
The $SU(2N)$ gauge sector has the $2N$ flavors and 
the quantum deformed moduli space \cite{Intriligator:1995au,Seiberg:1994bz},
$\Delta W= X(\det_{\rm all}\, M_{rs} -B\bar B -\Lambda^{4N})$, 
where $X$ is a Lagrange multiplier superfield,
$B$ and $\bar B$ are baryon and anti-baryon superfields,
which are singlets under $SU(N)$. 
If we assume that only $B$ and $\bar B$ develop their VEVs, 
then baryons and mesons become massive.
Thus the effective theory becomes the pure $SU(N)$ 
supersymmetric Yang-Mills theory and finally the theory is confined.

\subsection{RG flow of soft SUSY breaking terms in the duality cascade}
\label{sec:SUSY-breaking cascade 2}

Here, we study the RG flow of SUSY breaking terms 
in softly broken supersymmetric theories. We assume all soft SUSY breaking terms are free parameters as in section \ref{sec:SUSY-breaking}.
Applying the spurion method to the cascading theory, 
we investigate the RG flow of soft SUSY breaking terms 
through the duality cascade.
We consider the $SU(kN) \times SU((k-1)N)$ gauge theory 
with two pairs of chiral matter fields $Q_r$ and $\bar Q_s$ and their 
superpotential (\ref{eq:W-4}).
The beta-functions of their gaugino masses, $M_{1/2}^{(k)}$ and 
$M_{1/2}^{(k-1)}$, are written as 
\begin{eqnarray}
\mu \frac{d M_{1/2}^{(k)}}{d \mu} &=&  
-N (k+2+2(k-1)\gamma_Q) H'(\alpha_k) \alpha_k M_{1/2}^{(k)}
\nonumber \\
& &
- 2(k-1)N H(\alpha_k)
\frac{\partial \gamma_Q}{\partial \alpha_k} 
\alpha_k M_{1/2}^{(k)}  
\nonumber \\
& &
- 2(k-1)N H(\alpha_k)
\frac{\partial \gamma_Q}{\partial \alpha_{k-1}} 
\alpha_{k-1} M_{1/2}^{(k-1)}, 
\label{M(k)beta}
\\
\mu \frac{d M_{1/2}^{(k-1)}}{d \mu} &=&  
- N (k-3+2k\gamma_Q) H'(\alpha_{k-1}) \alpha_{k-1} M_{1/2}^{(k-1)}
\nonumber \\
& &
- 2kN H(\alpha_{k-1})
\frac{\partial \gamma_Q}{\partial \alpha_{k-1}} 
\alpha_{k-1} M_{1/2}^{(k-1)}  
\nonumber \\
& &
- 2kN H(\alpha_{k-1})
\frac{\partial \gamma_Q}{\partial \alpha_k} 
\alpha_{k} M_{1/2}^{(k)}, 
\label{M(k-1)beta}
\end{eqnarray}
where $H(\alpha)=\tilde{F}(\alpha)/\alpha \approx \alpha$ and 
$H'(\alpha) = d H/d\alpha$.

As in Section \ref{sec:rigid-SUSY}, 
we start the RG flow at the energy scale $\Lambda$ from 
the vicinity of the fixed point A, i.e. 
$(g_k,g_{k-1},\eta) =(0,g_{k-1}^*,0)$.
Around the fixed point A, we have $k-3+2k\gamma_Q \approx 0$.
As long as $g_{k-1}$ is large and stable, the second
term in (\ref{M(k-1)beta}) reduces 
the gaugino mass $M_{1/2}^{(k-1)}$ exponentially
as the energy scale $\mu$ decreases.
On the other hand, we find $\beta_{M_{1/2}^{(k)}}<0$ 
because 
$k+2+2(k-1)\gamma_Q > 0$ and $H(\alpha_k) \approx \alpha_k \approx 0$.
Thus, the gaugino mass $M_{1/2}^{(k)}$ increases as 
the energy scale $\mu$ decreases.
However, such increase of $M_{1/2}^{(k)}$  is not drastic 
during the weak coupling region of $\alpha_k$.

Next, the theory moves from the vicinity of the fixed point 
$(g_k,g_{k-1},\eta) =(0,g_{k-1}^*,0)$ toward another fixed point, 
$(g_k,g_{k-1},\eta) =(g_{k}^*,0,0)$, where 
$k+2+2(k-1)\gamma_Q \approx 0$.
Around the latter fixed point, we find 
$\beta_{M_{1/2}^{(k-1)}}>0$ because $k-3+2k\gamma_Q < 0$, 
$H(\alpha_{k -1}) \approx 0$ and $M_{1/2}^{(k)}$ becomes irrelevant as below.
That is, the gaugino mass  $M_{1/2}^{(k-1)}$ decreases
perturbatively as the energy scale $\mu$ decreases.

On the other hand, the gaugino mass $M_{1/2}^{(k)}$ would be
suppressed exponentially in turn due to the second term in
(\ref{M(k)beta}), as 
going toward the IR fixed point.
However we note that the third term cannot be neglected,
when $\alpha_k M_{1/2}^{(k)}$ is reduced to be
comparable with $\alpha_{k-1} M_{1/2}^{(k-1)}$.
Then the gaugino mass $M_{1/2}^{(k)}$ does not follow a simple
exponential suppression. Rather it converges to a certain
value determined by $\alpha_{k-1}$ and $M_{1/2}^{(k-1)}$
obtained at the renormalized scale.

If we admit using the one-loop anomalous dimension,
then the RG behavior discussed above could be
explicitly examined.
Here we shall look into the theory on the renormalized
trajectory given in Fig.~2.
In Fig.~3, the gaugino masses $M_{1/2}^{(k-1)}(\mu)$
and $M_{1/2}^{(k)}(\mu)$ of the theory with $k=5$
are plotted with respect to
the scale parameter $\ln (\mu/\mu_0)$.
At the scale $\mu_0$, the gauge couplings are chosen
as $(\alpha_k, \alpha_{k-1}) = (0.0128, 0.04)$, which
is a point on the renormalized trajectory rather close to 
the fixed point A in Fig.~2.
The initial values at $\mu = \mu_0$ are taken to be
1.0 for both gaugino masses.

\begin{figure}[hbt]
\begin{center}
\includegraphics[width=0.65\textwidth ,clip]{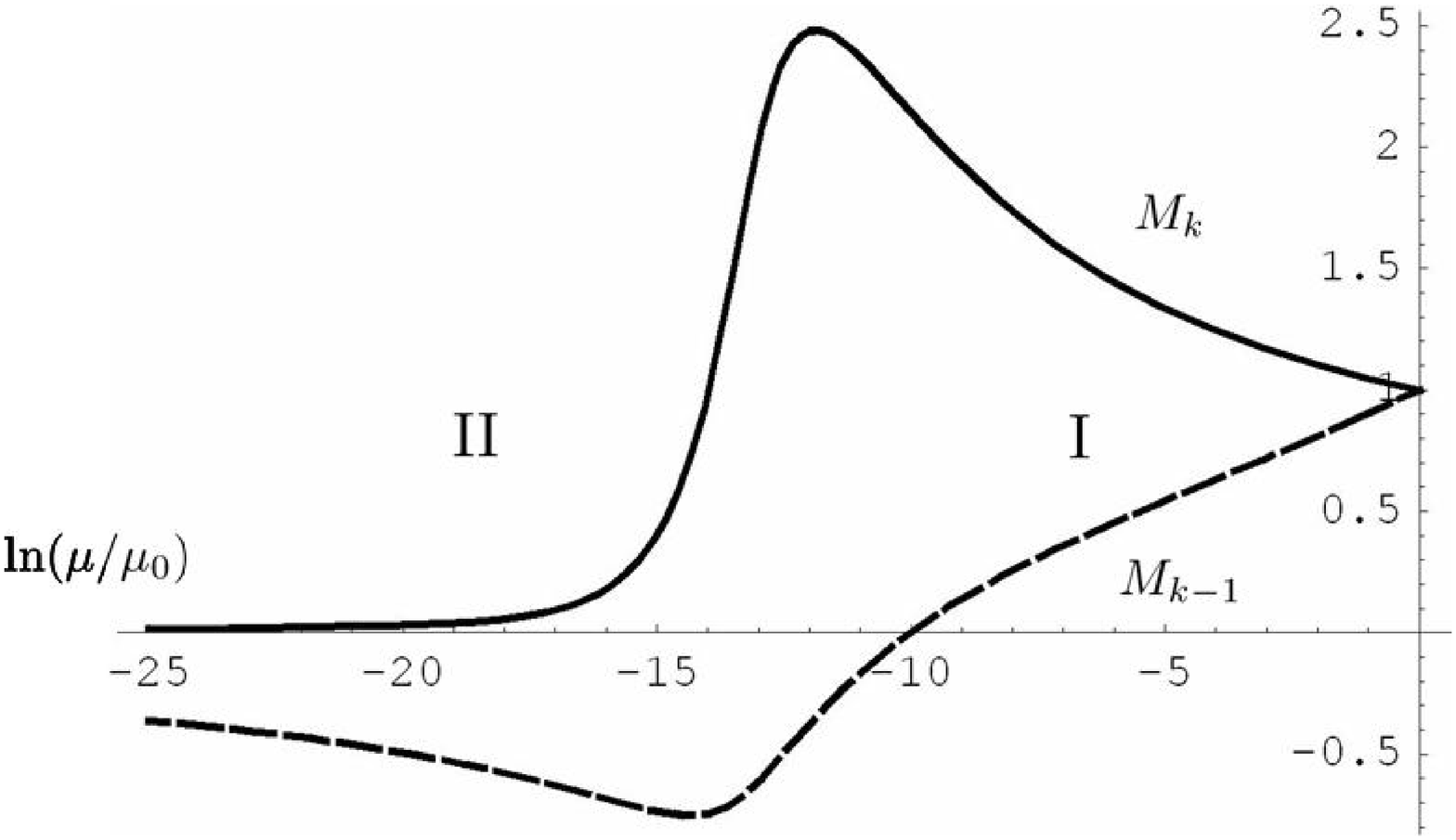}
\caption{RG running of the gaugino masses 
$M_{1/2}^{(k-1)}(\mu)$ and $M_{1/2}^{(k)}(\mu)$}
with respect to $\ln (\mu/\mu_0)$.
The gauge couplings are given at $\mu = \mu_0$
as $(\alpha_k, \alpha_{k-1})$$=$$(0.0128, 0.04)$
and  run along the renormalized trajectory.
\end{center}
\end{figure}

It is seen that $M_{1/2}^{(k-1)}$ is reduced as discussed,
but turns to be negative due to the third term in (\ref{M(k-1)beta}),
since $M_{1/2}^{(k)}$ glows slightly first.
In the region II, the gaugino mass $M_{1/2}^{(k)}$ turns out to be
suppressed strongly, while $M_{1/2}^{(k-1)}$
changes perturbatively.
In Fig.~4, the log-plot of the gaugino mass $M_{1/2}^{(k)}$
is shown by the bold line.
It is also seen that the suppression behavior deviates from
the exponential one in the end and
$M_{1/2}^{(k)}$ converges to a line.
Indeed, 
the convergence point of $\alpha_k M_{1/2}^{(k)}$ 
could be estimated as 
\begin{equation}\label{eq:converge-gaugino}
\alpha_k M_{1/2}^{(k)} \sim - 
\alpha_{k -1} M_{1/2}^{(k-1)}.
\end{equation}

\begin{figure}[hbt]
\begin{center}
\includegraphics[width=0.65\textwidth ,clip]{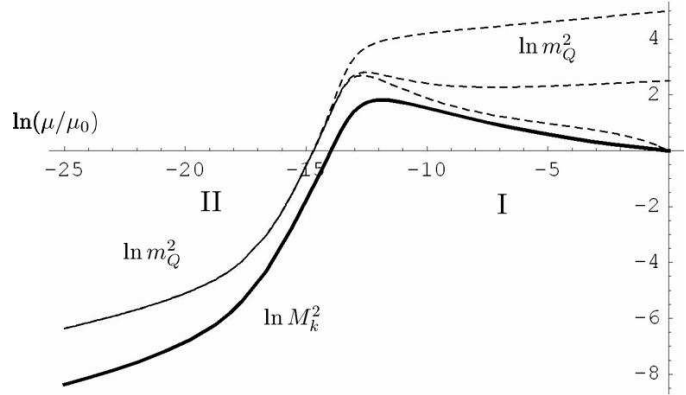}
\caption{RG running behaviors of the scalar mass 
$\ln m^2_Q$ and the gaugino mass 
$2 \ln M_{1/2}^{(k)}$ are shown by dotted lines and
the bold line, respectively.
}

\end{center}
\end{figure}

Similarly, we examine the RG running of 
the soft mass squared $m^2_Q$.
At the vicinity of the fixed points, $m^2_Q$ is 
also expected to be exponentially suppressed
as discussed in section \ref{sec:3-1}.
However existence of two gauge couplings
makes the situation more complicated.
The RG equation for $m^2_Q$ is given as
\begin{equation}
\mu \frac{d m^2_Q}{d \mu} = 
\left.
\gamma_Q(\tilde{\alpha}_k, \tilde{\alpha}_{k-1})
\right|_{\theta^2 \bar{\theta}^2}.
\end{equation}
Here, let us use the one-loop anomalous dimension
given by (9).
Then the RG equation is reduced to be
\begin{equation}
\mu \frac{d m^2_Q}{d \mu} = 
-k \alpha_k (2 |M_{1/2}^{(k)}|^2 + \Delta_k)
-(k-1) \alpha_{k-1} (2 |M_{1/2}^{(k-1)}|^2 + \Delta_{k-1}),
\label{RGeq:mQ2}
\end{equation}
where 
\begin{eqnarray}
\Delta_k &=&
H(\alpha_k)\left[
3k|M_{1/2}^{(k)}|^2 - 2(k-1)m^2_Q
\right], \\
\Delta_{k-1} &=&
H(\alpha_{k-1})
\left[
3(k-1)|M_{1/2}^{(k-1)}|^2 - 2k m^2_Q
\right].
\end{eqnarray}

In Fig.~4, the RG evolution of $m^2_Q$ of the same theory
on the renormalized trajectory is shown by
dotted lines. The initial values are taken as
$\ln m^2_Q = 0, 2.5, 5.0$ just for the illustration.
In the region I, we may neglect subleading terms of $\alpha_k$ and also
$M_{k-1}$, since it is suppressed. 
Then, Eq. (\ref{RGeq:mQ2}) is approximated to be
\begin{equation}
\mu \frac{d m^2_Q}{d \mu} \simeq
2k(k-1) (\alpha_{k-1}^*)^2 m^2_Q
- 2k\alpha_k |M_{1/2}^{(k)}|^2.
\end{equation}
This equation tells us that $m^2_Q$ is not just suppressed
but converges as
\begin{equation}
m^2_Q \to \frac{1}{(k-1) (\alpha_{k-1}^*)^2} \alpha_k |M_{1/2}^{(k)}|^2,
\end{equation}
since running of $\alpha_k |M_{1/2}^{(k)}|^2$ is rather slow.
In the case of $k=5$, the fixed point coupling
$g_{k-1}^*$ is not large and the convergence is not
so strong.
In the region II, running of $M_{1/2}^{(k)}$ changes to
exponential suppression. However, similarly to the behavior
in the region I, it converges in the IR limit as
\begin{equation}\label{eq:converge-scalar}
m^2_Q \to \frac{1}{k (\alpha_{k}^*)^2} 
\alpha_{k-1} |M_{1/2}^{(k-1)}|^2.
\end{equation}

We summarize the RG flow of the gaugino masses and soft scalar masses 
as the theory moves from the fixed point 
$(g_k,g_{k-1},\eta) =(0,g_{k-1}^*,0)$ toward the fixed point
$(g_k,g_{k-1},\eta) =(g_{k}^*,0,0)$.
At the first stage, i.e. the perturbative regime of $\alpha_k$, 
the gaugino mass $M_{1/2}^{(k-1)}$ is suppressed, while
$M_{1/2}^{(k)}$ and the soft scalar mass squared 
$m^2_Q$ increase perturbatively.
In entering the conformal regime of $\alpha_k$, 
both $M_{1/2}^{(k)}$ and $m^2_Q$ begin exponential damping,
while $M_{1/2}^{(k-1)}$ runs perturbatively.
In the IR limit, the gaugino mass $M_{1/2}^{(k)}$
and the soft scalar mass squared $m^2_Q$ are found to converge
to certain values determined by $\alpha_{k-1}$ and
$M_{1/2}^{(k-1)}$. 
Hence, these parameters evolute to be
of the same order and are fixed in the
IR limit irrespectively of their initial values.

In addition to the gaugino masses $M_{1/2}^{(k)}$, $M_{1/2}^{(k-1)}$ 
and scalar mass $m_Q$, the SUSY breaking terms corresponding to 
the superpotential (\ref{eq:W-4}) may be important, that is,
\begin{equation}\label{eq:Aterm-4}
W=h(1 - A_h \theta^2)~{\rm tr} \det_{r,s} (Q_r \bar Q_s).
\end{equation}
The RG flow behavior of the coupling $\mu hA_h$ is 
drastic following the anomalous dimensions of $Q_r$ and 
$\bar Q_s$.
Both RG flows of $\mu h$ and $\mu hA_h$ are almost the same.
That implies that their ratio $A_h$ does not change
drastically\footnote{Note that the RG flow of $\eta$ has no 
fixed point with a finite value of $\eta$.
In our case, the RG flow of $A_h$ will be ruled by
gauge couplings and gaugino masses which can be finite values. In the region II, 
$A_h$ will be affected by mainly $\alpha_{k-1}^nM_{1/2}^{(k-1)}$ in the dimensionful parameters.}.

The theory around the fixed point 
$(g_k,g_{k-1},\eta) =(g_{k}^*,0,0)$ would be well-described by 
its dual theory with the gauge coupling $g_{k-2}$ and 
the yukawa coupling $y$.
The dual theory has the gaugino mass $M_{1/2}^{(k-2)}$, 
soft scalar masses of $q_r$, $\bar q_s$ and 
$M_{rs}$ as $m_q$ and $m_M$, the A-term $a$  and the B-term $b$.
The latter two terms are associated with the superpotential 
(\ref{eq:W-dual}) and are written as 
\begin{equation}
W=y (1 - A_y\theta^2){\rm tr}~\bar q_r M_{rs} q_s 
+ m(1 - B\theta^2){\rm tr} \det_{r,s} M_{rs}.
\end{equation}
Here, we denote $a = yA_y$ and $b=mB$.
The exact matching relations of soft terms between dual theories 
are not clear, but we assume that 
$M_{1/2}^{(k)}(\Lambda_k) \sim M_{1/2}^{(k-2)}(\Lambda_k)$ and 
all of soft scalar masses are of the same order at $\Lambda_k$.
Furthermore, we assume that all of $A_h$, $A_y$ and $B$ are 
of the same order at $\Lambda_k$.

When the gauge coupling $g_{k-2}$ approaches 
toward its non-trivial fixed point, the gaugino mass 
$M_{1/2}^{(k-2)}$ and soft scalar mass squared $m^2_q$ are 
also exponentially suppressed.
This behavior is similar to that of $M_{1/2}^{(k)}$
and $m^2_Q$ discussed previously.
Moreover, in the dual theory the yukawa coupling $y$ 
approaches to the fixed point $y^*$.
In this case, a small deviation $\delta y = y - y^*$ 
is exponentially damping as (\ref{eq:RG-da}).
The spurion method leads that the A-term coupling $A_y$ is also
suppressed exponentially.
On the other hand, the RG behavior of $B$ is rather similar to 
one of $A_h$.
It is found that both RG flow behaviors of 
the mass $m/\mu$ and $b/\mu$ are almost the same and 
they are determined by large anomalous dimensions of 
$M_{rs}$.
However, their ratio $B$ does not change drastically\footnote{Note
 that the RG flow of $m/\mu$ has no fixed point with its fine value.
In this IR region, $B$ will be affected by mainly $\alpha_{k-1}^nM_{1/2}^{(k-1)}$ 
in the dimensionful paramters.}.

In the dual theory, not only $m^2_q$ but also
the sum of soft scalar masses squared, 
$m^2_q+m^2_{\bar q}+m^2_M$, is also suppressed
in the conformal region.
That implies that each of $m^2_q$ and $m^2_M$
is suppressed when $m^2_q =m^2_{\bar q}$, which is the 
relation we are assuming. 
However, we cannot neglect the effects through
$SU(N(k-1))$ gauge interaction such as the discussions of 
convergence points, (\ref{eq:converge-gaugino}) and 
(\ref{eq:converge-scalar}), 
in the original $SU(Nk) \times SU(N(k-1))$ theory.

The gaugino mass $M_{1/2}^{(k-2)}$, the A-parameter $A_y$ 
and the scalar masses squared $m^2_q$ and $m^2_M$ in the dual theory
are not just suppressed out, rather converge to certain
values given by $\alpha_{k-1}$ and 
$M_{1/2}^{(k-1)}$ in the IR limit again.
It is straightforward to solve the RG equations,
if we admit using the one-loop anomalous dimensions
of $q$ and $M$ just as performed above.
However, we shall avoid to present a similar 
analysis here.
It may be explicitly seen that both $m^2_q$ and 
$m^2_M$ as well as $M_{1/2}^{(k-2)}$ and $|A_y|^2$
converge the values of the same order
given by $\alpha_{k-1}|M_{1/2}^{(k-1)}|^2$.
The meson field $M$ belongs to the adjoint
representation of $SU(N(k-1))$ group and
suffers from the effects through
$SU(N(k-1))$ gauge interaction more.
Therefore, $m^2_M$ is found to be positive
and larger than $m^2_q$ in the IR
\footnote{Soft masses for singlet mesons $M^0_{rs}$
may be driven to be negative because of the yukawa couplings.}.


When the supersymmetric mass $m$ of the chiral fields $M_{rs}$ 
becomes large, we integrate out these fields.
Then, the low energy theory becomes the 
$SU((k-1)N) \times SU((k-2)N)$ gauge theory with two 
pairs of bifundamental and anti-bifundamental fields and 
the quartic superpotential, $W = \tilde h {\rm tr} \det q_r \bar q_s$.
The theory has soft SUSY breaking terms, i.e. the gaugino masses, 
$M_{1/2}^{(k-1)}$ and $M_{1/2}^{(k-2)}$, and soft scalar mass 
$m_q$.
In addition, we have the SUSY breaking term corresponding to 
the superpotential $W = \tilde h {\rm tr} \det q_r \bar q_s$, that is, 
\begin{equation}
W = \tilde h (1 - \theta^2 A_{\tilde h}) {\rm tr} \det q_r \bar q_s .
\end{equation}
The size of $A_{\tilde h}$ may be of the order of $B$ or 
$A_y$ at the decoupling scale of $M_{rs}$.
If these SUSY breaking terms are smaller than other mass scales 
such as the energy scale $\mu$ and the supersymmetric mass $m$, 
the above cascade continues as rigid SUSY theory in 
Section \ref{sec:rigid-SUSY}.
Through the cascade, 
the gaugino masses and soft scalar masses are damping 
except the perturbative regime, where the theory moves from 
the fixed point $(g_k,g_{k-1},\eta) =(0,g_{k-1}^*,0)$ toward the fixed point
$(g_k,g_{k-1},\eta) =(g_{k}^*,0,0)$.
When we integrate out $M_{rs}$, which are charged under 
the $SU((k-1)N)$ gauge group, threshold corrections would appear.
For example, the gaugino mass $M_{1/2}^{(k-1)}$ would receive 
such threshold corrections $\Delta M_{1/2}^{(k-1)}$, 
which would be estimated by $\Delta M_{1/2}^{(k-1)}={\cal O}(\alpha_{k-1}B)$.
That would be small, because $\alpha_{k-1}$ is small.
At any rate, if the cascade continues, the total gaugino mass 
$M_{1/2}^{(k-1)}$ would be suppressed at the next stage 
such as the gaugino mass $M_{1/2}^{(k)}$ is suppressed at the stage 
discussed above.

As discussed above, the cascade would continue 
unless SUSY breaking terms are comparable with 
other mass scales such as the energy scale $\mu$ 
and the supersymmetric mass $m$.
Gaugino masses and SUSY breaking scalar masses would be suppressed 
through the cascade except the regime I, where 
the gaugino mass $M_{1/2}^{(k)}$ would increase.
On the other hand, the SUSY breaking parameters, $B$ and $A_h$, 
would not be suppressed like the others.
Note that the B-term corresponds to the off-diagonal entries 
of mass squared matrix of the fields $M_{rs}$, 
that is, eigenvalues of mass squared would be written by 
$|m|^2+m^2_M\pm |mB|$.
A large value of $|B|$ would induce a tachyonic mode.
Then, the scalar component of superfields $M_{rs}$ 
may develop their VEVs and the 
gauge symmetry $SU((k-1)N)$ may be broken.
Also, through this symmetry breaking, the matter fields 
$q_r$ and $\bar q_s$ may gain mass terms due to the yukawa coupling 
with $M_{rs}$.
Then, the duality cascade would be terminated when 
mass parameters, $|m|^2$, $|mB|$ and $m_M^2$, are
comparable \footnote{Similarly, the singlet meson fields $M_{rs}^0$ 
may develop their VEVs depending on values of their various mass terms.
Their VEVs induce mass terms of dual quarks.
If such masses are large enough, the dual quarks would decouple and 
the flavor number would reduce to be outside of the conformal window.
Then, the cascade could end.
In addition, scalar components of $q_r$ and $\bar q_s$ may 
develop their VEVs depending on values the A-terms and 
their soft scalar masses as well as other parameters in the 
scalar potential.
Their VEVs break gauge symmetry and the cascade would end. 
}.
This type of ending of the duality cascade could happen 
only in the softly broken supersymmetric theories and 
such symmetry breaking would be important.
Thus, we will study such breaking more explicitly 
in the next section.
Similar symmetry breaking would be realized 
not only in the ``magnetic dual theory'', but also in 
the original ``electric theory'' with the 
quartic A-term (\ref{eq:Aterm-4}). 
If the quartic A-term is comparable with SUSY breaking scalar masses 
$m_Q$, the origin of scalar potential of $Q$ would be 
unstable and similar symmetry breaking would happen.
Such gauge symmetry breaking with reducing the flavor number 
may correspond to the symmetry breaking by VEVs of $M_{rs}$ 
with inducing dual quark masses.

Whether $M_{rs}$ include tachyonic modes depends on 
values of $|m|^2+m^2_M\pm |mB|$, i.e. their initial 
conditions as well as matching conditions.
In a certain parameter region, the scalar fields 
$M_{rs}$ may include tachyonic modes and symmetry 
breaking may happen.
In the other parameter regions, the cascade would continue
like the rigid supersymmetric theory.
For example, when the magnitude of SUSY breaking terms 
is much smaller than the supersymmetric mass $m$ and 
the energy scale $\mu$, the cascade would continue 
in almost the same way as the rigid supersymmetric theory.
Then, it would end with the pure $SU(N)$ supersymmetric Yang-Mills 
theory with non-vanishing gaugino mass.


\subsection{Symmetry breaking}
\label{sec:symm-br}

In the previous section, we have pointed out the possibility 
that a tachyonic mode in the meson fields $M_{rs}$ 
would appear because of soft SUSY breaking terms and 
its VEV would break the symmetry.
Here, we study this aspect more explicitly.

\subsubsection{$SU(kN) \times SU((k-1)N)$ model}

First, we study the $SU((k-2)N)\times SU((k-1)N)$ theory, which 
is dual to the $SU(N) \times SU((k-1)N)$ theory.
As discussed in the previous section, 
the dual theory includes the meson fields 
$M_{rs}$, which have the supersymmetric mass $m$, 
the SUSY breaking soft scalar masses $m_M$ and 
the B-term $mB$.
That is, their scalar potential $V$ is written by 
\begin{eqnarray}
V &=&  (|m|^2+m_M^2)\sum_{rs}|M_{rs}|^2 + 
(mB (M_{11}M_{22} - M_{12}M_{21}) + h.c.) + V_D^{(k-1)}, \nonumber \\
V_D^{(k-1)}  &=& \frac12g^2_{k-1}D^2_{(k-1)} ,
\end{eqnarray}
where $D_{(k-1)}$ denotes the D-term of the 
$SU((k-1)N)$ vector multiplet.
Here, we have assumed the $SU(2)$ invariance for the $(r,s)$ 
indices of $M_{rs}$.
The eigenvalues of mass squared matrix are given by 
\begin{equation}\label{eq:mass-eigen}
|m|^2+m_M^2 \pm |mB|.
\end{equation}
If $|m|^2 \gg |m_M^2|, |mB| $, the theory is almost supersymmetric 
and the duality cascade would continue.
(Note that $m$ is the supersymmetric mass and the others 
are masses induced by SUSY breaking.) 
However, if the masses (\ref{eq:mass-eigen}) include 
a negative eigenvalue, there appears a tachyonic mode at 
the origin of the field space $M_{rs}$.
Note that the D-flat direction corresponds to the VEV direction, 
where $M_{rs}$ are written by diagonal elements, 
that is, Cartan elements.
That implies that when a negative eigenvalue is included 
in (\ref{eq:mass-eigen}), such a direction would be 
unbounded from below in the tree-level scalar potential.
Thus, the meson fields $M_{rs}$ would develop their VEVs,
whose order would be equal to the cut-off scale 
of the $SU((k-2)N) \times SU((k-1)N)$ theory, i.e. 
$\Lambda_k$.
The VEVs of adjoint fields $M_{rs}$ break the gauge group $SU((k-1)N)$ 
to a smaller group and induce mass terms of 
$q_r$ and $\bar q_s$ through the Yukawa couplings 
$yq_rM_{rs}\bar q_s$.

\subsubsection{$\prod_i SU(N_i)$ quiver model }

\begin{figure}
\begin{center}
{\unitlength=1mm
\begin{picture}(100,45)
\thicklines
\put(23,28){\vector(2,1){25}}
\put(52,40){\vector(2,-1){25}}
\multiput(20,25)(-6,-6){4}{\line(-1,-1){4}}
\multiput(80,25)(6,-6){4}{\line(1,-1){4}}
\put(10,30){$SU(N_1)$}
\put(47,43){$SU(N_2)$}
\put(75,30){$SU(N_3)$}
\put(21,27){\circle{3}}
\put(50,40){\circle{3}}
\put(79,27){\circle{3}}
\end{picture}}
\caption{$\prod_i SU(N_i)$ quiver model}
\label{fig:suni}
\end{center}
\end{figure}
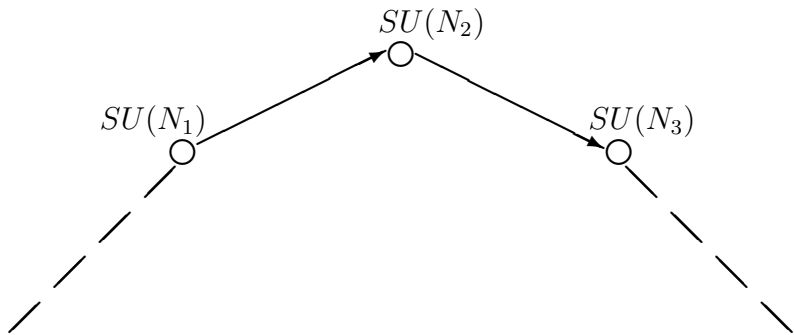

The above analysis can be extended to the $\prod_i SU(N_i)$ 
quiver gauge theory with their bifundamental matter fields.
We consider a subsector of the quiver theory, that is, 
the $SU(N_1) \times SU(N_2) \times SU(N_3)$ gauge theory 
with bifundamental matter fields, 
$(N_1,\bar N_2,1)$ and $(1,N_2,\bar N_3)$ as shown in Fig.~\ref{fig:suni}.
The $SU(N_1)$ and $SU(N_3)$ sectors would have other types of 
bifundamental matter fields, but we neglect them
\footnote{In each non-abelian gauge group, for example, we need vector-like matter fields 
in order to cancel anomaly. However, we assume that the theory is anomaly-free 
at every stage.}.
In addition, for simplicity we consider the case with 
$N_1 = N_3$.
Here, we dualize the $SU(N_2)$ sector.
Then, there appear the dual matter fields $q$ and $\bar q$ with
the representations $(\bar N_1,\tilde N_2,1)$ 
and $(1,\bar {\tilde N_2},N_3)$, where $\tilde N_2 = N_1 - N_2$.
In addition, the meson field $M$ with the 
representation $(N_1,1,\bar N_3)$ appears 
and has yukawa couplings among $q$ and $\bar q$.
The supersymmetric mass term of the meson field 
in the superpotential 
is not allowed, i.e. $m=mB=0$.
In this case, only the SUSY breaking soft scalar mass $m_M$ 
as well as the D-term potentials  appears 
in the scalar potential of the meson field $M$.
Thus, the scalar potential is simple.
The scalar mass squared $m_M^2$ tends to converge to a positive value 
as discussed in the previous section.
Thus, the symmetry breaking may not happen by 
the VEV of $M$ in this theory.

When $N_1 = N_3 =2$, supersymmetric mass terms of meson fields 
in the superpotential would be allowed.
Alternatively, when the model includes anti-meson fields 
$\bar M$, the supersymmetric mass term would be allowed 
in the superpotential.
In these models, the corresponding B-terms would also be allowed.
Furthermore, in the latter model, there are D-flat directions, 
i.e. $M= \pm \bar M$.
In this case, the scalar potential would be written by 
\begin{eqnarray}
V &=&  (|m|^2+m_M^2)|M|^2 + (|m|^2+m_{\bar M}^2)|\bar M|^2 + 
(mB M\bar M + h.c.) \nonumber \\ 
& & + V_D^{(N_1)}+ V_D^{(N_3)}, 
\end{eqnarray}
where $V_D^{(N_1)}$ and $V_D^{(N_3)}$  are D-term scalar potentials 
for the $SU(N_1)$ and $SU(N_3)$ vector multiplets.
This potential at the tree level is unbounded from below along the 
D-flat direction $M= \pm \bar M$ if 
\begin{equation}
2 |m|^2 + m_M^2 + m_{\bar M}^2 < 2|mB|.
\end{equation}
In addition, the meson fields include a tachyonic mode 
when 
\begin{equation}
(|m|^2+m_M^2) (|m|^2+m_{\bar M}^2) < |mB|^2,
\end{equation}
or 
\begin{equation}
(|m|^2+m_M^2) (|m|^2+m_{\bar M}^2) > |mB|^2 
{\rm~~and~~} 2 |m|^2 + m_M^2 + m_{\bar M}^2 < 0.
\end{equation}
Thus, various phenomena could happen depending on 
values of mass parameters, $m$, $m_M$, $m_{\bar M}$ and $mB$, 
that is, the unbounded-from-below direction, 
the symmetry breaking without the unbounded-from-below direction 
or no symmetry breaking.
Indeed, this situation is quite similar to what 
happens in the Higgs scalar potential of the MSSM.

\subsection{Illustrating model}
\label{sec:Illustrating model}

Here we give a simple example of theories, whose 
field contents are  similar to 
the MSSM or its extensions and where symmetry breaking 
would happen.


\begin{figure}
\begin{tabular}{ccc}
\begin{minipage}{0.4\hsize}
\includegraphics[width=\hsize ,clip]{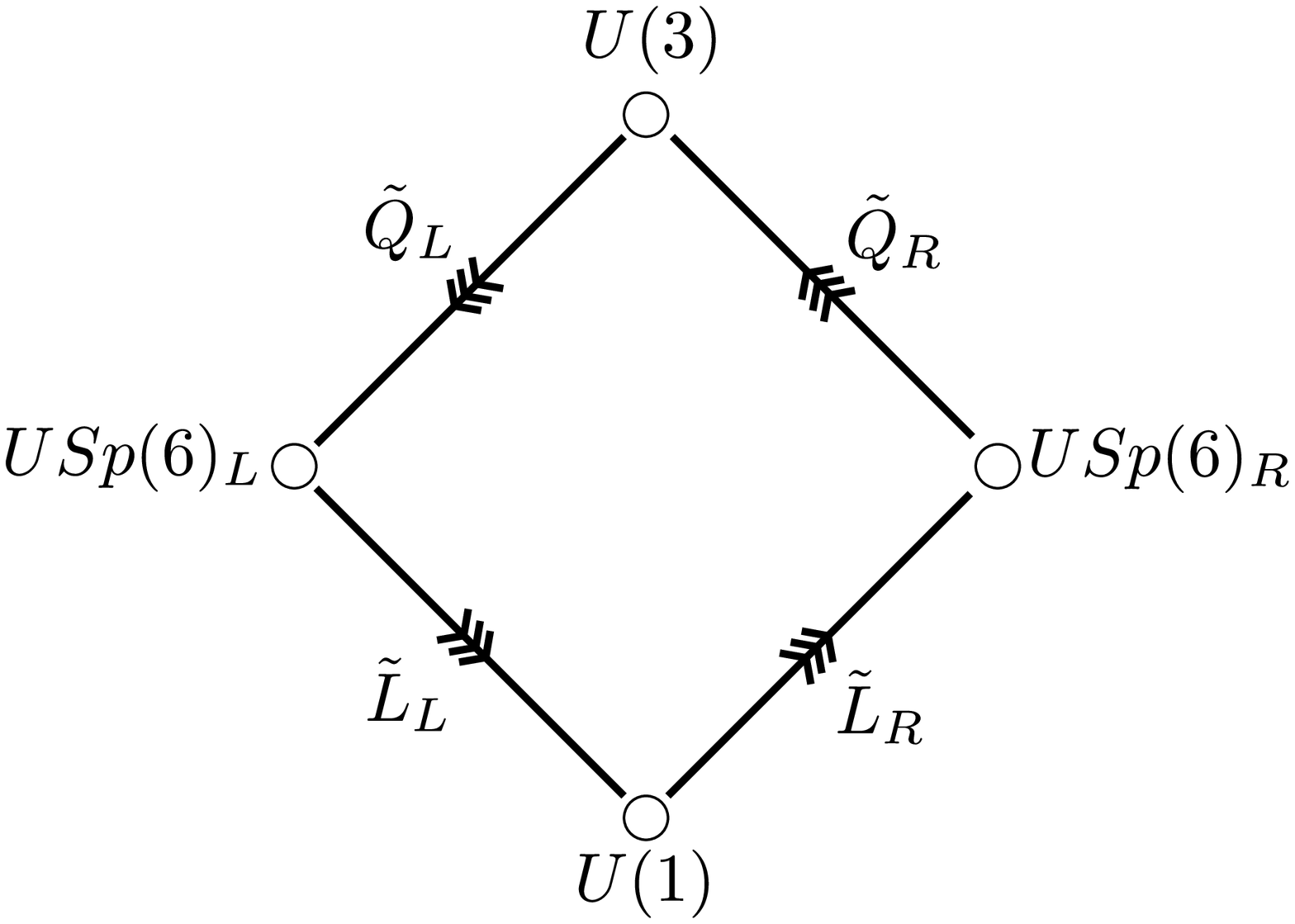}
\end{minipage}
&\begin{minipage}{0.05\hsize}$\rightarrow$\end{minipage}&
\begin{minipage}{0.4\hsize}
\includegraphics[width=\hsize ,clip]{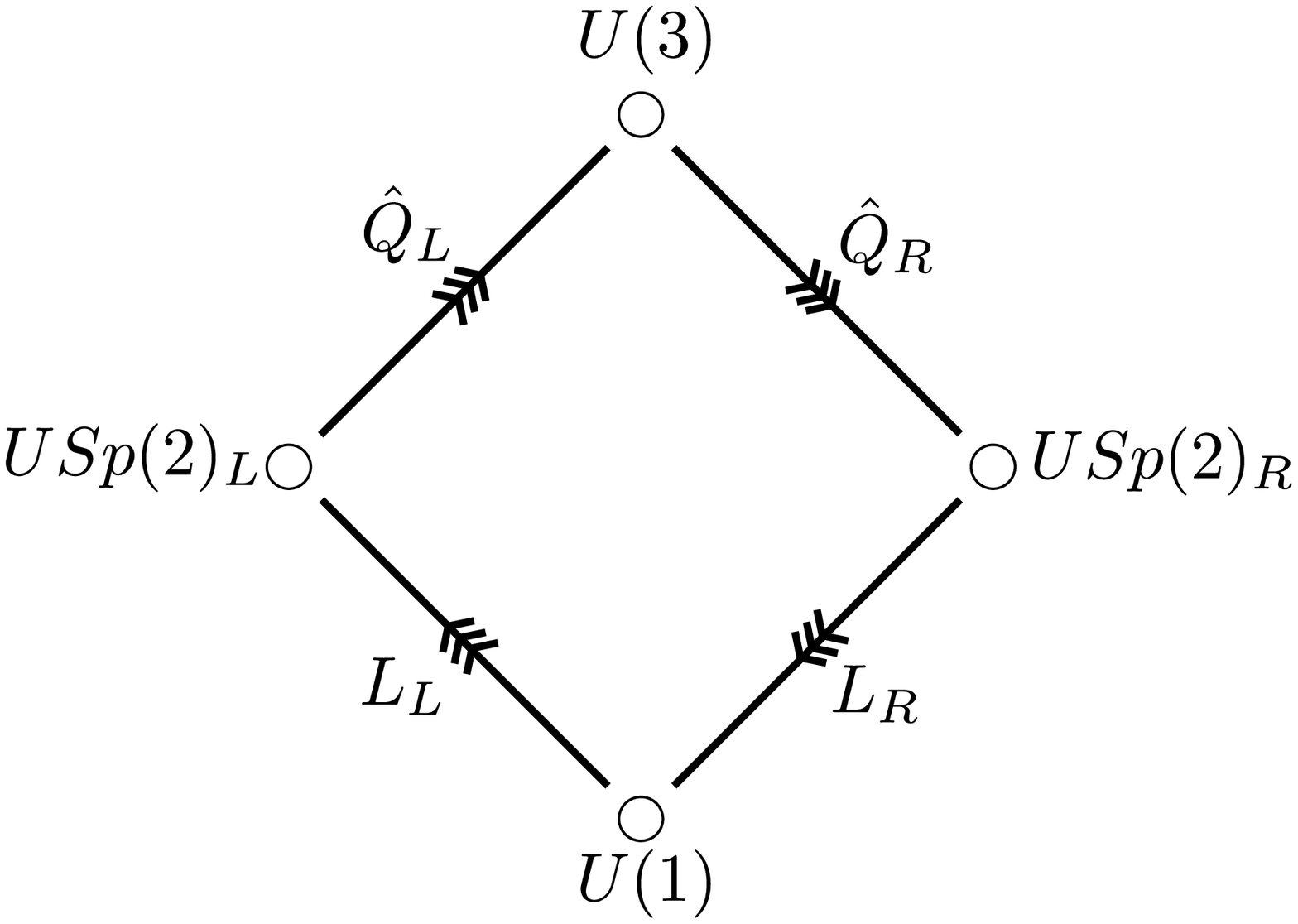}
\end{minipage}\\
&&\\
&\begin{minipage}{0.05\hsize}$\rightarrow$\end{minipage}&
\begin{minipage}{0.4\hsize}
\includegraphics[width=\hsize ,clip]{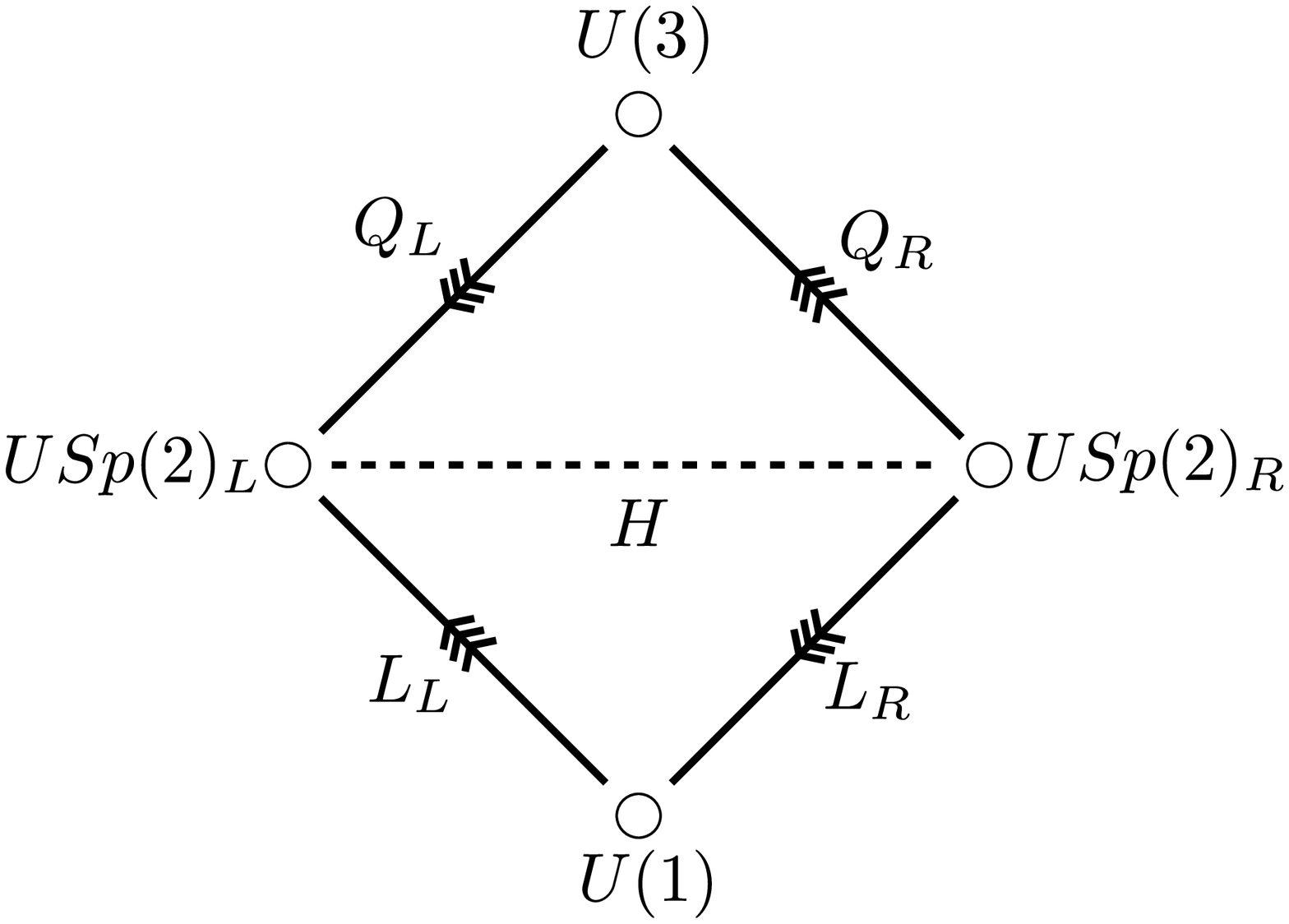}
\end{minipage}
\end{tabular}
\caption{Quiver diagrams of the illustrating model}
\label{fig:u3}
\end{figure}

We consider the gauge group 
$U(3)\times USp(6)_L \times USp(6)_R \times U(1)$ and 
three families of bifundamental fields, 
$\tilde Q_L:(3,6,1,0)$, $\tilde Q_R:(\bar 3,1,6,0)$, $\tilde L_L:(1,6,1,-1)$ 
and $\tilde L_R:(1,1,6,1)$ and the superpotential 
\begin{equation}
W = h \tilde Q_L \tilde Q_R \tilde L_L \tilde L_R.
\end{equation}

We expect that first the gauge couplings of $USp(6)_L \times USp(6)_R$
would approach to their non-trivial fixed point.
Then, the $USp(6)_L \times USp(6)_R$ sector is dualized, that is, 
the gauge group is $U(3)\times USp(2)_L \times USp(2)_R \times
U(1)$ as shown in Fig.~\ref{fig:u3}.
Note that $USp(2) \simeq SU(2)$.
In addition we would have matter fields, 
$\hat Q_L:(\bar 3,2,1,0)$, $\hat Q_R:(3,1,2,0)$, $L_L:(1,2,1,1)$ 
and $L_R:(1,1,2,-1)$.
Also, we would have several ``meson fields'' 
$M:(3,1,1,-1)$ and $\bar M:(\bar 3,1,1,1)$, which have mass terms 
$mM\bar M$ and Yukawa couplings with $\hat Q_L, \hat Q_R, L_L$ 
and $L_R$, but they can be integrated out because of heavy mass terms 
$mM\bar M$.
Then, we obtain the superpotential
\begin{equation}\label{eq:model-1-W-2}
W = \hat h \hat Q_L \hat Q_R  L_L        L_R.
\end{equation}

Next, we expect that the gauge coupling of $SU(3)$ approaches to 
the conformal fixed point.
Then, the $U(3)$ sector is dualized.
The gauge group is $U(3)\times USp(2)_L \times USp(2)_R \times
U(1)$ and we would have matter fields, 
$Q_L:(3,2,1,0)$, $Q_R:(\bar 3,1,2,0)$, $L_L:(1,2,1,1)$ 
and $L_R:(1,1,2,-1)$ as well as several ``Higgs fields'' 
$H:(1,2,2,0)$.
The superpotential is obtained as
\begin{equation}
W = y_Q Q_L Q_R H + y_L L_L L_R H + m HH.
\end{equation}
Note that the operator (\ref{eq:model-1-W-2}) 
corresponds to $y_L L_L L_R H$.
However, the gauge symmetry $U(3)\times USp(2)_L \times USp(2)_R \times
U(1)$ allows the mass terms $mHH$.
Thus, we assume that such terms would be induced and 
we have added such terms.
Then, if SUSY breaking terms induce a tachyonic mode of $H$, 
the symmetry $USp(2)_L \times USp(2)_R$ would be broken.

In this model, $USp(2)_L$ and $USp(2)_R$ symmetry 
breaking would happen at the same time.
Although the left-right asymmetry is required for 
a realistic model, it would be difficult to generate such 
left-right asymmetry in this model.
Some modification is necessary for a realistic model. 
At any rate, this model is an illustrating model 
for symmetry breaking.
Such symmetry breaking by SUSY breaking terms in 
the duality cascade may be important, e.g. 
to realize the standard model at the bottom of the 
cascade.

\subsection{Conclusion in section \ref{sec:duality cascade}}
\label{sec:conclusion dual}

We have studied the RG flow of softly broken 
supersymmetric theories showing the duality cascade. 
We find that conformal dynamics realizes approximate R-symmetry in the duality cascade as the discussion in section 4 as follows.
Gaugino masses, A-term and scalar masses are suppressed 
in most regime of the RG flow although they increase 
in a certain perturbative regime.
After exponential damping, the gaugino mass 
$M^{(k)}_{1/2}$ corresponding to the strongly 
coupled sector converges to a certain value, 
which is determined by the gauge coupling $\alpha_{k-1}$ 
and the gaugino mass $M_{1/2}^{(k-1)}$ in 
the weakly coupled sector.
The scalar mass would also converge to the same order value.
At the next stage of the cascade, 
the strongly and weakly coupled sectors are 
interchanged with each other and 
the gaugino mass $M_{1/2}^{(k-1)}$ 
would be suppressed.
Thus, through the sequential cascade, 
the magnitude of gaugino masses and scalar masses would be 
suppressed.

The B-term may be important.
In a certain parameter region, the B-term would 
induce tachyonic modes of $M_{rs}$ and symmetry 
breaking would happen.
Such an aspect would be important to realistic model building.

The RG flow of SUSY breaking terms in the cascading theory 
is quite non-trivial 
as the RG flow of gauge couplings.
The gravity dual of the cascade rigid supersymmetric theory has 
been studied extensively.
However, our analysis implies that the dilaton is also running as
$e^{-\phi} \sim \alpha_k^{-1} +  \alpha_{k-1}^{-1}$ \cite{Klebanov:2000hb},
but the supergravity solution of the D3-brane does not admit this running behavior
and most of the supergravity dual theories concentrate on the constant dilaton backgrounds.
In this sense, the suppression of the gaugino masses would be a quite different mechanism
from the suppression due to the warp factor as already pointed out in
\cite{DeWolfe:2002nn,Kachru:2003aw,Giddings:2005ff,Douglas:2007tu}.
The region of RG flow in our study might 
be outside of the supergravity approximation,
but it would be quite interesting to study the 
gravity dual side corresponding to the RG flow of 
SUSY breaking terms including the dilaton running.

We have considered the scenario that supersymmetry is 
broken at high energy and investigated the RG flow of 
SUSY breaking terms.
Alternatively, we could consider another scenario that 
supersymmetry would be broken at some stage of 
the cascade.
For example, supersymmetry is broken dynamically 
through the cascade and such breaking is mediated to 
the visible sector.
Such a study would also be important.

\section{Summary}

In section 2, we studied SUSY breaking in local and global supersymmetric theory. The NS argument suggests that $U(1)_R$ symmetry plays an important role in building the models which cause SUSY breaking. We discussed the features of R-symmetric models which have SUSY breaking vacua in section \ref{sec:global-SUSY}. We suggest that it is the sufficient condition for SUSY breaking that the number of fields with R-charge 2 is larger than the number of $U(1)_R$-invariant independent operators which couple with the fields with R-charge 2. In the model, there is no supersymmetric vacuum and the minimum of the potential is given by the F-terms of the fields with R-charge 2. On the other hand, there are runaway directions in the model with fields whose R-charges are negative and/or more than 2.

In section \ref{sec:rsugra explicit R-breaking in global SUSY}, we studied the effect of explicit R-symmetry breaking terms in global supersymmetric models. Based on the argument by ISS, 
we have shown that certain classes of explicit R-symmetry breaking 
terms can restore SUSY, and the original SUSY breaking vacuum 
can become metastable.

In section \ref{sec:rsugra}, we discussed SUSY breaking in local supersymmetric models.        
It is a challenging 
issue in supergravity models to realize 
the almost vanishing vacuum energy.
The vacuum energy may be tuned to vanish, e.g., by 
the constant superpotential term, which 
is a sizable R-symmetry breaking term.
That would affect all of vacuum structure such as 
metastability of SUSY breaking vacua and 
presence of SUSY preserving vacua.
We studied this vacuum structure by 
using the generalized O'Rraifeartaigh model added certain classes of R-symmetry breaking 
terms such that the vanishing vacuum energy is 
realized. We found that the metastability of SUSY breaking vacua can be avoided by limiting classes of R-symmetry breaking 
terms.

In section 3, we argued that conformal dynamics causes SUSY breaking vacua based on the argument in section 2. We find that fields with R-charge 2 play a key role in SUSY breaking and their quadratic and trilinear couplings destabilize the SUSY breaking vacua in explicit R-symmetry breaking models. Conformal dynamics can realize the suppression of unpleasant terms, so that it causes long-lived metastable SUSY breaking vacua.
This is because conformal dynamics makes approximate R-symmetry recovered at a low energy scale. We found that approximate R-symmetry can be realized even in softly SUSY breaking theories in section 4. In fact, gaugino mass and A-term, which are soft SUSY breaking and explicit R-symmetry breaking terms, are suppressed corresponding to a gauge coupling and yukawa couplings going to IR fixed points. These suppressions appear even in more complicated, such as the duality cascade. Conformal dynamics allows Seiberg dual near an IR fixed point, and we call sequential Seiberg dual the duality cascade. In section \ref{sec:duality cascade}, we found that gaugino masses and A-term are suppressed as long as the cascade continues. However, B-term remains to be a finite value at a low energy scale, so there is a possibility that the B-term causes gauge symmetry breaking corresponding to EW symmetry breaking. We suggest a scenario that a SM-like model appear at the bottom of the duality cascade, based on the above arguments.

Conformal dynamics plays important roles in 
various aspects of (supersymmetric) field theories 
and particle phenomenology, and other good aspects of models with conformal dynamics are also suggested. For example, conformal dynamics realizes
not only suppressions of flavor-dependent SUSY breaking terms  \cite{Luty:2001jh,Dine:2004dv,Sundrum:2004un,Ibe:2005pj,Schmaltz:2006qs,
Murayama:2007ge,Cohen:2006qc}, but also realistic hierarchies of quark and lepton masses 
\cite{Kobayashi:2001kz,Nelson:2000sn,Kobayashi:2001is}. Our models, which we introduce in section 3, 4 and 5, are also expected to realize the interesting aspects for phenomenology. However, we have not yet discussed application of our model, especially in section 5, for phenomenology explicitly. This is our future work.

\vspace{1cm}
\noindent
{\bf Acknowledgement}

The author would like to thank H. Abe, T. Higaki, T. Kobayashi, K. Ohta and H. Terao for fruitful collaborations.
The work of Y.O. is supported by the Japan 
society of promotion of science (No. 20$\cdot$324).

\appendix
\section{Supersymmetric masses involving R-axion}
\label{app:rmass}
In this appendix, we show some general results for the 
SUSY masses for the scalar component of an 
R-axion multiplet. For this analysis, it is convenient 
to redefine the R-charged superfield $Y$ by 
\begin{eqnarray}
R &=& \frac{2}{q_Y} \ln Y, 
\nonumber
\end{eqnarray}
where $R$ can be interpreted as the R-axion superfield. 
(Note that $R=-aT$ in Eq.~(\ref{eq:def:t}).) 
In this basis, the K\"ahler potential and the superpotential 
(\ref{eq:rsp1}) is written as 
\begin{eqnarray}
K &=& K(R+\bar{R},\chi_i,\bar\chi_i), 
\nonumber \\
W &=& e^R \zeta(\chi_i). 
\label{eq:rsp4}
\end{eqnarray}
{}From Eq.~(\ref{eq:rsp4}), we find 
$W^{-1}\partial_Y^m W=1$ where $m=1,2,.\ldots$, and obtain 
\begin{eqnarray}
G_{RR} &=& K_{RR}+W^{-1}W_{RR}-(W^{-1}W_R)^2 
\ = \ K_{RR} \ = \ K_{R\bar{R}}, 
\nonumber \\
G_{\chi_i R} &=& K_{\chi_i R}+W^{-1}W_{\chi_i R}
-(W^{-1}W_{\chi_i})(W^{-1}W_R) 
\ = \ K_{\chi_i R} \ = \ K_{\chi_i \bar{R}}. 
\nonumber
\end{eqnarray}
Substituting these into the general formulae for the second 
derivatives at the SUSY point, 
\begin{eqnarray}
V_{I\bar{J}} \Big|_{D_KW=0} 
&=& e^G (G^{M \bar{N}}G_{MI}G_{\bar{N}\bar{J}}-2G_{I\bar{J}}), 
\nonumber \\
V_{IJ} \Big|_{D_KW=0} 
&=& -e^G G_{IJ}, 
\nonumber
\end{eqnarray}
we find 
\begin{eqnarray}
V_{R\bar{R}} \Big|_{D_KW=0} &=& 
V_{RR} \Big|_{D_KW=0} \ = \ -K_{R\bar{R}}\,m_{3/2}^2, 
\label{eq:vrr} \\
V_{\chi_i \bar{R}} \Big|_{D_KW=0} &=& 
V_{\chi_i R} \Big|_{D_KW=0} \ = \ -K_{\chi_i \bar{R}}\,m_{3/2}^2, 
\label{eq:vpr}
\end{eqnarray}
where $m_{3/2}^2=e^G$ is the gravitino mass square.

{}From Eq.~(\ref{eq:vrr}), the mass squared eigenvalues of 
$({\rm Re}\,R,\,{\rm Im}\,R)$ can be calculated as 
$0$ and $-2m_{3/2}^2$ with the canonical kinetic terms 
normalized by $K_{R\bar{R}}>0$. 
The first massless eigenmode corresponds to the R-axion scalar 
associated to the spontaneously broken global $U(1)_R$ symmetry. 
The second negative-definite eigenvalue indicates that the special 
SUSY solution (\ref{eq:susyvc2}) is at best a saddle point 
solution. Note that the gravitino mass $m_{3/2}$ is nonvanishing at 
this point and the vacuum energy is negative. We also find from 
Eq.~(\ref{eq:vpr}) that the mixing-mass between $R$ and $\chi_i$ 
is vanishing if the K\"ahler (kinetic) mixing is vanishing, 
$K_{\chi_i \bar{R}}=0$. In this case, the R-axion direction is 
completely separated from the other fields $\chi_i$, that is,
the above mass eigenvalues of R-axion multiplet become exact 
in this case. 

Finally we comment that the second derivatives (\ref{eq:vrr}) and 
(\ref{eq:vpr}) are all vanishing at the SUSY point (\ref{eq:susyvc1}) 
where $m_{3/2}=0$. In this case, both the real and the imaginary 
scalar component of R-axion multiplet remain massless. 
Note that Eq.~(\ref{eq:susyvc2}) may also allow a solution 
even in this case if $\zeta$ is {\it not} a generic function.

\section{Supergraph formalism}
\label{app:superfield}

We discuss superfield propagators and Feynman rules for supergraphs, based on Ref.\cite{Grisaru:1979}.
Let us consider the Langarangian of $\Phi$, chiral superfield, as follows.

\beq
\begin{split}
L=& \int d^4 \theta \{\frac{1}{2} \begin{pmatrix} \Phi & \Phi^{\dagger } \end{pmatrix}  M \begin{pmatrix} \Phi \\ \Phi^{\dagger} \end{pmatrix} \\
M=& \begin{pmatrix} -\frac{m}{4 \square } D^2 & 1 \\ 1 & -\frac{m}{4 \square } \ov{D}^2 \end{pmatrix} +L_{int}  
\end{split}
\eeq

We define the generating function for superfield Green functions,

\beq
\begin{split}
Z[J,J^{\dagger}] = &\bigl\langle 0 \bigr| T exp [ i \int d^4 \theta d^4 x \{ J(z) \left( - \frac{1}{4} \frac{D^2}{ \square}  \right) \Phi(z)  \\
  &+J^{\dagger} (z) \left( - \frac{1}{4} \frac{D^2}{ \square}  \right) \Phi^{\dagger} (z)   \}] \bigl| 0 \bigl \rangle
\end{split}
\eeq

, where $z$ is defined as $z=(x,~\theta,~\ov{\theta})$ and we use the below relation,
\beq
\frac{\delta }{\delta J(x, \theta, \ov{\theta}) } J(x', \theta', \ov{\theta}') =- \frac{1}{4}\ov{D}^2 \delta ^4 (\theta- \theta')\delta^4(x-x').
\eeq
$D$ and $\ov{D}$ are covariant derivative, and also satisfy the identities,
\beq
\begin{split}
\int d^4x d^2 \theta (- \frac{1}{4}\ov{D}^2 f(z)) =& \int d^4x d^4 \theta f(z) \\
\int d^4x d^2 \theta  f(z)  =& \int d^4x d^4 \theta ( - \frac{1}{4} \frac{ D^2}{\square} )f(z).
\end{split}
\label{id:covariant}
\eeq

We define $Z_0[J,~J^{\dagger }]$ as the generating functional for free superfield Green's functions, and $Z[J,~J^{\dagger }]$ is described as
\beq
Z[J,~J^{\dagger }] = exp \{ i \int d^4 x L_{int} \left( \frac{\delta }{\delta J}, \frac{\delta }{\delta J^{\dagger}} \right) \} Z_0 [J,~J^{\dagger }].
\eeq
We find $Z_0[J,~J^{\dagger }]$,
\beq
Z_0[J,~J^{\dagger }] = exp \{ -\frac{i}{2} \int d^4 x d^4 \theta d^4 x' d^4 \theta' \begin{pmatrix} J & J^{\dagger} \end{pmatrix} \Delta_{GRS} (z,z') \begin{pmatrix} J \\ J^{\dagger} \end{pmatrix}  \},
\eeq
where $\Delta_{GRS}$ is the propagator introduced by Grisaru, Ro\v{c}ek and Segel \cite{Grisaru:1979}:
\beq
\Delta_{GRS} (z-z') = \frac{1}{ \square -m^2} \begin{pmatrix} -\frac{m}{4 \square } D^2 & 1 \\ 1 & -\frac{m}{4 \square } \ov{D}^2 \end{pmatrix} \delta (z-z').
\eeq 
We consider renormalizable superpotential, $W(\Phi)= \frac{m}{2} \Phi^2 +  \frac{\lambda}{6}  \Phi^3$, so that $L_{int}$ is given by 
\beq
L_{int}\left( \frac{\delta }{\delta J}, \frac{\delta }{\delta J^{\dagger}} \right) = \int d^4x d^2 \theta \left( \frac{1}{i} \frac{\ov{D}^2 D^2}{16 \square} \frac{\delta }{\delta J} \right) +~h.c.~,
\eeq 
where we use 
\beq
\begin{split}
\ov{D} D^2 \ov {D}^2 = & 16 \square \ov{D}^2 \\
D^2 \ov{D}^2 D^2 = & 16 \square D^2. 
\end{split}
\label{eq:covariant}
\eeq  
The vertexes can be obtained form the following formula by using (\ref{id:covariant}),
\beq
\begin{split}
\int d^4x d^4 \theta & L_{int}\left( \frac{\delta }{\delta J}, \frac{\delta }{\delta J^{\dagger}} \right) J(z_1)J(z_2)J(z_3) = - \frac{\lambda}{6}\int d^4x d^2 \theta \left( \frac{\delta}{ \delta J(z)} \right)^3 J(z_1)J(z_2)J(z_3) \\
=&- \lambda \int d^4x d^4 \theta \delta^8(z_1-z) \{- \frac{1}{4}\ov{D_2}^2 \delta^8(z_2-z) \} \{- \frac{1}{4}\ov{D_3}^2 \delta^8(z_3-z) \}.
\end{split}
\eeq
On the other hand, the propagator of vector superfield is described as 
\beq
\Delta_V(z,~z')^{AB}= - \frac{1}{\square} \delta^{AB} \delta(z-z'). 
\eeq
 
Now we find out Feynman rules for supergraph. 
Each vertex includes a factor of $-\frac{1}{4}\ov{D}^2$ or $-\frac{1}{4} D^2$ acting on each chiral (or antichiral) superfield, but we omit one $-\frac{1}{4}\ov{D}^2$ (or $-\frac{1}{4} D^2$) and integrate $\int d^4x d^4 \theta$. 
The amputated one-particle-irreducible graphs in the effective action should have each amputated external line, so $-\frac{1}{4}\ov{D}^2$ (or $-\frac{1}{4} D^2$) should be omitted at a vertex for each external chiral (or antichiral) superfield.

The Feynman rules and the GRS propagators give a loop graph with n-th vertexes an expression of the following form:
\beq
(D_1^2)^{l_1}(\ov{D_1})^{k_1}\delta^4(\theta_1-\theta_2)(D_2^2)^{l_2}(\ov{D_2})^{k_2}\delta^4(\theta_2-\theta_3) ....(D_n^2)^{l_n}(\ov{D_n})^{k_n}\delta^4(\theta_n-\theta_1),
\label{eq:supergraph-loop}
\eeq
where $l_i,~k_i$ are $0$ or $1$, and we use (\ref{eq:covariant}). It is integrated over $ d^4 \theta_1 ..d^4 \theta_{n}$. We find the n-th point Green functions are always given by the form of

\beq
\int d^4 \theta \Pi_{i=1}^n d^4x_i G_n(x_1,..., x_n) f(\Phi,D \Phi,..).
\eeq 
The function f depends on superfields and covariant derivatives, so that this result leads to non-renormalization of the superpotential, since the form of each loop graph is $\int d^4 \theta$. 

Furthermore, we discuss power counting rules to look into the degree of the divergence. 
All renormalizable vertexes go as $D^4 \sim p^2$, and external chiral lines go as $1/D^2 \sim 1/p$. Loops go as $d^4p/D^4 \sim p^2$, because 4 $D$ are used to cancel a loop's delta function, such as $\delta^4(\theta_n-\theta_1)$ in (\ref{eq:supergraph-loop}).  
Eventually, the degree of the divergence, $\omega $, is given by
\beq
\omega =2L-2P-C+2V-E=2-C-E,
\eeq
where $L$, $P$, $C$, V, and $E$ denote the number of Loops, propagators $((\Delta_{GRS})_{11,22})$, chiral propagators$((\Delta_{GRS})_{12,21})$, vertexes, and chiral external lines respectively, and we use $L=2P-V+1$.
This result means that we find the divergence is at most logarithmic.\footnote{D-term does not receive radiative corrections, as far as gauge symmetry is preserved.} 

SUSY breaking terms can be incorporated into the superspace perturbation by using spurion superfields we introduce in the Sect.\ref{app:spurion}.
In the spurion formalism, the spurion superfields, which include the SUSY breaking terms as the components, are treated as external fields in perturbation. This does not destroy the above argument, so that the absence of quadratic divergence is ensured.

\section{Spurion Technique}
\label{app:spurion}
We discuss the spurion method, based on Ref. \cite{Jack:1997pa,Terao:2001jw}.
The soft SUSY breaking terms, which does not cause quadratic divergence, are restricted to the following, 
\beq
\begin{split}
L_{soft} = &-\int d^4 \theta (m^2_{ij} \theta^2 \ov{\theta}^2) \Phi^{i\dagger}\Phi^j - \int d^2 \theta (M \theta^2)\frac{1}{g^2} Tr(W^{\al} W_{\al}) \\ 
&-\int d^2 \theta \frac{1}{2} (\mu_{ij} \theta^2) \Phi^i \Phi^j - \int d^2 \theta \frac{1}{6} (h_{ijk} \theta^2)\Phi^i \Phi^j \Phi^k + h.c.
\end{split}
\label{eq:soft}
\eeq
It is possible that we check these are soft SUSY breaking terms according to appendix {app:superfield}.
First, we concentrate on the contributions of yukawa couplings, A-term $h_{ijk}$ and scalar mass terms $m_i^2$.   
Based on the non-renormalization of the superpotential, the effective Lagrangian in this case is given by,
\beq
\begin{split}
L_{eff}=& \int d^4 \theta \tilde{Z}_i (\theta, \ov{\theta}) \Phi_i^{\dagger} \Phi_i +\int d^2 \theta \frac{1}{2} (m_{0ij}-\mu_{ij} \theta^2) \Phi^i \Phi^j    \\
&+ \int d^2 \theta \frac{1}{6} (y_{0ijk}-h_{0ijk} \theta^2)\Phi^i \Phi^j \Phi^k + h.c.  
\end{split}
\eeq
Since the renormalization of (anti-)chiral superfields must be also (anti-)chiral,
$\tilde{Z}_i (\theta, \ov{\theta})$ should be described as the following form,
\beq
\tilde{Z}_i (\theta, \ov{\theta}) = Z_i(\ov{\theta})^{\dagger} (1-m^2_i\theta^2 \ov{\theta}^2)Z_i(\theta).
\eeq
Under this description, the renormalized yukawa coupling superfields can be defined as 
\beq
Y_{ijk}(\theta) = y_{ijk}-h_{ijk}\theta^2 =Z_i^{-1}(\theta)Z_j^{-1}(\theta)Z_k^{-1}(\theta) (y_{0ijk}-h_{0ijk} \theta^2).
\eeq
These correspond to field redefinitions of (anti-)chiral superfields. $Y_{ijk}(\theta)$ are treated as external fields in perturbation.  
Furthermore, we find that the superfield propagators in the softly broken theories are modified from $\Delta_{GRS}$,
\beq
\Delta_{soft}^{ij}=(1+\frac{1}{2}m_i^2\theta^2 \ov{\theta}^2) \Delta_{GRS}^{ij}(1+\frac{1}{2}m_j^2\theta^2 \ov{\theta}^2).
\eeq
In other words, this modification means that yukawa couplings are modified as follows,
\beq
\tilde{y}_{ijk}=Y_{ijk}+ \frac{1}{2}(m_i^2+m_j^2+m_k^2)y_{ijk} \theta^2 \ov{\theta}^2.
\label{eq:spurion yukawa}
\eeq
Eventually, the dependence of $\tilde{Z}_i(\theta,~\ov{\theta})$ is given by,
\beq
\tilde{Z}_i(\theta,~\ov{\theta})=\tilde{Z}_i(\tilde{y}_{ijk},\bar{\tilde{y}}_{ijk}).
\label{eq:softly broken wave}
\eeq
We define $Z_i$ as the wave function renormalization factors of $\Phi_i$ in supersymmetric models without soft terms, so that $\tilde{Z}_i$ satisfy $\tilde{Z}_i(y_{ijk},\bar{y}_{ijk})=Z_i(y_{ijk},\bar{y}_{ijk})$. 
    
We can also discuss gauge coupling and gaugino mass, according to the above argument. 
Before the discussion about the soft term, we review the RG flow of gauge coupling.
The holomorphic gauge coupling $S$ is renormalized only at 1-loop \cite{Weinberg:1998uv}.
The RG equation for the holomorphic coupling is given by  
\beq
\mu \frac{d S}{ d \mu} = \frac{1}{16 \pi^2}(3T_G- \sum_iT_i),
\eeq
where $T_G=C_2(G)$ and $T_i=T(R_i)$ for the gauge representation $R_i$ of the chiral superfield $\Phi_i$.
The relation between the holomorphic gauge coupling and the physical gauge coupling $\al$ is given by
\beq
\begin{split}
&8 \pi^2 (S+S^{\dagger}) - \sum_{i}T_i \ln Z_i =F_g (\al) \\
&F_g(\al)=\frac{1}{\al}+T_G \ln \al +\sum_{n > 0} a_n \al^n,
\end{split}
\label{eq:exact gauge coupling}
\eeq
where $\al=\frac{g^2}{8 \pi^2}$ and $a_n$ are scheme dependent constants. The NSVZ scheme corresponds to the case of $a_n=0$ for all $n$. Eventually, we find the relation (\ref{eq:exact gauge coupling}) gives the exact beta function, 
\beq
\mu \frac{d}{d \mu} \al = \beta_{\al}= \frac{1}{F'_g(\al)} \{ 3T_G - \sum_{i}T_i (1- \gamma_i) \},
\eeq
where $\gamma_i$ denotes the anomalous dimension for $\Phi_i$,
\beq
\gamma_i= - \mu \frac{d \ln Z_i}{d \mu}.
\eeq

Now we discuss the soft SUSY breaking term, gaugino mass, according to the discussion about the scalar masses and A-term.
In softly broken theory, the holomorphic gauge coupling is modified as follows,
\beq
\tilde{S}= \frac{1}{g_h^2}(1-2M_{1/2} \theta^2). 
\eeq
Furthermore, (\ref{eq:softly broken wave}) and (\ref{eq:exact gauge coupling}) gives the following description,
\beq
8 \pi^2 (\tilde{S}+\tilde{S}^{\dagger}) - \sum_{i}T_i \ln \tilde{Z_i} =F_g (\tilde{\al}).
\eeq
where $\tilde{\al}$ is defined as follows,
\beq
\tilde{\al}=\al(1+ M_{1/2} \theta^2 + \ov{M}_{1/2} \ov{\theta}^2 +(2|M_{1/2}|^2+ \Delta_g)\theta^2 \ov{\theta}^2),
\label{eq:spurion gauge}
\eeq
and $\Delta_g$ is given by 
\beq
\Delta_g= \frac{1}{\al F'_g(\al)} \{ \sum_i T_i m_i^2 -(\al^2 F'_g(\al))'|M_{1/2}|^2 \}.
\eeq
On the other hand, we find the dependence of $\tilde{Z}_i$ by applying the argument about yukawa coupling, 
\beq
\tilde{Z}_i(\theta, \ov{\theta})= \tilde{Z}_i( \al, \tilde{y}_{ijk}, \bar{\tilde{y}}_{ijk}).
\eeq
Finally, we find RG equations of soft SUSY breaking terms we introduce in Sect.4. 
We define the beta functions of gauge coupling and yukawa coupling as Sect.4,
\begin{equation}
\label{eq:spurion-RG2}
\mu \frac{d \tilde \alpha}{d \mu} = \beta_\alpha(\tilde \alpha, \tilde
y_{ijk},  \bar{\tilde{y}}_{ijk}), \qquad 
\mu \frac{d \tilde y_{ijk}}{d \mu} = \beta_{y_{ijk}}(\tilde \alpha, \tilde
y_{ijk},  \bar{\tilde{y}}_{ijk}).
\end{equation}
The beta-function of the gaugino mass $M_{1/2}$ is obtained as 
\begin{equation}
\mu \frac{dM_{1/2}}{d\mu} = \left( M_{1/2}\alpha 
\frac{\partial}{\partial \alpha}  - a_{ijk}  
\frac{\partial}{\partial y_{ijk}}  \right) \left(  
\frac{\beta_\alpha}{\alpha}\right)
\equiv
D_1 \left(  
\frac{\beta_\alpha}{\alpha}\right).
\end{equation}
The RG equation for the A-term is
\beq
\mu \frac{d h_{ijk}}{d\mu} = \frac{1}{2} (\gamma_i+\gamma_j+\gamma_k)h_{ijk}-(D_1\gamma_i+ D_1\gamma_j+ D_1\gamma_k)y_{ijk}. 
\eeq 
The RG equation for the soft scalar mass $m_i$ of a
chiral superfield $\phi_i$ is obtained as
\begin{equation}
\mu \frac{d m^2_i}{d \mu} =
\left. 
\gamma_i(\tilde{\alpha}, \tilde{y}_{ijk}, \bar{\tilde{y}}_{ijk})
\right|_{\theta^2 \bar{\theta}^2}.
\end{equation}
This leads the following,
\begin{eqnarray}
\mu \frac{d m^2_i}{d \mu} &=& D_2 \gamma_i \ ,\\
D_2 &=& D_1 \bar{D}_1 
+ (|M_{1/2}|^2 + \Delta_g)\alpha \frac{\partial}{\partial \alpha} 
\nonumber \\
& &
+ \frac{1}{2}(m^2_i + m^2_j + m^2_k)
\left(
y_{ijk }\frac{\partial}{\partial y_{ijk}}
+
\bar{y}_{ijk }\frac{\partial}{\partial \bar{y}_{ijk}}
\right) .
\end{eqnarray}

The above results, such as the coupling superfields given by (\ref{eq:spurion gauge}) and (\ref{eq:spurion yukawa}) are also supported by the symmetry argument \cite{ArkaniHamed:1998kj,Giudice:1997ni}. Once we suppose that the coupling superfields are dynamical, then the softly broken theories have a global $U(1)_{\Phi_i}$ symmetry corresponding to each chiral superfield $\Phi_i$, 
\beq
\begin{split}
\Phi_i \rightarrow & e^{T_i} \Phi_i, \\
\tilde{Z}_i \rightarrow & e^{-T_i^{\dagger}} \tilde{Z}_i e^{-T_i} \\
Y_{ijk} \rightarrow & e^{-T_i} Y_{ijk} \\
S \rightarrow & S - \frac{T}{8 \pi^2} \sum_i T_i,
\end{split}
\eeq
where $T_i$ is also a chiral superfield. $Y_{ijk} \tilde{Z}_i^{-1}Y^{\dagger}_{ijk}$ is given as an invariant form. This leads the deformation of yukawa couping in (\ref{eq:spurion yukawa}). The symmetry also gives the deformation of gauge coupling in (\ref{eq:spurion gauge}).

\end{document}